\documentclass[english,aps,reprint,prb,groupedaddress,nofootinbib]{revtex4-1}

\usepackage[T1]{fontenc}
\usepackage[utf8]{inputenc}
\usepackage{amsfonts,amstext,amssymb,amsmath,amsthm,mathtools}
\usepackage{physics,braket}
\usepackage{tabularx,multirow}
\usepackage{hyperref}

\setcitestyle{numbers,square}

\newcommand{\Heff}{H_{\textrm{eff}}}
\newcommand{\en}{\varepsilon}
\newcommand{\Den}{\Delta\en}

\newcommand{\td}{\tilde{t}}
\newcommand{\hd}{\tilde{h}}

\newcommand{\Fig}[2]{%
	\parbox[c]{\hsize}{\includegraphics[width=#1\columnwidth]{Figures/#2}}
}

\DeclarePairedDelimiter\pair{\langle}{\rangle}
\DeclarePairedDelimiter\ceil{\lceil}{\rceil}
\DeclarePairedDelimiter\floor{\lfloor}{\rfloor}

\begin{document}

\title{Perturbative approach to tunneling and quantum interferences in spin clusters}

\author{Ivo A. Maceira$^{1}$, Frédéric Mila$^{1}$, Markus Müller$^{2}$}
\affiliation{$^{1}$Institute of Physics, Ecole Polytechnique F\'ed\'erale de
Lausanne (EPFL), CH-1015 Lausanne, Switzerland}
\affiliation{$^{2}$Paul Scherrer Institute, Condensed Matter Theory, PSI Villigen, Switzerland}

\date{\today}

\begin{abstract}
	Collective tunneling is a ubiquitous phenomenon in finite-size spin clusters that shows up in systems as diverse as molecular magnets or spin clusters adsorbed at surfaces. The basic problem we explore is to understand how small flipping terms can cooperate to flip a large spin to the opposite direction or a cluster of interacting elementary Ising spins into the time-reversed state. These high order processes will involve at least two channels, a single spin-flip channel due to a transverse field and a two-spin flip channel due to exchange or other pairwise interactions, or due to single-ion anisotropies. In view of the complexity of high order perturbation theory, non perturbative approaches based on large-spin path integrals were developed when this problem was first addressed in the context of single spin models. In the present paper, we show that high-order perturbation theory can in fact be formulated and evaluated with the help of simple recurrence relations, leading to a compact theory of tunnelling in macroscopic spins, in one-dimensional clusters, as well as in small higher-dimensional clusters. This is demonstrated explicitly in the case of the Ising model with a transverse field and transverse exchange, and in the case of macroscopic spins with uniaxial anisotropy. Our approach provides a transparent theory of level crossings, where the tunneling between time reversed configurations vanishes as a function of the external field. Those crossings result from the destructive quantum interferences between competing flipping channels. Destructive interferences are expected to be present as soon as the two-spin flip channels have an overall positive amplitude and thus compete with the intrinsically negative second-order processes due to the transverse field. Our theory consistently predicts $N$ crossings in chains of $N$ Ising spins, $2S$ crossings in single spins of magnitude $S$, and yields explicit analytical formulae for the level crossings of open chains and macroscopic spins. Disorder can be easily implemented in this perturbative formalism. Leading disorder effects can be treated analytically for spin rings. We find that at the smallest transverse field crossing the suppression of tunneling is most robust with respect to disorder and fluctuations in the parameters. We briefly discuss the implications of our findings for the use of realistic spin clusters on surfaces to store information.
\end{abstract}

\maketitle


Quantum tunneling in magnetic clusters has been intensively studied in the nineties as a special case of macroscopic quantum tunneling~\cite{Leggett1987}. Quantum tunneling between two states with very different quantum numbers, e.g. $S_z=S$ and $-S$ for large spin $S$, is in general a very high-order process since elementary terms such as a transverse field or exchange processes only change this quantum number by 1 or 2, and high-order perturbation calculations of the tunneling were limited to systems with a single tunneling channel~\cite{Hartmann-Boutron1995,Hartmann-Boutron1996,vanHemmen1986}. Other approaches included a WKB approximation in the semi-classical limit~\cite{vanHemmen1986,vanHemmen1998}, but the most successful approach proved to be an instanton and path integral formulation~\cite{Enz1986,Chudnovsky1988,Loss1992,Delft1992,Garg1993,Chudnovsky1994}. The predictions of these theories, for instance the difference between half-integer and integer spin~\cite{Loss1992,Delft1992}, or the presence of oscillations of the tunneling as a function of a transverse field~\cite{Garg1993}, have been beautifully confirmed by experiments on ferromagnetic molecules that measured Landau-Zener transition probabilities, which are sensitive to the tunneling between nearly degenerate levels~\cite{Wernsdorfer1999}. The tunnel splitting was found to oscillate with the transverse field, and the position of the minima of the tunneling amplitudes were shown to alternate depending on whether the difference between the $S_z$ components of the initial and final spin states was even or odd. For a review, see Refs.~\cite{Sessoli2003,Sessoli2006}.

More recently, it has become possible to create and control very small clusters of magnetic adatoms deposited on surfaces~\cite{Khajetoorians2011,Eigler2012,Khajetoorians2012}, where the exchange couplings between adatoms can be tailored by their positioning~\cite{Brovko2008}. One promising idea is to use arrays of small antiferromagnetic (AF) chains or ladders of Ising spins as a means to store bit information in a very compact way~\cite{Eigler2012}. Applying a large enough voltage pulse with the STM tip, tunneling between the AFM Ising ground states can be induced, providing a way to switch between bit states. The limitation of the associated memory comes from spontaneous tunneling, thermal or quantum, between the two AF ground states. For low enough temperature, the switching rate between the Ising AFM ground states saturates to the quantum tunneling rate, which decays exponentially with system size. Thus, one way to preserve the bit state longer is to increase the size of the cluster, at the cost of lower density of information. If however quantum tunneling depends strongly on an applied transverse field, with marked minima as in the case of molecular magnets, one could reduce the rate of quantum tunneling without increasing the cluster size.

The first indication that this could indeed be the case comes from a recent experiment that demonstrated that anisotropic chains of $N$ spins-1/2 can have ground state crossings as a function of an applied magnetic field~\cite{Toskovic2016}. The crossings occur between the two lowest levels which form a quasi-degenerate subspace (which becomes exactly degenerate in the thermodynamic limit). At these crossings, the quantum tunneling between the two states is completely suppressed. For small spin chains, this can easily be demonstrated numerically~\cite{Dmitriev2002}. A general theory of these level crossings has not yet been developed, however. One step in this direction has been achieved in Ref.~\cite{Gregoire2017}. This approach relies on a mapping of the spin model onto a fermionic chain using Jordan-Wigner transformation, and it is thus limited to open chains. A mean-field decoupling of this fermionic model maps it onto the Kitaev model of a one-dimensional p-wave superconductor~\cite{Kitaev2001}. This leads to the prediction of exactly $N$ crossings, where $N$ is the number of sites of the chain. In this framework, the crossings are naturally interpreted in terms of the oscillating, exponentially weak coupling between the two Majorana edge states on either end of the chain.

In this paper, we develop a perturbative approach that provides a unified and general framework for all these phenomena. It relies on a reformulation of degenerate perturbation theory to lowest non-trivial order in terms of a simple recurrence between flipping amplitudes. This approach leads to a general expression of the tunneling amplitude as a homogeneous polynomial in the amplitude of the transverse exchange processes and of the square of the transverse field. In this approach, the destructive quantum interferences that lead to level crossings appear as a natural consequence of the competition between tunnel processes with positive and negative amplitudes. In the case of a single exchange channel, level crossings will be present as soon as the amplitude of this process is positive (AFM) since it will then compete with the second order process due to the transverse field, which is intrinsically negative. Our approach leads to a number of analytical results in the limit of small exchange processes and transverse field (e.g. the exact solution for the open chain, or for a macroscopic spin), to very good asymptotic expressions for large closed rings and the effect of weak disorder, as well as to general qualitative conclusions (e.g. concerning the number of level crossings in a ring).

In higher dimension, our method still works, as we show on small rectangular, triangular, and cubic clusters, but it becomes rapidly very complex because of the large number of cluster shapes generated in the recurrence. The interest of the method is to show for the example of small clusters that the interference leads to a number of ground state crossings equal to the number of Ising spins, independently of the geometry, suggesting that this remains true for larger systems.

Our formalism turns out to be particularly convenient to study the effect of disorder (such as heterogeneities among different clusters, spatial $g$-factor variations, weak random fields, etc.) on the suppression of tunneling amplitudes. As we shall see, the suppression of tunneling in the lowest transverse fields is the least affected by disorder. This leads us to the conclusion that in order to achieve a maximally robust suppression of tunneling over a range of potentially fluctuating parameters of clusters, one should use the lowest transverse field for which a ground state crossing occurs in the disorder-free limit.

The paper is organized as follows: In Sec.~\ref{sec:models} we introduce the spin models used throughout the rest of the paper. In Sec.~\ref{sec:perturbation_theory_for_collective_tunneling} we present the iterative perturbation theory method applied to Ising models. In Sec.~\ref{sec:1d_models}, we apply the method to a 1D ring and chain, and in Sec.~\ref{sec:2d_and_3d_clusters} to small 2D and 3D clusters. In Sec.~\ref{sec:low_disorder} we introduce disorder on the pertubative couplings of a ring and obtain the mean square displacement of the transverse field zeros, as well as the second moment of the tunneling to leading order in the disorder strength. In Sec.~\ref{sec:tunneling_in_single_spin_models} we apply our method to an anisotropic single spin model and obtain the tunneling amplitude, including an exact result for the crossing field values. In Sec.~\ref{sec:other_systems_with_competing_tunneling_channels} we discuss other systems where one can observe interference between different tunneling paths, and consequently crossings. Finally, in Sec.~\ref{sec:final_remarks} we discuss our results and their experimental consequences.

\section{Models}%
\label{sec:models}

All the models we consider are spin models and can be written in the form
\begin{equation}
	H = H_0 + V,
\end{equation}
where $H_0$ is a dominant diagonal term with a doubly-degenerate ground state where the two states transform into each other by flipping all spins, and
\begin{equation}
	V = \lambda V_1 + \lambda^2 V_2,
\end{equation}
where $V_1$ and $V_2$ are perturbations that respectively flip one or two spins. $\lambda$ is only an auxiliary parameter which we introduce to organize the perturbation expansion. It will be set to $1$ later on. We thus require that the matrix elements of $V_{1,2}$ are much smaller than the norm of the terms in $H_0$. We use the following notation for the two lowest energy states of $H$,
\begin{equation}
	H\ket{\Psi_{\pm}} = E_{\pm}\ket{\Psi_{\pm}}.
\end{equation}
Our main goal is to calculate in leading order in $\lambda$ the energy splitting $\Delta E \equiv E_+ - E_-$.

All the Hamiltonians we consider share the same symmetry, which physically corresponds to a spin reflection across the $x$-$y$ plane, $S^z\to -S^z$. We write the symmetry formally in a way that applies to all models, as
\begin{equation}
	R \equiv T e^{i \pi S_z},\quad R^2=1,\quad [H,R]=0,
	\label{eq:R_symmetry_definition}
\end{equation}
where $S_z$ is the total spin projection along $z$, so that $\exp(i \pi S_z)$ rotates all spins by $\pi$ around their $z$ axis, and $T$ is the time-reversal operator which inverts all spins.

Let us call the two ground states of $H_0$ as $\ket{\emptyset},\ket{\Sigma}$. The operator $R$ flips all spins and transforms one ground state into the other,
\begin{equation}
	R\ket{\emptyset} = \ket{\Sigma},\quad R\ket{\Sigma} = \ket{\emptyset}.
\end{equation}
Taking this into account together with the fact that $R$ is anti-linear, we may write the (unnormalized) ground states as simultaneous eigenstates of $H$ and $R$, as
\begin{equation}
	\ket{\Psi_{\pm}} = e^{-i\phi_{\pm}/2} \ket{\emptyset} \pm e^{i\phi_{\pm}/2} \ket{\Sigma} + \dots,
	\label{eq:ising_eigenstates_zero_order}
\end{equation}
with two unknown phases $\phi_{\pm}$.

In the following subsections we introduce the spin models used throughout the paper. We start with an Ising model with transverse terms, followed by other Ising models which are extensions of that model. One can solve the extended models straightforwardly, only requiring a transformation of the couplings to be able to use the solution of the first Ising model. Finally, we introduce an anisotropic single spin model, a conceptually simpler model than the Ising models, which was studied in the literature to analyze ground state degeneracies, having many similarities to a ferromagnetic Ising model.

\subsection{Ising model with transverse field and exchange}%
\label{sub:ising_model}

\begin{figure}
	\centering
	\includegraphics[width=\columnwidth]{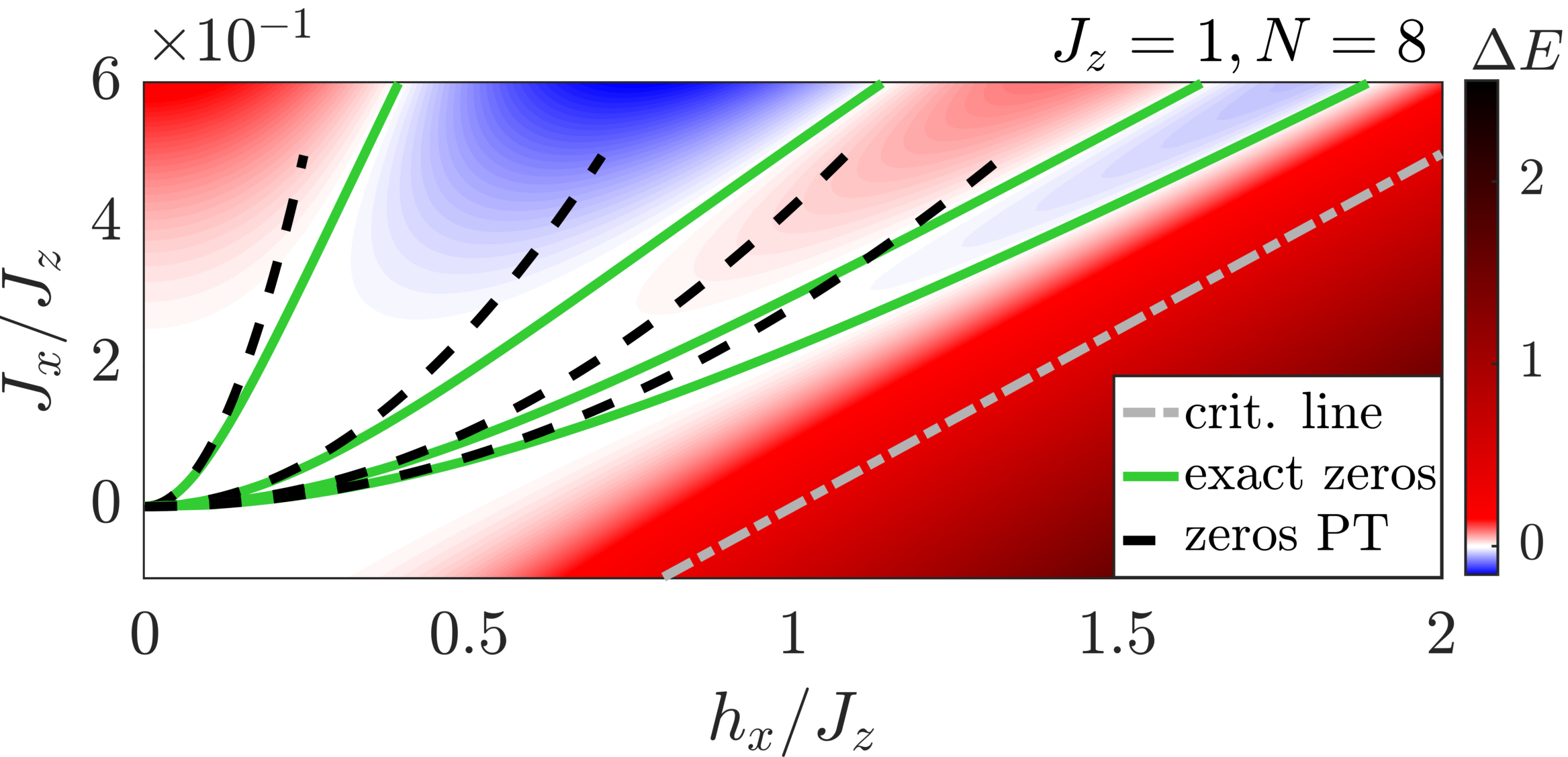}
	\caption{Color plot of the energy difference between the two lowest energy levels of the model~(\ref{eq:H_ising_model}) on a spin chain of 8 spins, obtained with exact diagonalization. The full lines indicate the locations where the ground state is degenerate, while the dashed lines indicate our perturbative prediction (Eq.~(\ref{eq:closed_zeros})), which works very well for small fluctuation parameters $J_x$ and $h_x$. Calculating the energy gap close to the phase transition $h_x=J_z, J_x=0$ perturbatively in the small parameters $h_x-J_z$ and $J_x$ yields $\Delta \approx 2(h_x -J_z -2J_x )$~\cite{Peschel1981}. This gives the approximate location of the quantum critical line between the Ising-ordered and the paramagnetic phases as $J_x \approx (h_x - J_z)/2$, as shown in the plot. This is still accurate for large $J_x$ since a similar calculation close to the phase transition at $(h_x=2J_x, J_z=0)$ gives the same gap~\cite{Rujan1981}. For the full phase diagram, see Ref.~\cite{Hassler2012}. There are no zeros when $J_x<0$, and the ordered phase shrinks with increasingly negative $J_x$. Both reflect the fact that the transverse terms reinforce each other for $J_x<0$, while they compete for $J_x>0$.}
	\label{fig:exact_zeros_chain}
\end{figure}

We first consider the Ising model in a small transverse field and subject to weak transverse exchange interactions~\cite{Gregoire2017}:
\begin{equation}
	\begin{aligned}
		H_0 &= J_z \sum_{\pair{ij} } \sigma^z_i\sigma^z_j,\\
		V_1 &= -h_x \sum_{i} \sigma^x_i, \quad V_2 = J_x \sum_{\pair{ij} }\sigma^x_i\sigma^x_j.
	\end{aligned}
	\label{eq:H_ising_model}
\end{equation}
Here $\sigma$ are the Pauli matrices and $N$ is the number of Ising spins. We keep the lattice general for now, however, we do require that the lattice be classically unfrustrated, such that the ground state manifold of $H_0$ is only doubly degenerate. In the following sections we will carry out more detailed calculations by restricting ourselves to specific lattices and Ising interactions.

We denote the Ising eigenstates and energies as
\begin{equation}
	H_0\ket{m} = \en_m\ket{m},
	\label{eq:H_0_eigenvalue_equation}
\end{equation}
where $\ket{m}$ are the classical Ising configurations with the spins either up or down along the axis $z$. We label the states $\ket{m}$ by the set $m$ of spins that are flipped with respect to one of the Ising ground states. By definition, our reference ground state corresponds to the empty set $m=\emptyset$, and the other ground state to $m=\Sigma$, the set of all spins. The energy of these two states is $\en_\emptyset (= \en_\Sigma)$.

Before proceeding with introducing various extensions, let us comment on the ground state degeneracies on the spin chain with even $N$. When $J_x>0$ and for finite size, there are lines in the ($J_x, h_x$)-plane where the ground state is degenerate, as shown in Fig.~\ref{fig:exact_zeros_chain}. In our perturbative regime, these lines scale as $h_x \sim \sqrt{J_x\abs{J_z}}$. The lines continue for larger $J_x,h_x$, but the scaling becomes linear. In fact, when $J_x, h_x \gg J_z$, all lines except one approach the critical line $J_x = h_x/2$ of the $J_z=0$ classical model (an AFM Ising with longitudinal field), separating the AFM-ordered phase along $x$ from the PM phase. The classical ground state on this critical line is highly degenerate, but the $J_z$ exchange coupling lifts the degeneracy and induces many ground state crossings close to the classical critical line. The line corresponding to the smallest field for given $J_x$ instead approaches the classical critical line $J_x = h_x$. On that line it costs no energy to flip the terminal spin of the chain which is anti-aligned to the external field in the AFM phase. This separate degeneracy line is present only for chains with even $N$.

\subsection{Ising model with a staggered field}%
\label{sub:staggered_field}

\begin{figure}
	\centering
	\includegraphics[width=\columnwidth]{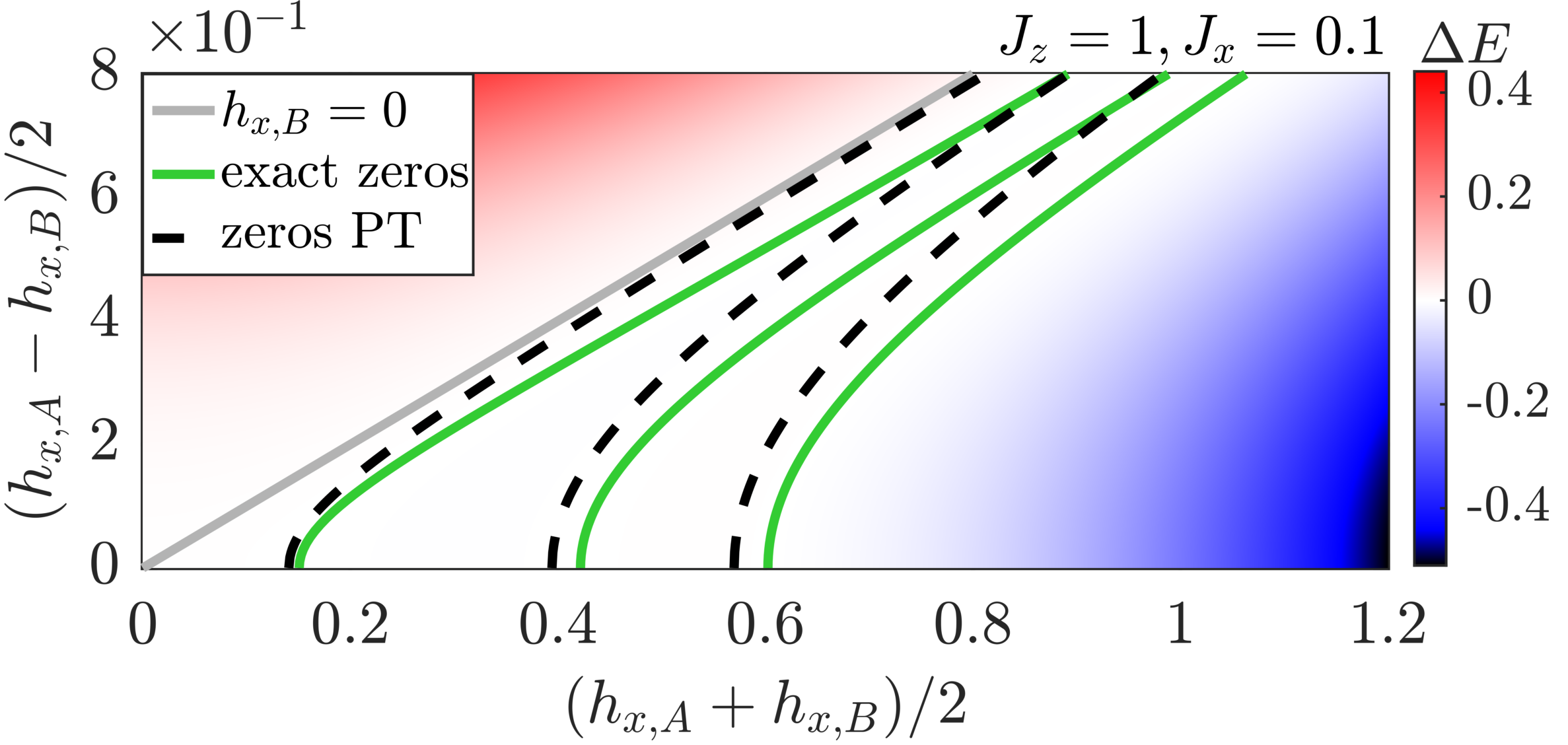}
	\caption{Color plot of the exact energy difference between the two lowest energy levels of the model defined in Sec.~(\ref{sub:staggered_field}) on a spin chain of 6 spins, obtained with exact diagonalization. The full green lines indicate the locations where the ground state is degenerate, while the dashed lines indicate the prediction from perturbation theory (Eqs.~(\ref{eq:h_n_staggered}, \ref{eq:closed_zeros})). The diagonal $h_{x,B} = 0$ is marked to show that the zeros only occur in the region where the field and the exchange term $V_2$ favor opposite ground states.}
	\label{fig:exact_zeros_staggered}
\end{figure}

The first extension we consider is a staggering of the magnetic field on bipartite lattices with the same number of sites on either sublattice, but different transverse fields $h_{x,A}$ and $h_{x,B}$ acting on the two sublattices $A$ and $B$, respectively. Thus, the modified perturbation is
\begin{equation}
	V_1 = - \left( h_{x,A} \sum_{i \in A} \sigma^x_i + h_{x,B} \sum_{i \in B} \sigma^x_i \right),
\end{equation}
with unchanged $V_2$. In Fig.~\ref{fig:exact_zeros_staggered} we show the ground state energy splitting of this model as a function of the staggered fields. The zero lines scale as
$
h_{x,A}h_{x,B} \sim J_x \abs{J_z},
$
but only when
$
h_{x,A} h_{x,B} J_x>0
$
do ground state degeneracies occur, that is, only when the field and the exchange coupling favor opposite ground state configurations. In general we can state that tunneling suppression occurs when the transverse fluctuations are competing or "frustrated".

\subsection{Ising model with general transverse couplings}%
\label{sub:generic_transverse_couplings}

\begin{figure}
	\centering
	\includegraphics[width=\columnwidth]{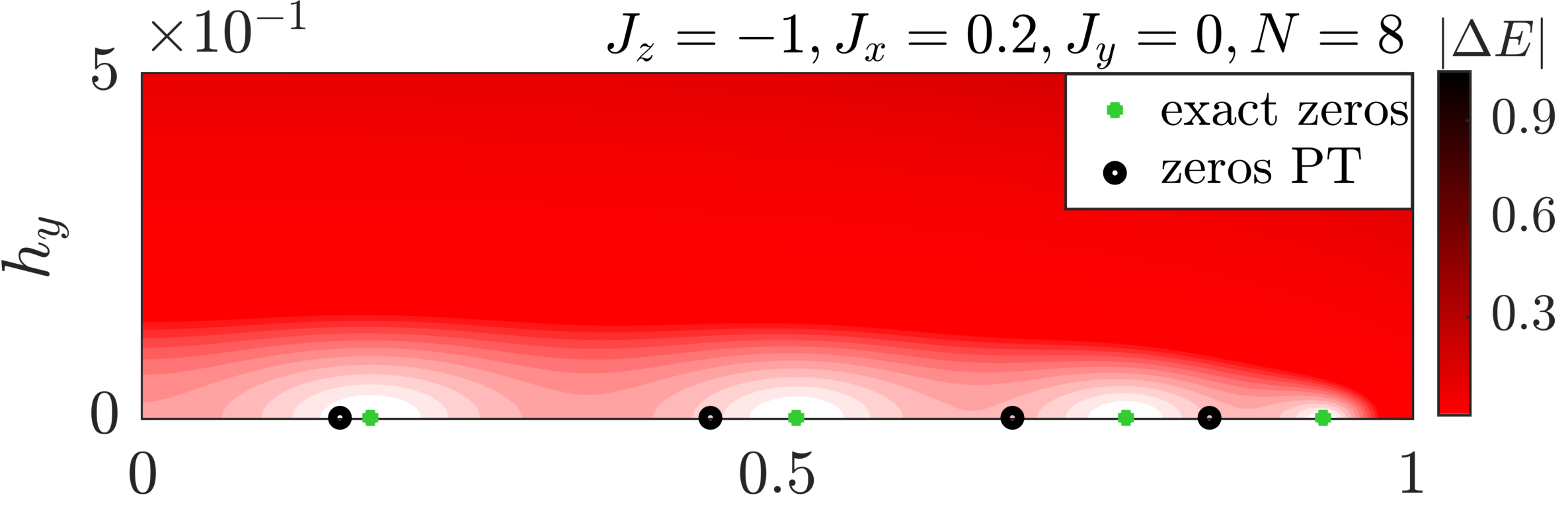}
	\includegraphics[width=\columnwidth]{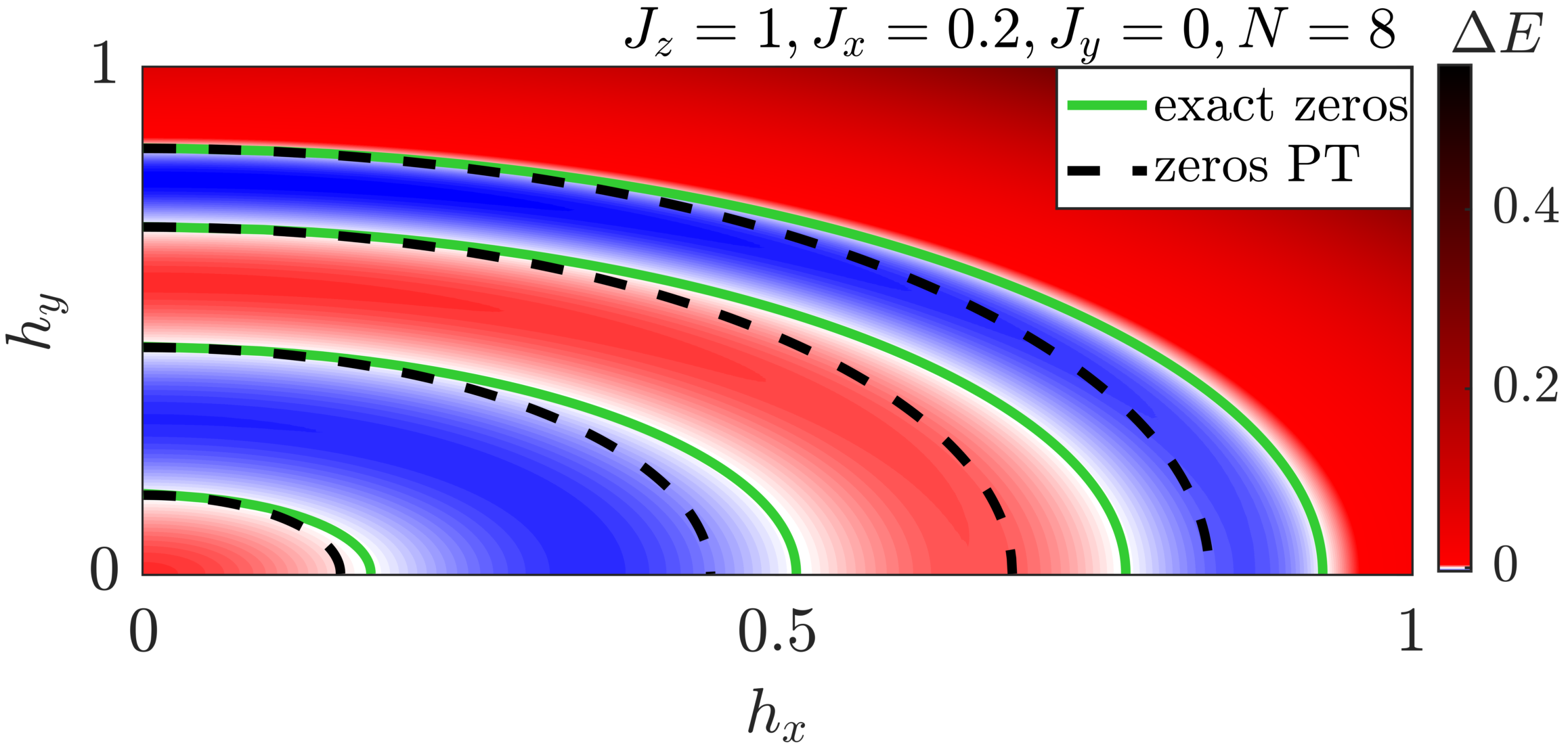}
	\caption{Color plot of the exact energy difference between the two lowest energy levels of the model defined in Eq.~(\ref{eq:H_ising_extended_model}) on a spin chain of 8 spins, obtained with exact diagonalization. We set $J_y=0$, $J_x=0.2$ so we may study the energy splitting in the $h_x$-$h_y$ plane. The ground state crossings are drastically different depending on the sign of $J_z$. If the ground state is ferromagnetic (above), there are only degeneracies when $h_y = 0$ (These are the same zeros as in Fig.~\ref{fig:exact_zeros_chain} along the $J_x=0.2$ line). This is predicted by our perturbative calculations, where we obtain the zeros marked in black as given by Eq.~(\ref{eq:zeros_ising_ferro_x}). In the antiferromagnetic case (below), we obtain lines of zeros in the plane which connect the crossings of the model~(\ref{eq:H_ising_model}) (along the $x$ axis) to the exact crossings of the Kitaev chain model given by Eq.~(\ref{eq:Kitaev_zeros}) (along the $y$ axis). These lines are consistent with the perturbative solution (dashed line), where we find that the zeros do not depend on the orientation of the field in the $x$-$y$ plane (Eq.~(\ref{eq:zeros_ising_antiferro})).}
	\label{fig:exact_zeros_h_plane}
\end{figure}

We will also extend the model by having transverse couplings along both transverse axes,
\begin{equation}
	\begin{aligned}
		V_1 &= - \sum_{i} \left( h_x \sigma^x_i + h_y \sigma^y_i \right),\\
		V_2 &= \sum_{\langle ij \rangle} (J_{x} \sigma_i^{x} \sigma_j^{x} + J_{y} \sigma_i^{y} \sigma_j^{y}).
	\end{aligned}
	\label{eq:H_ising_extended_model}
\end{equation}
On a 1D chain of spins with open boundary conditions, this model has an interesting limit in which it is exactly solvable. Taking
$
J_{y}=h_{x}=0,
$
one can map the model with a Jordan-Wigner transformation~\cite{Pfeuty1970} onto the Kitaev chain model~\cite{Kitaev2001}, a free fermion model which one can solve exactly~\cite{Gregoire2017,Ivo2018} to obtain field values in which not only the ground state is degenerate, but all eigenstates are:
\begin{equation}
	h_y^{(n)} = 2\sqrt{J_x J_z}\cos\left( \frac{\floor{N/2} + 1 - n}{N+1} \pi \right),
	\label{eq:Kitaev_zeros}
\end{equation}
where
$
n = 1,\dots,\floor{N/2}.
$
If we take $J_x>0$, these zeros only appear if $J_z>0$, as one can see in Fig.~\ref{fig:exact_zeros_h_plane}. Interestingly, our perturbative approach yields exactly the same expression in the appropriate limit (Eqs.~(\ref{eq:closed_zeros}, \ref{eq:zeros_ising_antiferro})), with no higher order corrections. This is presumably a consequence of the fact that the zero energy Majorana fermions do not backscatter, so that our leading order approximation, which is equivalent to a forward scattering approximation for the fermions, becomes exact~\cite{Fisher1995}.

\subsection[Single spin model]{Single spin model}%
\label{sub:single_spin_model}

Instead of studying a cluster of $N$ spins explicitly with all its internal couplings, one may consider the cluster as a big effective spin $S$ and consider the tunneling from the up to the down state of this composite spin. The precise Hamiltonian for this equivalent big spin would in general be very complex. However, qualitative features can be expected to be captured by simple effective interactions which can be written as low powers of the total spin operators $S_{x,y,z}$.

A transverse field on a ferromagnetic cluster of spins naturally translates into a transverse field acting on the big spin. Also, the ferromagnetic Ising configurations are the projections with largest total $S_z$ of the largest spin one can form with $N$ $(S=1/2)$-spins. Thus, the ground state of a Hamiltonian of the form
$
-S_z^2
$
corresponds to a ferromagnetic Ising ground state of the original cluster. Indeed, we shall see later that this model captures qualitatively the features of a ferromagnetic Ising model ($J_z < 0$) as considered in Sec.~\ref{sub:generic_transverse_couplings}. In contrast, it is not clear whether such an approximate mapping is meaningful for antiferromagnetic clusters.

Here we will reconsider quadratic single spin Hamiltonians equivalent to those considered earlier in the literature~\cite{Garg1993,Hartmann-Boutron1996,vanHemmen1998,Wernsdorfer1999}. In particular, we take the most general quadratic single spin model with anisotropy and a field transverse to the easy axis. Upon choosing axes that bring the quadratic part to a diagonal form, we are left with three quadratic couplings $J_{x,y,z}$. We are free to choose the easy, medium and hard axes to be, respectively, along the $z$-, $y$- and $x$-axes (i.e., $J_z \le J_y \le J_x$). We are also free to set one of these couplings to zero, since the ground state splitting will be independent of a constant term
$
J \vec{S}\cdot\vec{S}
$
that one can add to the Hamiltonian. We therefore consider the Hamiltonian
\begin{equation}
	\begin{aligned}
		H_0 &= J_z S_z^2,\\
		V_1 &= - (h_x S_x + h_y S_y),\quad V_2 = J_x S_x^2,
	\end{aligned}
	\label{eq:H_single_spin}
\end{equation}
where we chose $J_y = 0$, implying that $J_z \le 0$ and $J_x \ge 0$.
When performing perturbative calculations, we assume that
\begin{equation}
	\abs{J_z} \gg \abs{J_x},\abs{h_x},\abs{h_y}.
	\label{eq:perturbative_regime}
\end{equation}
This model has $2S = N$ values of the magnetic field where the ground state is doubly degenerate, as demonstrated by A. Garg~\cite{Garg1993} using path integral and instanton calculations justified at large $S$, but without restricting to the perturbative regime~(\ref{eq:perturbative_regime}). In the considered large $S$ limit, the crossings can be interpreted as the negative interference of two tunneling instantons having different Berry phases. We shall derive a very similar result within perturbation theory, but without relying on the size of the spin (or $N$) and without taking a saddle point approximation.

We will show in Sec.~\ref{sec:tunneling_in_single_spin_models} that here again the underlying mechanism behind the zeros is the competition of multiple tunneling paths with oscillating signs. Finally, using a different approach we will derive non-perturbatively the location of the $2S$ equally spaced transverse field zeros without relying on any approximation.

\section{Perturbation theory for collective tunneling}%
\label{sec:perturbation_theory_for_collective_tunneling}

In this section we will present the perturbative method applied to Ising models defined in Sec.~\ref{sec:models}. The method is derived in more detail in appendix~\ref{app:deduction_of_method}.

Standard techniques to carry out degenerate perturbation theory at arbitrarily high orders construct a perturbative expansion for an effective Hamiltonian $\Heff$ which only acts on the Hilbert space spanned by unperturbed states and yields the exact splitting of the ground state manifold due to perturbations. Let us call the set of unperturbed ground state configurations
$
g = \{ \emptyset, \Sigma \}.
$
Defining $P$ as the projector onto the subspace $g$, one constructs $\Heff$ which projects out all the excited states, i.e.,
$
P\Heff = \Heff P = \Heff.
$
However, $\Heff$ is not fully specified by these requirements. A first full series expansion for a possible choice of $\Heff$ was obtained in \cite{Kato1949} for a general Hamiltonian with a discrete spectrum. It leads to a generalized eigenvalue equation that must be solved for the split ground state energies. A variation of \cite{Kato1949} was later given in \cite{Bloch1958}, where the eigenvalue equations are simpler, but the operator $\Heff$ will in general turn out to be non-Hermitian. Here we use the latter approach. The eigenvalue equations of this effective Hamiltonian are
\begin{equation}
	\Heff P \ket{\Psi_{\pm}} = E_{\pm} P\ket{\Psi_{\pm}},
	\label{eq:Heff_eigenvalue_equation}
\end{equation}
where $\ket{\Psi_{\pm}}$ are the lowest energy eigenstates of $H$. Owing to the anti-linear symmetry $R$ (Eq.~(\ref{eq:R_symmetry_definition})), which lets us write $\ket{\Psi_{\pm}}$ as in Eq.~(\ref{eq:ising_eigenstates_zero_order}), we deduce that the energy splitting is given by
\begin{equation}
	\Delta E = (e^{i\phi_{+}} + e^{i\phi_{-}})\braket{\emptyset|\Heff|\Sigma},
	\label{eq:DEn_matrix_element}
\end{equation}
in terms of the off-diagonal matrix element of the effective Hamiltonian. Our perturbative method allows us calculate the matrix element $\bra{\emptyset} \Heff \ket{\Sigma}$ to leading order in $\lambda$. Since $V_1$ and $V_2$ respectively flip one or two spins when acting on $\ket{m}$, and given that
$
V = \lambda V_1 + \lambda^2 V_2,
$
the power of $\lambda$ of a tunneling path between $\ket{\emptyset}$ and $\ket{\Sigma}$ corresponds to the number of spin flips that occurred. Since the shortest paths have exactly $N$ spin flips (each spin flips once and only once), it follows that $\lambda^{N}$ is the lowest order that will occur, and thus
\begin{equation}
	\bra{\emptyset} \Heff \ket{\Sigma} = t_N \lambda^{N} + O\left(\lambda^{N+1}\right),
	\label{eq:t_definition}
\end{equation}
where we have defined $t_N$ as the leading order term.

In the limit $\lambda \rightarrow 0$, the phases in Eq.~(\ref{eq:ising_eigenstates_zero_order}) must vanish, $\phi_{\pm} \rightarrow 0$. This follows from the fact that in this limit the ground states approach the two linear combinations
\begin{equation}
	\ket{\Psi_{\pm}} \to \ket{\emptyset} \pm \ket{\Sigma} + O(\lambda).
\end{equation}
One thus finds
\begin{equation}
	\abs{\Delta E} = 2 \abs{t_N} + O(\lambda^{N+1}).
\end{equation}
As shown in Section~\ref{app:deduction_of_method}, the tunneling amplitude is given by
\begin{equation}
	t_N = \sum_{n=\ceil{\frac{N}{2}}}^{N} \sum_{\{l_i\}} \bra{\emptyset}V_{l_1} S \dots V_{l_{n-1}} S V_{l_{n}} \ket{\Sigma}
	\label{eq:t_development}
\end{equation}
where the
$
l_{i=1,...n} \in \left\{ 1,2\right\}
$
obey
$
l_1 + l_2 + \dots + l_n = N,
$
and
\begin{equation}
	S \equiv -\sum_{m \not\in g} \frac{\ket{m}\bra{m}}{\Den_m},
	\label{eq:S_definition}
\end{equation}
where $\Den_m \equiv \en_m - \en_{\emptyset}$ is the unperturbed excitation energy of the state $\ket{m}$. Let us denote by $\abs{m}$ the cardinality of the set $m$, that is, the number of spins that are flipped relative to the ground state $\emptyset$, so that $\abs{\Sigma} = N$. The tunneling $t_N$ can be calculated recursively. To this end we introduce $c_{m}$ as an intermediate tunneling coefficient, analogous to $t_N$ in Eq.~(\ref{eq:t_development}),
\begin{equation}
	c_{m} \equiv \sum_{n=\ceil{\frac{\abs{m}}{2}}}^{\abs{m}} \sum_{\{l_i\}} \bra{\emptyset}V_{l_1} S \dots V_{l_{n-1}} S V_{l_{n}} S \ket{m},
	\label{eq:c_m_definition}
\end{equation}
where we impose the condition
$
l_1 + l_2 + \dots l_n = \abs{m},
$
on the $l_i$, which restricts the sum to the terms that contribute to leading order ($\lambda^{\abs{m}}$) to the tunneling between $\ket{\emptyset}$ and $\ket{m}$. Finally, by inserting the identity as $\sum_{m'} \ket{m'}\bra{m'}$ before the last two factors and expanding $V_{l_n} S \ket{m}$, one obtains a recursion relation that connects $c_m$ to coefficients of smaller clusters $m' \subset m$, yielding the cluster recursion relations
\begin{equation}
	\begin{aligned}
		c_{m} = &-\frac{1}{\Den_m}
		\left( \sum_{i \in m } \braket{m \setminus \{i\}|V_1|m} c_{m \setminus \{i\}}\right.\\
				&\left.+\sum_{i,j \in m }\braket{m \setminus \{i,j\}|V_2|m}c_{m \setminus \{i,j\}}\right),
	\end{aligned}
	\label{eq:c_m_general_recursion}
\end{equation}\\
where we sum over flipped spins $i \in m$, or pairs of flipped spins $\{i,j\} \in m$. An analogous expression follows for $t_N$,
\begin{equation}
	\begin{aligned}
		t_N = &\sum_{i \in \Sigma } \braket{\Sigma \setminus \{i\}|V_1|\Sigma} c_{\Sigma \setminus \{i\}}\\
		+&\sum_{i,j \in \Sigma }\braket{\Sigma \setminus \{i,j\}|V_2|\Sigma}c_{\Sigma \setminus \{i,j\}}.
	\end{aligned}
	\label{eq:t_N_general_recursion}
\end{equation}
The recursions~(\ref{eq:c_m_general_recursion}, \ref{eq:t_N_general_recursion}) hold for general perturbations $V_{1,2}$ which flip single spin or pairs of spins, respectively. It is straightforward to generalize the recursions~(\ref{eq:c_m_general_recursion}) to include higher order terms of the form $\lambda^k V_k$, which flip $k$ spins.

The iterative procedure can be further simplified due to the independence of disconnected clusters. Let us consider a cluster $m$ that is composed of several mutually disconnected clusters $m_i$ of spins flipped relative to $\emptyset$. We call a set of clusters disconnected if the excitation energy of the set is the sum of the excitation energies of the individual clusters, i.e.
$
\Den_m = \Den_{m_1} + \Den_{m_2} + \dots \Den_{m_n}.
$
In Sec.~\ref{app:cluster_independence}, we show that for such separable sets, the intermediate tunneling coefficient is the product of tunneling coefficients for their disconnected components,
\begin{equation}
	c_{m} = c_{m_1}c_{m_2} \dots c_{m_n}.
	\label{eq:c_m_independence}
\end{equation}
To calculate $t_N$ for a given cluster one proceeds as follows: One identifies all inequivalent connected subclusters $m$ of the considered spin cluster, associates a coefficient $c_{m}$ to each of them, and uses Eqs.~(\ref{eq:c_m_general_recursion}, \ref{eq:c_m_independence}) to calculate the coefficients recursively for increasing cluster sizes $\abs{m}$. We say that two clusters are equivalent if their coefficients $c_m$ are the same, which is for instance the case if the two clusters are symmetry related.

If the unperturbed Hamiltonian is the Ising model with nearest neighbor couplings, the excitation energy $\Den_{m}$ of a cluster is $2 \abs{J_z}$ times the number of bonds that connect flipped spins $(\in m)$ to unflipped spins $(\in \Sigma\setminus m)$. In other words, the excitation energy is proportional to the total length of all domain walls between $m$ and its complement.

In general, the recursions~(\ref{eq:c_m_general_recursion}, \ref{eq:t_N_general_recursion}) can be rather complicated to solve, especially if one has to consider a large number of inequivalent clusters. However, we will see in the following subsections that they take a simplified form when applied to some of our models, and even result in recursion relations with closed form solutions in some cases. Applying the method to the first Ising model~(\ref{eq:H_ising_model}), our next step, will be very instructive.

\subsection{Tunneling in Ising models with transverse field and exchange}%
\label{sub:tunneling_in_the_ising_model_with_transverse_field_and_exchange}

Let us consider the model~(\ref{eq:H_ising_model}) and let us apply the recursion relations~(\ref{eq:c_m_general_recursion}, \ref{eq:t_N_general_recursion}) to it. The first thing to note is that the matrix elements in the recursion relations simplify greatly. They are given by%
\begin{equation}
\label{V1}
		\braket{m \setminus \{i\}|V_1|m} = -h_x
\end{equation}
and
\begin{equation}
\braket{m \setminus \{i,j\}|V_2|m} = J_x
\end{equation}
if $i,j$ are nearest neighbors, and vanish otherwise. Then, the recursion relations become
\begin{align}
	\label{eq:c_m_recursion}
	c_m &= \frac{1}{\Den_m} \left( h_x \sum_{i \in m } c_{m \setminus \{i\}} -J_x\sum_{\mathclap{\pair{ij} \in m}}c_{m \setminus \{i,j\}}\right),\\
	\label{eq:t_N_recursion}
	t_N &= -h_x \sum_{ \mathclap{i \in \Sigma} } c_{ \Sigma \setminus \{i\} } + J_x \sum_{ \mathclap{\pair{ij} \in \Sigma} } c_{ \Sigma \setminus \{i,j\} }.
\end{align}
In this model, two clusters are equivalent if they are identical including their environment up to their first neighbors. As an example of this, consider a cluster of flipped spins in the bulk of a lattice with open boundary conditions. Any translation by a lattice unit whereby the cluster does not touch the boundaries results in an equivalent cluster. When a boundary is reached instead, the first neighbors of the cluster change and we will find a different cluster coefficient.

Since $t_N$ is the sum over all tunnel paths that flip every spin exactly once, it is clear that the resulting expression is a polynomial in $h_x$ and $J_x$, each term being proportional to a product $J_x^k h_x^{N-2k}$ with $0 \leq k \leq \floor{N/2}$. The general form for the tunneling coefficient of a cluster of $N$ spins, regardless of the lattice, will thus take the form
\begin{align}
	\label{eq:t_polynomial_form}
		t_N &= \sum_{k = 0}^{\floor{\frac{N}{2}}} a_k
		\frac{J_x^k h_x^{N-2k}}{\abs{J_z}^{N-k-1}}\\
	\label{eq:t_factorized_form}
			&\propto \frac{h_x^{{\rm mod}(N,2)}}
			{\abs{J_z}^{2\floor{\frac{N}{2}}-1}}
			\prod_{n=1}^{\floor{\frac{N}{2}}} ( J_x \abs{J_z} - \alpha_n h_x^2),
\end{align}
with some lattice-dependent real coefficients $a_k$, and (potentially complex) roots $\alpha_n$.

Due to the minus sign in the projector $S$ (Eq.~(\ref{eq:S_definition})), the sign of $a_k$ reflects the number of flipping terms $V_{1,2}$ that are applied on the corresponding tunneling paths. The sign thus alternates with $k$:
\begin{equation}
	\text{sgn}(a_k) = (-1)^{k-1},
\end{equation}
where we have also taken into account the negative sign of the matrix element of $V_1$ in Eq.~(\ref{V1}).
If $J_x < 0$, all monomials contribute with the same sign, and thus $|t_N|$ grows monotonously with the field $h_x$, with no zero crossing. However, if $J_x > 0$, there is a negative interference between paths with different numbers of perturbative steps, and $t_N$ may oscillate as a function of the field. In this case, we can have ground state crossings. If the polynomial of Eq.~(\ref{eq:t_factorized_form}) has positive real roots $\alpha_n$, the crossings occur at the fields
\begin{equation}
	h_x^{(n)} = \pm \alpha_n^{-1/2} \sqrt{J_x \abs{J_z}}.
	\label{eq:h_n_definition}
\end{equation}
Thus there may be up to $N$ values of the transverse field (or $\floor{N/2}$, if one restricts to $h_x \ge 0$) where $t_N=0$, depending on the number of real $\alpha_n$. (Note that real $\alpha_n$ are necessarily positive, since negative $\alpha_n$ would imply zeros for $J_x>0$, which is excluded).

Later on we will solve the recursion relations for specific spin clusters and lattices, where we do find that in all cases considered the $\alpha_n$ are real and positive. For now, let us assume that indeed all $\alpha_n > 0$, so that we have either $N$ degeneracy points or none (except for the trivial one $h_x=0$ for odd $N$), depending on the sign of $J_x$.

\subsection{Tunneling in Ising models with a staggered field}%
\label{sub:tunneling_in_ising_with_staggered_field}

We consider the model presented in Section~\ref{sub:staggered_field} where the field of the model~(\ref{eq:H_ising_model}) is staggered. One could calculate the matrix elements and write down the recursion relations for this model, but this turns out to be unnecessary. Let us first guess the polynomial form of $t_N$ for this model. The tunneling paths due to transverse exchange only contribute with a factor $J_x^{N/2}$. For tunneling paths involving spins flipped by transverse fields, those must come in equal numbers on the two sublattices, and thus $t_N$ must take the form
\begin{equation}
	t_N = \frac{J_x^{N/2}}{|J_z|^{N/2-1}} \sum_{k=0}^{N/2} a_k\left(\frac{h_{x,A}h_{x,B}}{J_x|J_z|}\right)^{N/2-k},
\end{equation}
where the coefficients $a_k$ must be the same as those of the polynomial for a homogeneous field $h_x$, cf. Eq.~(\ref{eq:t_polynomial_form}). Thus one simply should substitute $h_x^2 \to h_{x,A} h_{x,B}$ in that equation.
It follows that ground state degeneracy occurs whenever
\begin{equation}
	h_{x,A}h_{x,B} = \alpha_n^{-1} J_x \abs{J_z}.
	\label{eq:h_n_staggered}
\end{equation}
The condition
$
h_{x,A} h_{x,B} J_x>0
$
is a general prerequisite for such degeneracies. If all $\alpha_n$ are positive, we recover the behavior observed numerically in Fig.~\ref{fig:exact_zeros_staggered}

\subsection{Tunneling in Ising models with general transverse couplings}%
\label{sub:tunneling_general_transverse_couplings}

In the case of the more general Ising models~(\ref{eq:H_ising_extended_model}), we found the behavior of the ground state crossings to depend on the Ising ground state (Fig.~\ref{fig:exact_zeros_h_plane}). In particular, we distinguish whether the dominant Ising interactions are ferromagnetic $(J_z<0)$ or antiferromagnetic $(J_z>0)$, respectively.

\subsubsection{Ising ferromagnets}

Since the ground state is ferromagnetic, and since we only flip each spin once, the matrix element $\braket{m \setminus \{i,j\}|V_2|m}$  appears in  the recursion relations only in the form
\begin{equation}
	\braket{\uparrow\uparrow|V_2|\downarrow \downarrow} = J_x - J_y,
\end{equation}
for two neighboring spins, while $\braket{m \setminus \{i\}|V_1|m}$ only appears as
\begin{equation}
	\bra{\uparrow } V_1 \ket{\downarrow} = -h_x+i h_y.
\end{equation}
This implies that the resulting recursion and the tunneling amplitude $t_N$ will be the same as Eqs.~(\ref{eq:c_m_recursion}, \ref{eq:t_polynomial_form}), up to the substitution
\begin{align}
	\begin{array}{ll} J_x &\to J_x-J_y, \\
	h_x  &\to h_x-i h_y,
	\end{array} \quad {\textrm{(Ferro)}}
\end{align}
and it suffices to solve the model~(\ref{eq:H_ising_model}).

Performing the substitution in Eq.~(\ref{eq:h_n_definition}), one finds that the tunneling only vanishes if either $h_y=0$ and $J_x>J_y$, in which case there are zeros at the fields
\begin{equation}
	h_x^{(n)} = \pm \left(\frac{(J_x-J_y) |J_z|} {\alpha_n}\right)^{1/2},
	\label{eq:zeros_ising_ferro_x}
\end{equation}
or if $h_x=0$ and $J_y>J_x$, at fields
\begin{equation}
	h_y^{(n)} = \pm \left(\frac{(J_y-J_x) |J_z|} {\alpha_n}\right)^{1/2}.
	\label{eq:zeros_ising_ferro_y}
\end{equation}
In other words, the transverse field has to be applied in the spin direction which corresponds to the stronger antiferromagnetic (or weaker ferromagnetic) exchange. This is analogous to the result found in the single spin case by Garg~\cite{Garg1993}, which we will review in Section~\ref{sec:tunneling_in_single_spin_models} below.

\subsubsection{Ising antiferromagnets}

For antiferromagnetic Ising models case, let us consider a bipartite lattice, so that the ground states have opposite spins on each sublattice.  $V_2$ only acts in the form
\begin{equation}
	\braket{ \uparrow \downarrow |V_2| \downarrow \uparrow } = J_x + J_y,
\end{equation}
while $V_1$  flips spins from down to up on one sublattice and from up to down on the other sublattice,
\begin{align}
	\braket{ \uparrow |V_1| \downarrow } = -h_x+i h_y,\quad \braket{ \downarrow |V_1| \uparrow } = -h_x-i h_y.
\end{align}
We take each sublattice to have the same number of spins. Since the lattice is bipartite and $V_2$ flips one spin from each sublattice, there must be an equal number of single flips due to $V_1$ on either sublattice. This implies that we can simply substitute
\begin{align}
\label{AFsubst}
	\begin{array}{ll}
	J_x &\to J_x+J_y,\\
	h_x^2 &\to h_x^2+h_y^2,
	\end{array} \quad {\textrm{(Antiferro)}}
\end{align}
in Eqs.~(\ref{eq:c_m_recursion}, \ref{eq:t_polynomial_form}) to obtain the result for the generic, bipartite antiferromagnetic Ising models.
Interestingly, the direction of the homogeneous transverse field in the $x-y$ plane is irrelevant. That is, zeros of the tunneling amplitude  occur in circles in the transverse field plane, provided that  the transverse exchange is predominantly antiferromagnetic ($J_x+J_y>0$). The tunneling vanishes for transverse fields of magnitude
\begin{equation}
	h^{(n)} = \left(\frac{(J_x+J_y) |J_z|} {\alpha_n}\right)^{1/2},
	\label{eq:zeros_ising_antiferro}
\end{equation}
regardless of its angle in the $x-y$ plane. While there is no angle dependence to leading order, the radial symmetry is broken by higher order corrections as confirmed numerically in Fig.~\ref{fig:exact_zeros_h_plane}.

Note that the tunneling matrix element $t_N$ is in general complex, and thus the condition $t_N=0$ determines a manifold of codimension 2 in the parameter space of transverse couplings. Thus, by fixing the exchange couplings $J_x, J_y$ and looking for zeros in the transverse field plane, one will generically only find isolated points, as it happens in the case of ferromagnetic clusters. A qualitatively different situation arises in antiferromagnetic clusters because there, owing to the substitution (\ref{AFsubst}), the tunneling amplitude is always real, such that zeros organize in a manifold of codimension 1, i.e., closed lines in the transverse field plane.

\section{Tunneling in 1D systems}%
\label{sec:1d_models}

In this section we apply our method to the 1D model~(\ref{eq:H_ising_model}). The recursion relations for 1D systems are rather simple because any connected cluster of flipped spins is uniquely defined by its length and position. We first consider a ring of $N$ spins, where the exact solution of the recursion allows us to extract explicit asymptotic expressions for large $N$. Then we consider an open chain, where we even obtain a closed analytical expression for $t_N$ for any $N$.

\subsection{Closed chain: a ring of spins}%
\label{sub:spin_ring}

For a ring of spins, all connected clusters of a given length are equivalent. We denote by $c_l$ the intermediate tunneling coefficient associated with a cluster of length $l$. The application of Eq.~(\ref{eq:c_m_recursion}) is straightforward: By unflipping a single spin from the cluster, we obtain one of $l$ possible states with $l-1$ flipped spins. If the unflipped spin is at the edge of the cluster, its contribution to $c_{l}$ is
$
h_x c_{l-1}/(4\abs{J_z}),
$
since the excitation energy of the cluster is $4\abs{J_z}$ due to the two domain walls at its ends. If the unflipped spin is in the bulk then the state consists of two clusters. We thus use Eq.~(\ref{eq:c_m_independence}) to write the contribution of that state as
$
h_x c_n c_{l-n-1}
$
for some $0<n<l$. Defining
$
c_0 \equiv 1
$
we can combine the edge and the bulk terms, and using an analogous reasoning for the term related to $J_x$ we finally have
\begin{equation}
	c_l = \frac{h_x}{4\abs{J_z}}\sum_{n=0}^{l-1}c_n c_{l-n-1} -
	\frac{J_x}{4\abs{J_z}}\sum_{n=0}^{l-2}c_n c_{l-n-2}.
	\label{eq:c_l_periodic}
\end{equation}
To obtain the polynomial for the full tunneling coefficient $t_N$, a similar recursion can be used. It slightly differs from the above due to the periodic boundary conditions. Unflipping a pair or a single spin in the ring, we are left with a single cluster of length $N-2$ or $N-1$. The unflipped spin(s) can be at $N$ positions, so that we find
\begin{equation}
	t_N = N( -h_x c_{N-1} + J_x c_{N-2} ).
	\label{eq:poly_periodic}
\end{equation}
In general the location of the field-zeros $h_x^{(n)}$ depends on $N$. Interestingly, it turns out that the pair of largest zeros, $\pm h_x^{(\floor{N/2})}$, is common to chains of any size and takes the value $h_x^{(\floor{N/2})} =  2\sqrt{J_x \vert J_z \vert}$. To show that this is indeed so, we start from Eq.~(\ref{eq:poly_periodic}). The condition to have
$
t_{l+1} = 0
$
requires
$
c_l = (J_x/h_x) c_{l-1}.
$
If this is to hold for all $l$ and given that $c_0 = 1$, we must have
$
c_l = {(J_x/h_x)}^l.
$
Using this in Eq.~(\ref{eq:c_l_periodic}) and simplifying we find that this relation is indeed satisfied if
$
h_x = \pm 2\sqrt{J_x \vert J_z \vert}.
$
For this value of $h_x$ the tunneling $t_N$ thus vanishes for any $N$. We will retrieve this result from a direct calculation of $t_N$ below.

\begin{figure}[t]
	\centering
	\includegraphics[width=0.42\linewidth]{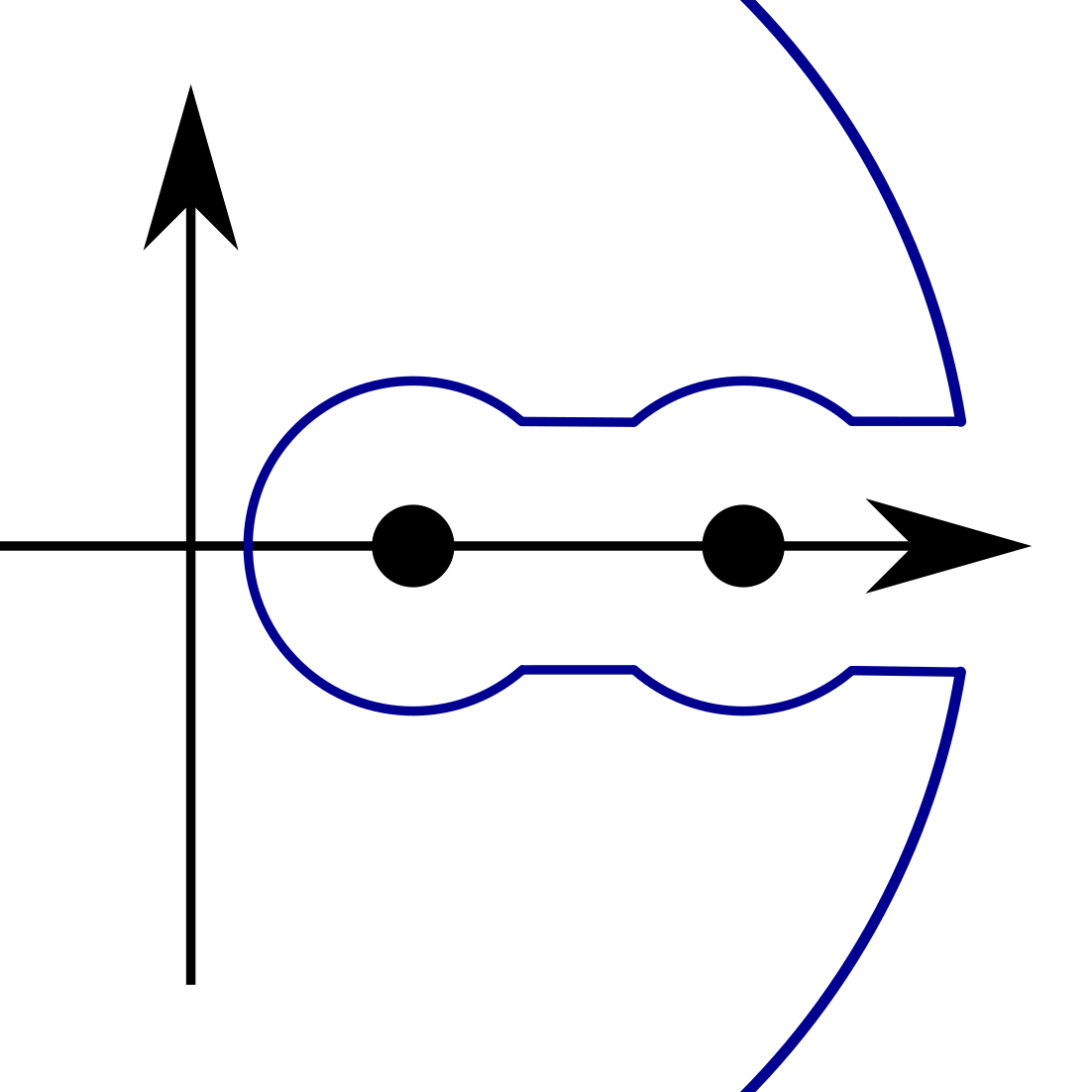}
	\put(-37,42){$z_+$}
	\put(-68,42){$z_-$}
	\put(-21,31){$1$}
	\put(-52,31){$2$}
	\put(-52,66){$3$}
	\put(-21,66){$4$}
	\put(-15,43){$x$}
	\put(-96,85){$y$}
	\includegraphics[width=0.42\linewidth]{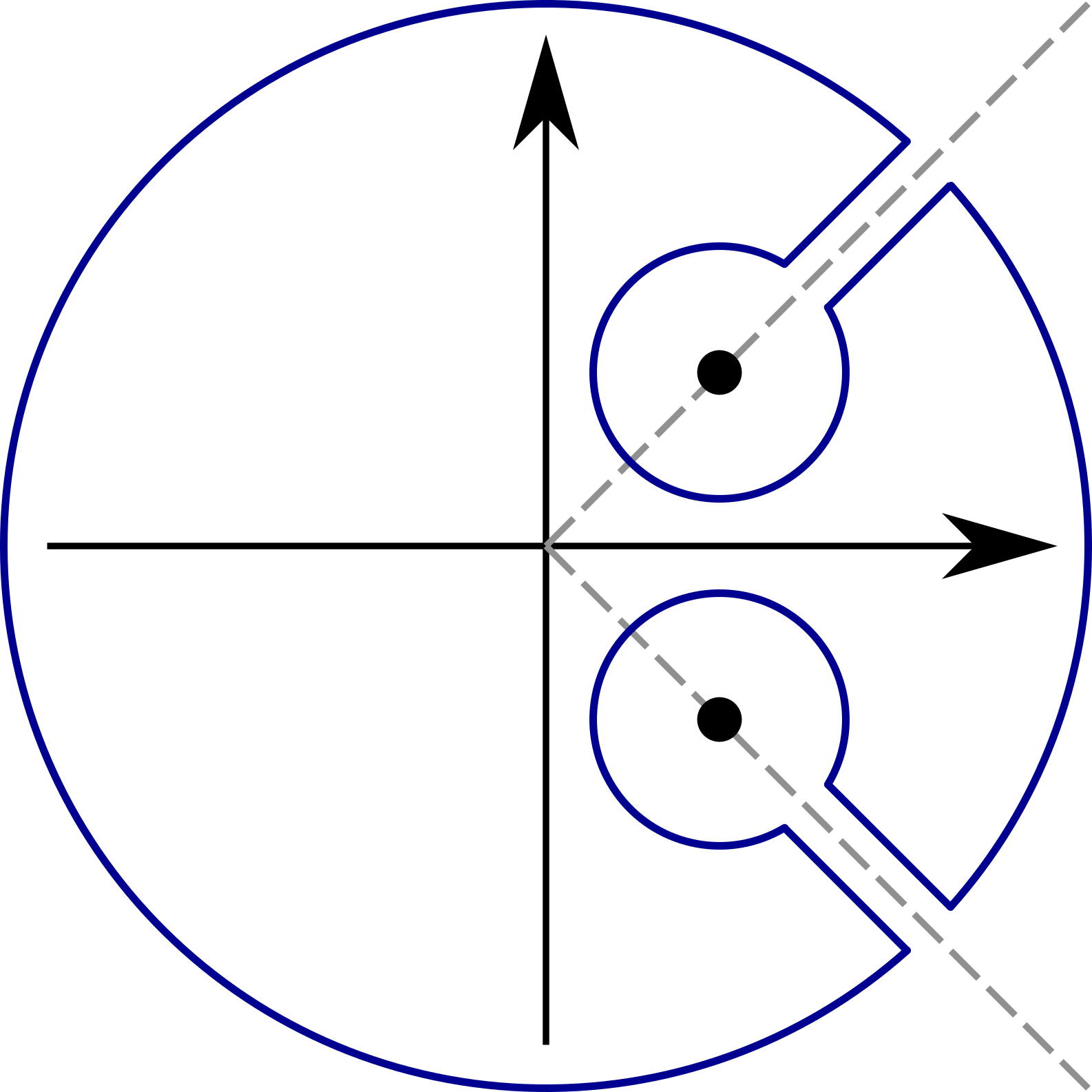}
	\put(-41,27){$z_-$}
	\put(-38,60){$z_+$}
	\put(-29,13){$1$}
	\put(-19,26){$2$}
	\put(-19,73){$3$}
	\put(-29,86){$4$}
	\put(-13,41){$x$}
	\put(-58,88){$y$}
	\caption{Keyhole contours of integration.}
	\label{fig:contours_t_N}
\end{figure}
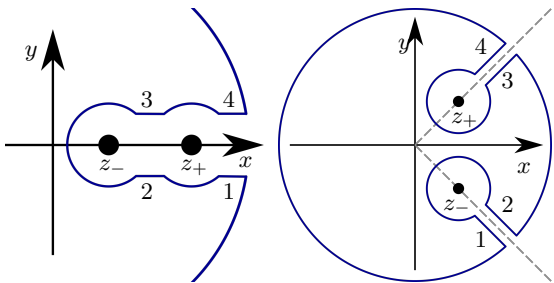

In order to calculate $t_N$ for any value of $h_x$, we define the generating functions
\begin{eqnarray}
	C(z) &=& \sum_{l = 0}^{\infty} z^l c_l, \\
	 T(z) &=& 2\abs{J_z} + \sum_{l = 1}^{\infty} z^l \frac{t_l}{l}, \nonumber
	\label{eq:C_T_definitions}
\end{eqnarray}
where $z$ is a complex variable. We then multiply Eqs.~(\ref{eq:c_l_periodic}, \ref{eq:poly_periodic}) by $z^l$, and sum them from $l=2$ to $l=\infty$. Solving for $C(z)$~\footnote{After summing over $l$ in Eq.~(\ref{eq:c_l_periodic}), one obtains terms with double summations of the form $\sum_{l=0}^{\infty} \sum_{n=0}^{l} $, which are equal to $\sum_{n=0}^{\infty} \sum_{l=n}^{\infty}$. Those lead to a term $C(z)^2$ on the right-hand side. Solving the resulting quadratic equation for $C(z)$, one has to choose the root that satisfies $\lim_{z\rightarrow 0} C(z) = c_0 = 1$. Using the same steps in Eq.~(\ref{eq:poly_periodic}) one obtains $T(z)$.} and eventually for $T(z)$, we obtain the closed expression
\begin{equation}
	\label{eq:T_z}
	T(z) = \sqrt{4J_x\abs{J_z} (z-z_+)(z-z_-)},
\end{equation}
where
\begin{equation}
	\label{eq:z_plus_minus}
	z_{\pm} = \frac{h_x \pm \sqrt{h_x^2 - 4 J_x \abs{J_z}}}{2J_x}.
\end{equation}
The above formula for $T(z)$ represents the power series (\ref{eq:C_T_definitions}) with its domain of convergence at small enough $z$, but analytically continues it beyond.
We can now calculate $t_N$ by contour integration of $T(z)/z^{N+1}$ around its pole at $z = 0$. We have the exact expression
\begin{equation}
	\frac{t_N}{N} = \frac{1}{2 \pi i} \oint_{z = 0} \frac{T(z)}{z^{N+1}} dz.
	\label{eq:t_N_contour}
\end{equation}
We now deform the contour, pushing it to infinity, but avoiding the branch cuts ending at $z_{\pm}$. This is best done using keyhole contours (Fig.~\ref{fig:contours_t_N}). The precise contour used depends on whether $z_\pm$ are real or a pair of complex conjugate numbers, which we discuss separately. We restrict ourselves to $J_x>0$, since only in that case $t_N$ exhibits interesting oscillations.

\paragraph{$h_x^2 > 4J_x \abs{J_z}$ -}

In this case, both $z_{\pm}$ are real and either both positive or both negative, depending on the sign of $h_x$. Without loss of generality, we take $h_x>0$. The appropriate contour is shown in Fig.~\ref{fig:contours_t_N} on the left. Pushing the radius of the large circle to infinity and letting the radius of the small circles around $z_\pm$ shrink to zero, the integrals $1$ and $4$ cancel, while the integrals $2$ and $3$ between the branch points add up to
\begin{equation}
	t_N = - \frac{2N \sqrt{J_x \abs{J_z}}}{\pi} \int_{z_-}^{z_+} \frac{\sqrt{(z_+-x)(x-z_-)}}{x^{N+1}} dx.
	\label{eq:int_periodic_real_branch}
\end{equation}
Note that upon changing $h_x$ within the domain $h_x^2 > 4J_x \abs{J_z}$, the integrand remains positive, and thus $t_N$ never becomes zero.

\paragraph{$h_x^2 = 4J_x \abs{J_z}$ -}

At the border of the above domain one has $z_{+} = z_{-}$. From Eq.~(\ref{eq:int_periodic_real_branch}) one sees that the tunneling becomes zero at this point, independently of $N$, as we have already found previously. Since there are no zeros at higher fields, this corresponds to the largest field zero.

\paragraph{$h_x^2 < 4J_x \abs{J_z}$ -}

Here the branch points become a pair of complex conjugate numbers. We have
\begin{equation}
	z_{\pm} = r_0 e^{\pm i \theta_0},
\end{equation}
where
\begin{equation}
	r_0 = \sqrt{\frac{\abs{J_z}}{J_x}},\quad
	\cos(\theta_0) = \frac{h_x}{\sqrt{4 J_x\abs{J_z}}}.
\end{equation}
We consider the contour shown in Fig.~\ref{fig:contours_t_N} on the right, where the branch points $z_{\pm}$ are avoided with keyholes oriented radially along the lines
$
z = r e^{\pm i \theta_0}.
$
Shrinking the small circles to zero, and expanding the large circle to infinity, the expression for $t_N$ simplifies to the contribution of two line integrals along the radial branch cuts, resulting in the exact expression:
\begin{equation}
	\begin{aligned}
	& t_N = \frac{4N\sqrt{J_x\abs{J_z}}}{\pi} \times \\
	& \mathrm{Im}\left(
		e^{i\theta_0(\frac{1}{2}-N)}
		\int_{r_0}^{\infty}
		\frac{ \sqrt{ ( r - r_0 ) ( r e^{i \theta_0} - r_0 e^{-i \theta_0}} ) } {r^{N+1}}dr
	\right).
	\end{aligned}
\end{equation}
At large $N$, we can make progress by replacing
$
r e^{i \theta_0} - r_0 e^{-i \theta_0}
$
by its value at $r = r_0$ (which is valid as long as $\theta_0\gg 1/N$). The remaining integral can be written in terms of Gamma functions. To leading order at large $N$ one obtains
\begin{equation}
	t_N \approx \abs{J_z} \left(\frac{ J_x}{\abs{J_z}}\right)^{N/2}
	\sqrt{\frac{8 \sin \theta_0}{\pi N}}
	\sin\left(\theta_0\left(\frac{1}{2}-N\right) + \frac{\pi}{4}\right).
	\label{eq:periodic_solution}
\end{equation}
Note that this expression has, however, $N+2$ zeros as a function of $h_x$: $N$ zeros arise from the vanishing of the high frequency sine at fields given by
\begin{equation}
	h_x^{(n)} = \pm 2\sqrt{J_x \abs{J_z}}
	\cos\left(\frac{\floor{N/2}+\frac{1}{4}-n}{N-\frac{1}{2}} \pi \right),
	\label{eq:periodic_zeros}
\end{equation}
with $n = 1,\dots,\floor{N/2}$. Two further zeros are due to the vanishing of
$
\sin \theta_0.
$
Those reproduce correctly the pair of largest field zeros, $h_x = \pm 2\sqrt{J_x \abs{J_z}}$, which we have already identified above.
The two zeros $h_x^{(n)}$ with $n=\floor{N/2}$ are instead artefacts that are introduced by approximating the numerator in the integrand with its value at $r=r_0$. This  approximation is not controlled  in that field regime since there one has $\theta_0 = O(1/N)$. These two zeros thus have to be discarded, and we are left with $N$ zeros, as it should be.

\begin{figure}[t]
	\centering
	\includegraphics[width=0.49\columnwidth]{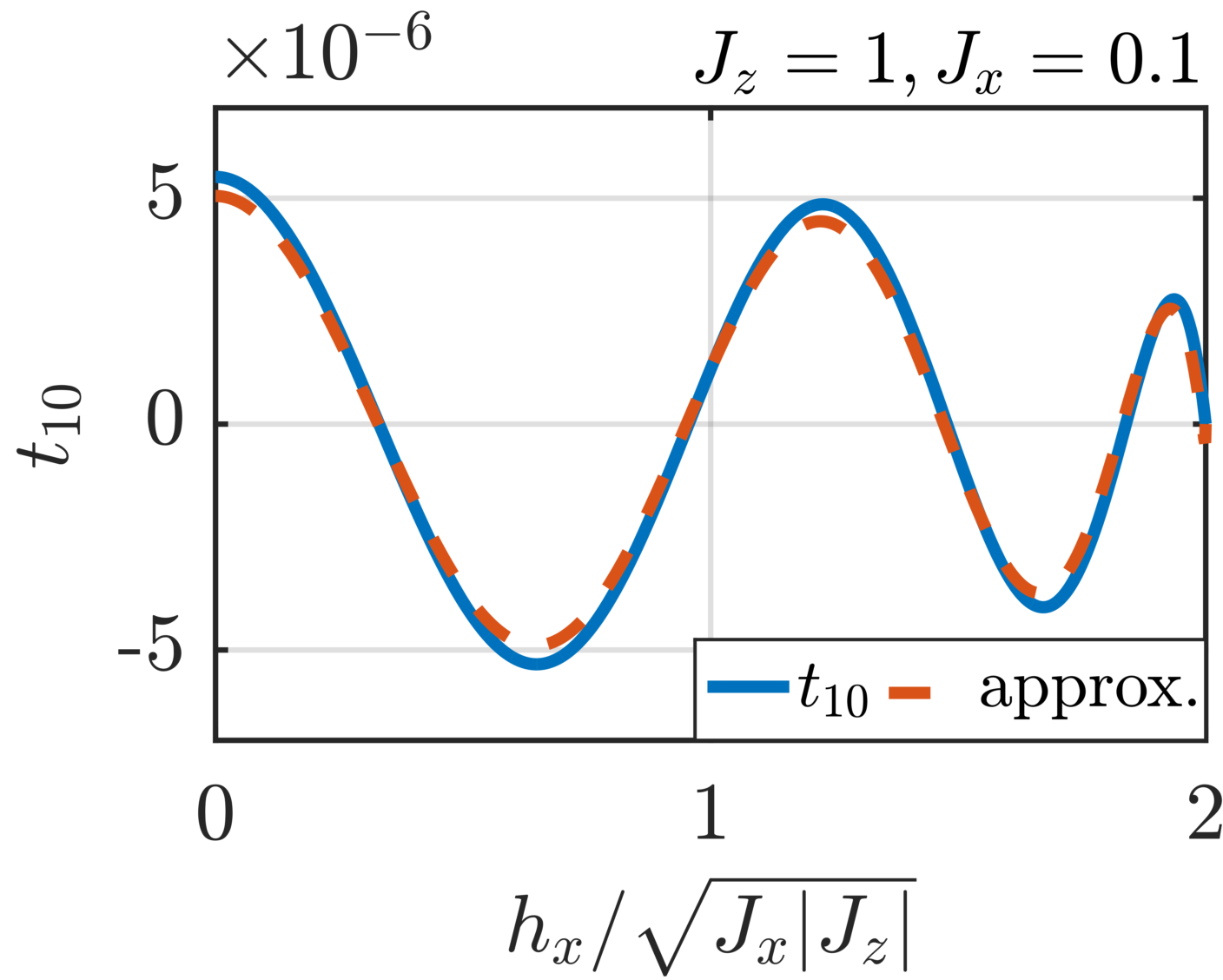}
	\includegraphics[width=0.49\columnwidth]{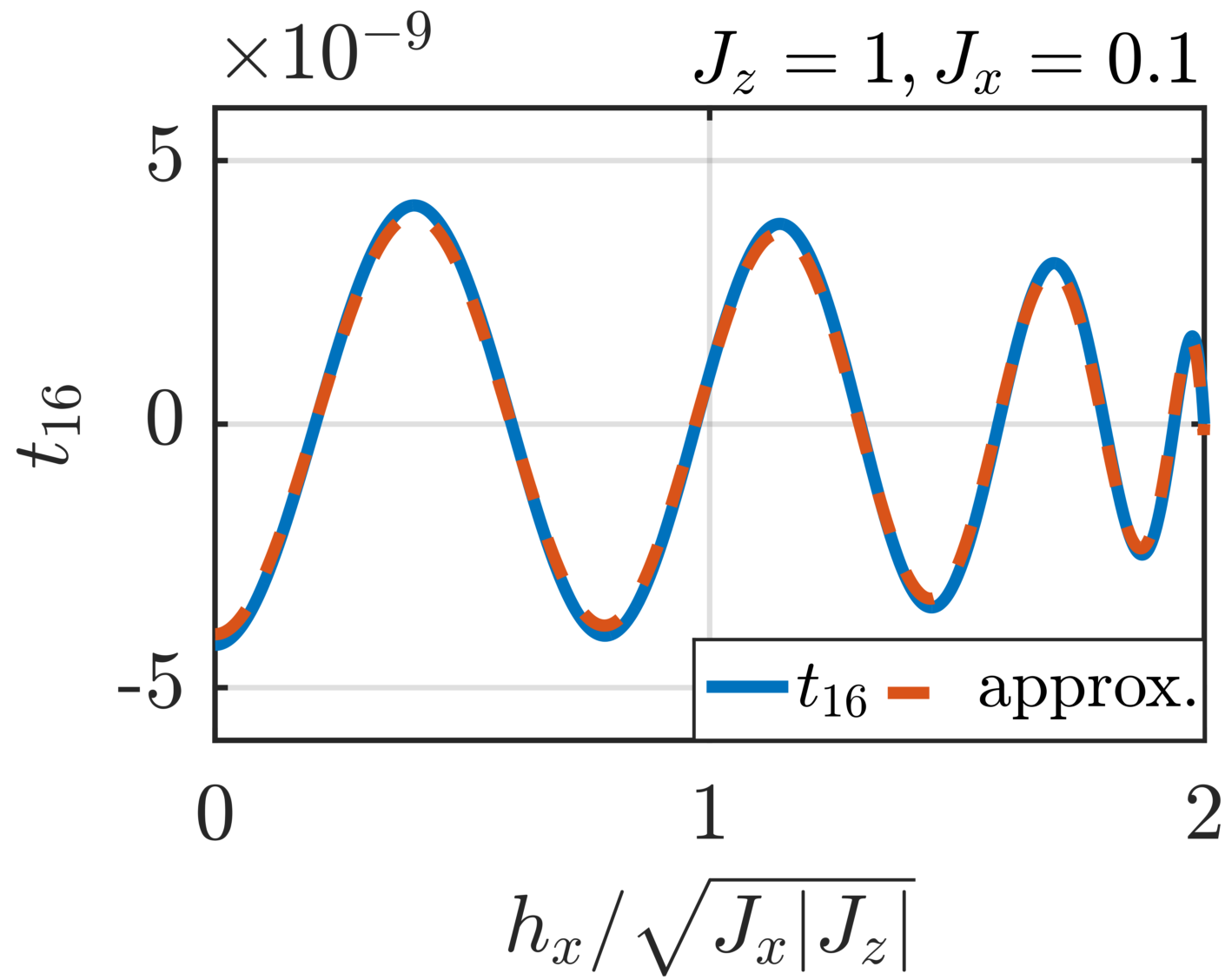}
	\caption{Comparison of the exact tunneling polynomial of a ring of spins with the asymptotic expression of Eq.~(\ref{eq:periodic_solution}), for $N = 10$ on the left and $N = 16$ on the right. The exact $t_N$ is obtained by explicitly solving the recursion relations in Eqs.~(\ref{eq:c_l_periodic}, \ref{eq:poly_periodic}). The asymptotic expression approximates the exact polynomial better and better as the system size increases.}
	\label{fig:DPT_int_periodic}
\end{figure}

In Fig.~\ref{fig:DPT_int_periodic} we compare the asymptotic Eq.~(\ref{eq:periodic_solution}) with the exact polynomial for $t_N$ obtained from explicitly solving Eqs.~(\ref{eq:c_l_periodic}, \ref{eq:poly_periodic}). The agreement is very good even for moderate $N$, and it further improves with system size.

%

\subsection{Open Ising spin chains}%
\label{sub:spin_chain}

In open chains, connected, flipped clusters that touch an end of the chain create only one domain wall. Accordingly, their excitation energy is only $2|J_z|$, half that of a bulk cluster. We define $d_l$ as the intermediate tunneling coefficient associated to such an edge cluster of size $l$, while $c_l$ is again that associated to bulk clusters. These coefficients satisfy similar recursion relations as before. The one for $c_l$ is unchanged, while for $d_l$ one finds
\begin{equation}
	d_l = \frac{h_x}{2\abs{J_z}} \sum_{n=0}^{l-1} d_n c_{l - n -1} -
	\frac{J_x}{2\abs{J_z}} \sum_{n=0}^{l-2} d_n c_{l - n - 2},
\end{equation}
where we again defined $c_0 \equiv 1$ and $d_0 \equiv 1$. The fully flipped state can only be created from edge clusters. We thus have
\begin{equation}
	t_N = - h_x\sum_{n=0}^{N-1} d_n d_{l - n - 1} +
	J_x\sum_{n=0}^{N-2} d_n d_{l - n - 2}.
\end{equation}
To proceed we again use the previously defined generating function $C(z)$ and define
\begin{equation}
	D(z) = \sum_{l = 0}^{\infty} z^l d_l, \quad \hat{T}(z) = \sum_{l = 1}^{\infty} z^l t_l.
\end{equation}
Again, multiplying the recursion relations by $z^l$, summing over $l$, and solving for $\hat{T}(z)$, we obtain
\begin{equation}
	\hat{T}(z) = \abs{J_z}\frac{z(z-h_x/J_x)}{(z-z_+)(z-z_-)},
\end{equation}
where the singularities $z_{\pm}$ are still given by Eq.~(\ref{eq:z_plus_minus}). However, here they appear as poles of $\hat{T}(z)$, not as branch points. In this case, the contour integral around $z=0$ can be transformed into a simple contour around the two poles, which yields the exact result for $t_N$ as a sum of two residues:
\begin{align}
\label{eq:open_solution}
\nonumber t_N &= \underset{z=0}{\rm{Res}} \left(\frac{\hat{T}(z)}{z^{N+1}}\right) = -\underset{z=z_+}{\rm{Res}} \left(\frac{\hat{T}(z)}{z^{N+1}}\right) -\underset{z=z_-}{\rm{Res}} \left(\frac{\hat{T}(z)}{z^{N+1}}\right) \\
			 &= \frac{\abs{J_z}}{(z_+ - z_-)}\left( \frac{z_-}{z_+^N} - \frac{z_+}{z_-^N} \right)\\
\nonumber &= \left\{
	\begin{array}{lr}
		\sim -\abs{J_z}{\left(\frac{h_x}{\abs{J_z}}\right)}^{N}, & \frac{h_x^2}{4\abs{J_z}} \gg J_x, \\
		-\abs{J_z}(N+1){\left(\frac{h_x}{2\abs{J_z}}\right)}^{N}, & \frac{h_x^2}{4\abs{J_z}} = J_x, \\
		-\abs{J_z}\left(\frac{J_x}{\abs{J_z}}\right)^{N/2} \frac{\sin\left[ (N+1) \theta_0 \right]}{\sin \left( \theta_0 \right)} , & \frac{h_x^2}{4\abs{J_z}} < J_x.
\end{array} \right.
\end{align}
Like for the closed chain, $t_N$ oscillates when
$
h_x^2 < 4J_x\abs{J_z}.
$
The high frequency sine function in the oscillatory regime has $N+2$ zeros, but when
$
h_x^2 = 4J_x \abs{J_z}
$
the denominator vanishes too, and $t_N$ does not vanish. We are thus left with $N$ zeros at the fields
\begin{equation}
	h_x^{(n)} = \pm 2\sqrt{J_x \abs{J_z}}\cos\left(\frac{\floor{N/2}+1-n}{N+1}\pi\right),
	\label{eq:closed_zeros}
\end{equation}
with $n = 1,\dots,\floor{N/2}$.

Upon comparing the position of the zeros for open boundary conditions, Eq.~(\ref{eq:closed_zeros}), with those for periodic boundary conditions, Eq.~(\ref{eq:periodic_zeros}), one finds that closing the chain shifts all fields $h_x^{(n)}$ to higher values. This is expected since a closed chain contains one more bond $J_x$, so that $h_x$ must slightly increase to compensate the increased exchange contribution to the tunneling.

\section[2D and 3D clusters]{2D and 3D clusters}%
\label{sec:2d_and_3d_clusters}

\begin{table}
	\centering
	\begin{tabularx}{\columnwidth}{XlXl}
		\Fig{0.08}{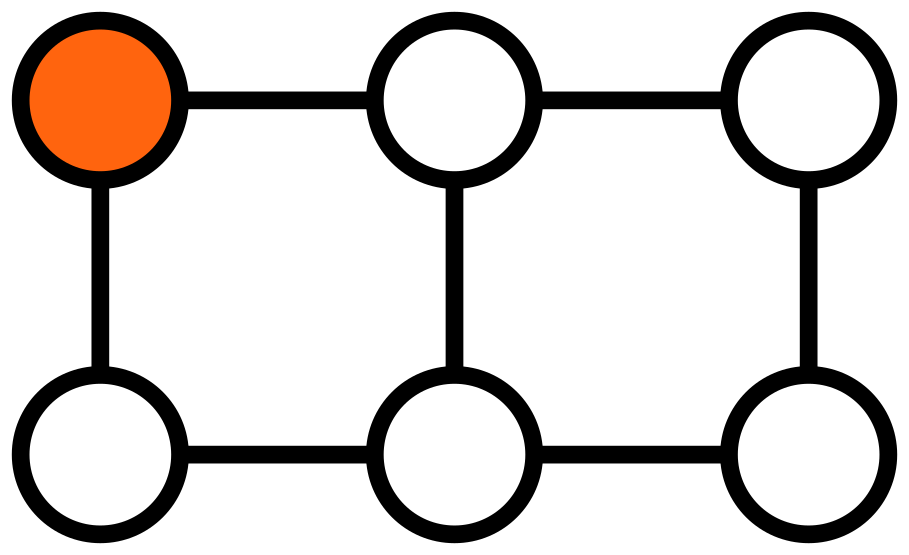} & $ c_{1,1} = \frac{h_x }{4 \abs{J_z}} $ & \Fig{0.08}{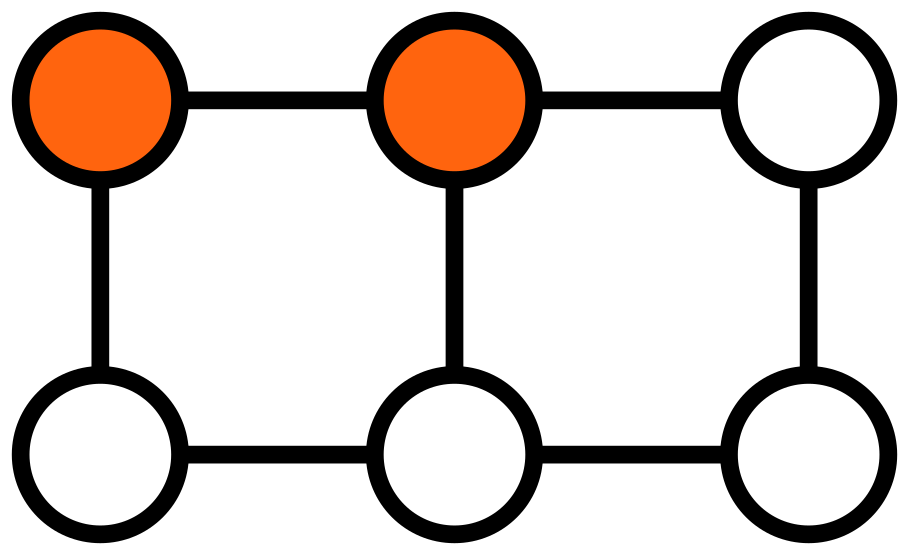} & $ c_{2,1} = \frac{h_x( c_{1,1} + c_{1,2} ) -J_x }{6 \abs{J_z}} $ \\
		\Fig{0.08}{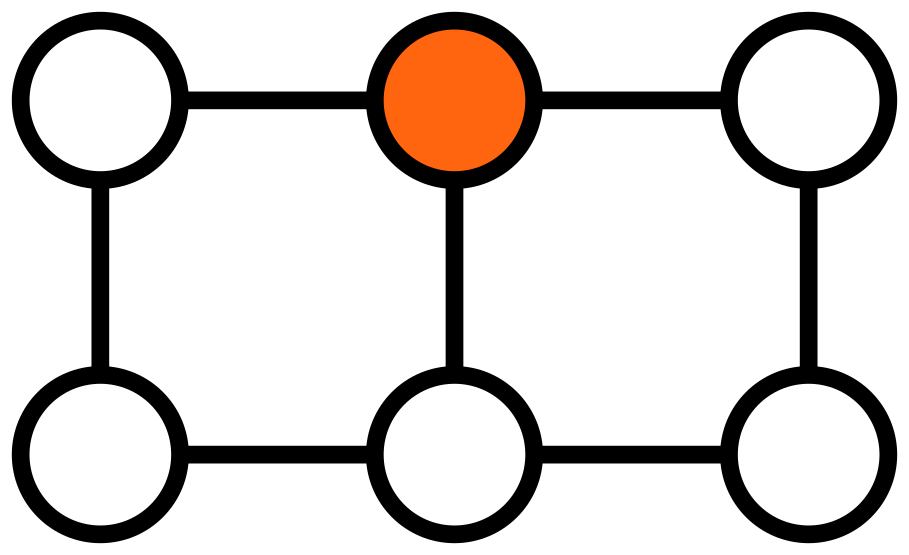} & $ c_{1,2} = \frac{h_x }{6 \abs{J_z}} $ & \Fig{0.08}{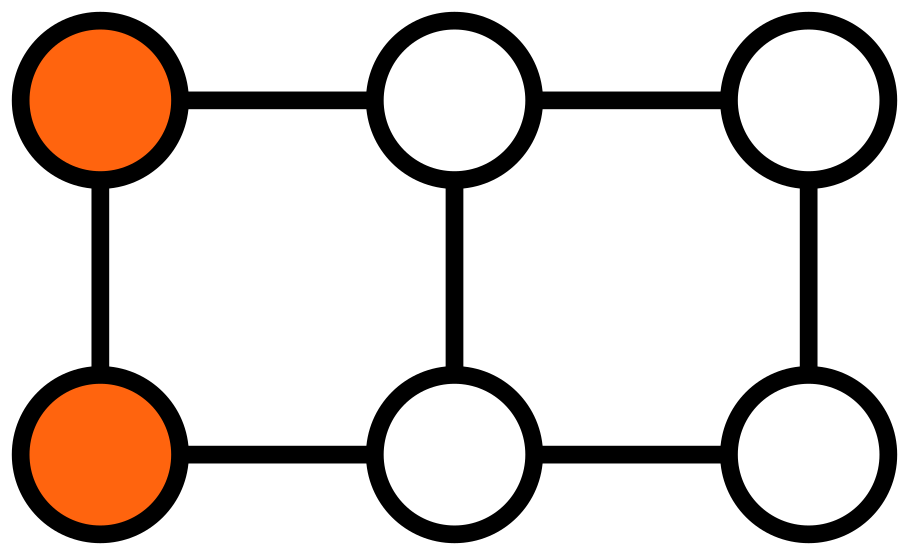} & $ c_{2,2} = \frac{h_x( 2 c_{1,1} ) -J_x }{4 \abs{J_z}} $ \\
		\Fig{0.08}{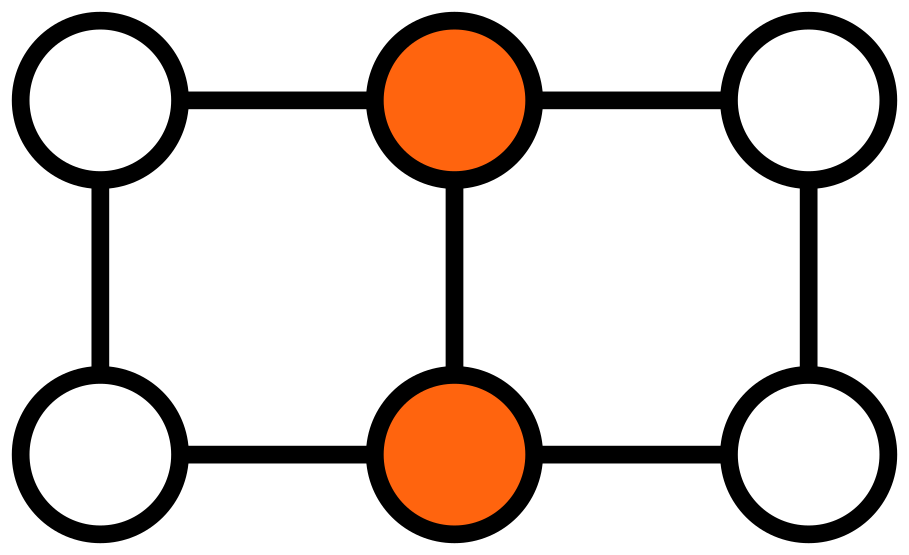} & \multicolumn{3}{l}{$ c_{2,3} = \frac{h_x( 2 c_{1,2} ) -J_x }{8 \abs{J_z}} $} \\
		\Fig{0.08}{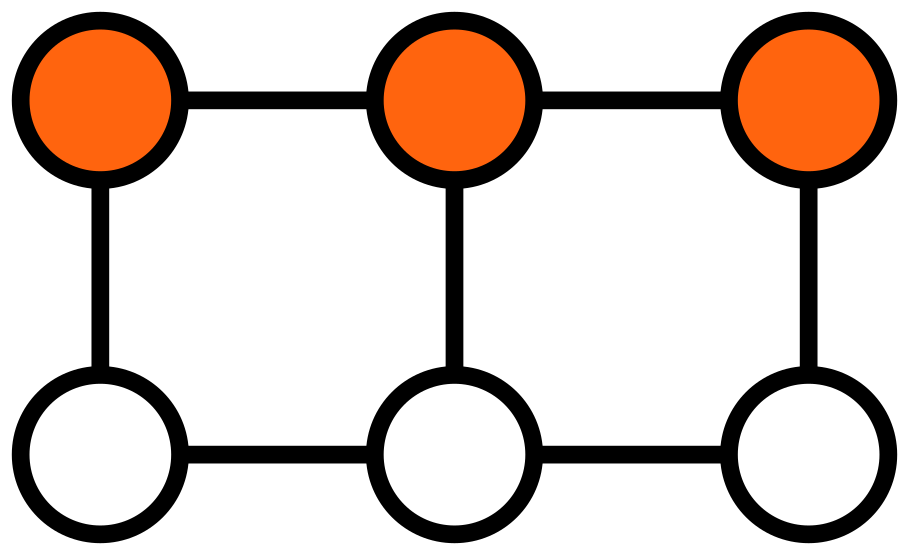} & \multicolumn{3}{l}{$ c_{3,1} = \frac{h_x( 2 c_{2,1} + c_{1,1}^2 ) -J_x( 2 c_{1,1} ) }{6 \abs{J_z}} $} \\
		\Fig{0.08}{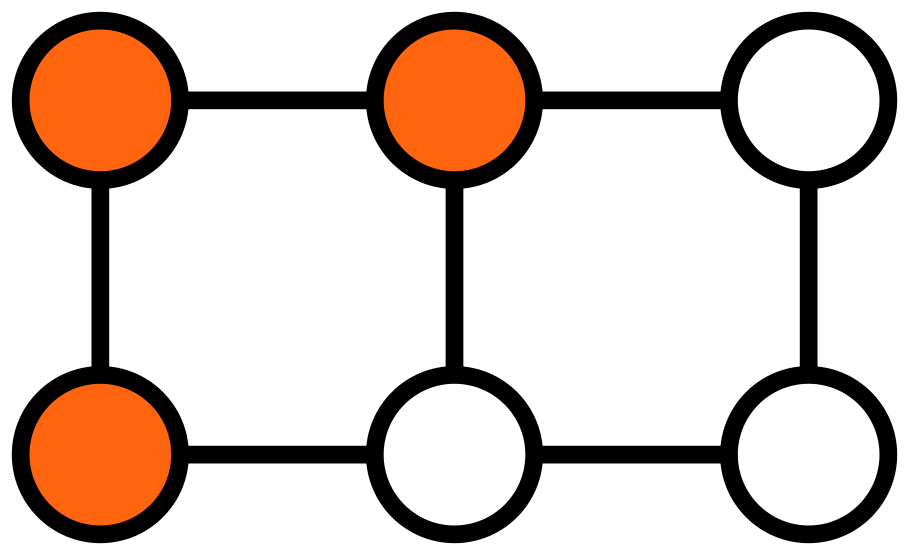} & \multicolumn{3}{l}{$ c_{3,2} = \frac{h_x( c_{2,1} + c_{2,2} + c_{1,1} c_{1,2} ) -J_x( c_{1,1} + c_{1,2} ) }{6 \abs{J_z}} $} \\
		\Fig{0.08}{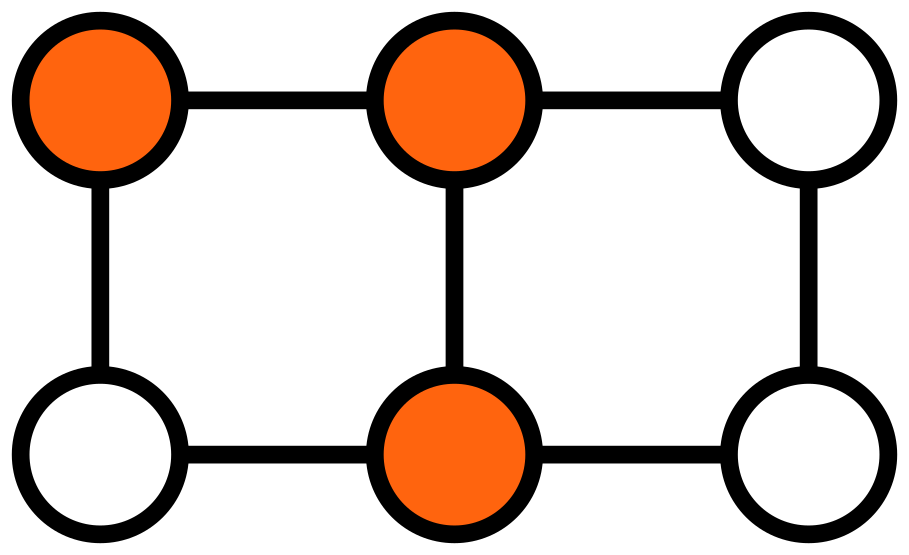} & \multicolumn{3}{l}{$ c_{3,3} = \frac{h_x( c_{2,1} + c_{2,3} + c_{1,1} c_{1,2} ) -J_x( c_{1,1} + c_{1,2} ) }{8 \abs{J_z}} $} \\
		\Fig{0.08}{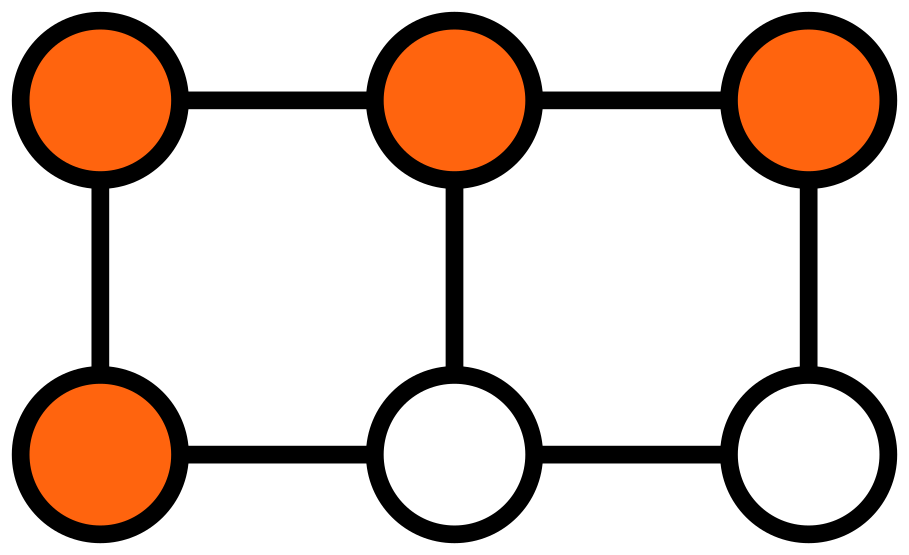} & \multicolumn{3}{l}{$ c_{4,1} = \frac{h_x( c_{3,1} + c_{3,2} + c_{2,1} c_{1,1} + c_{2,2} c_{1,1} ) -J_x( c_{2,1} + c_{1,1}^2 + c_{2,2} ) }{6 \abs{J_z}} $} \\
		\Fig{0.08}{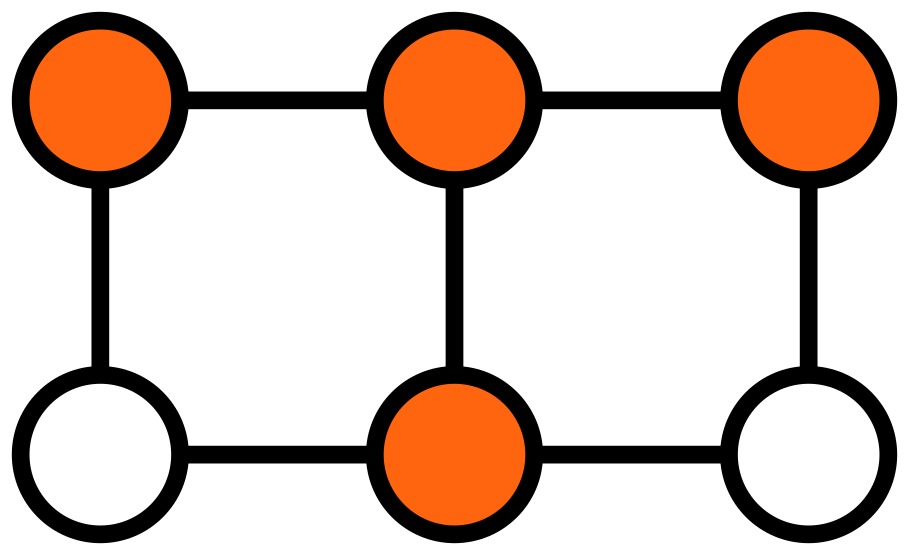} & \multicolumn{3}{l}{$ c_{4,2} = \frac{h_x( c_{3,1} + 2 c_{3,3} + c_{1,1}^2 c_{1,2} ) -J_x( 2 c_{1,1} c_{1,2} + c_{1,1}^2 ) }{8 \abs{J_z}} $} \\
		\Fig{0.08}{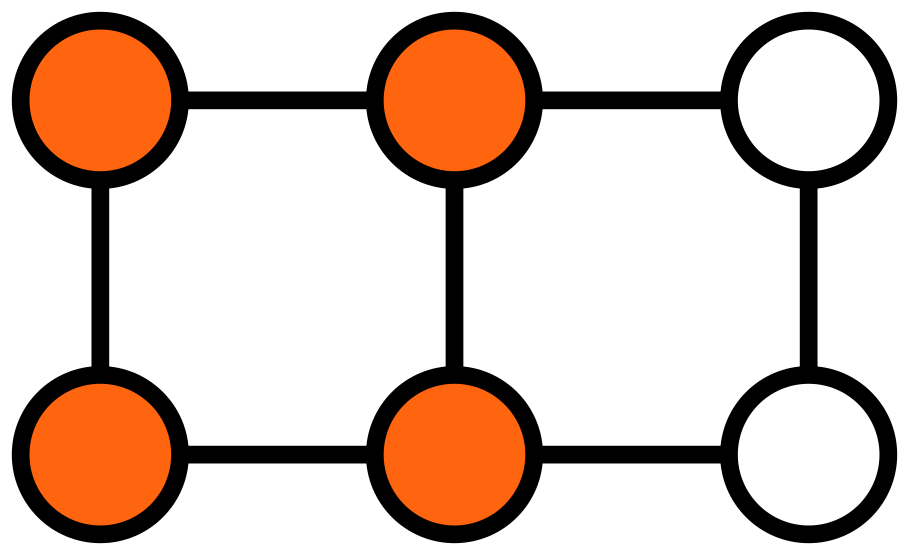} & \multicolumn{3}{l}{$ c_{4,3} = \frac{h_x( 2 c_{3,2} + 2 c_{3,3} ) -J_x( 2 c_{2,1} + c_{2,2} + c_{2,3} ) }{4 \abs{J_z}} $} \\
		\Fig{0.08}{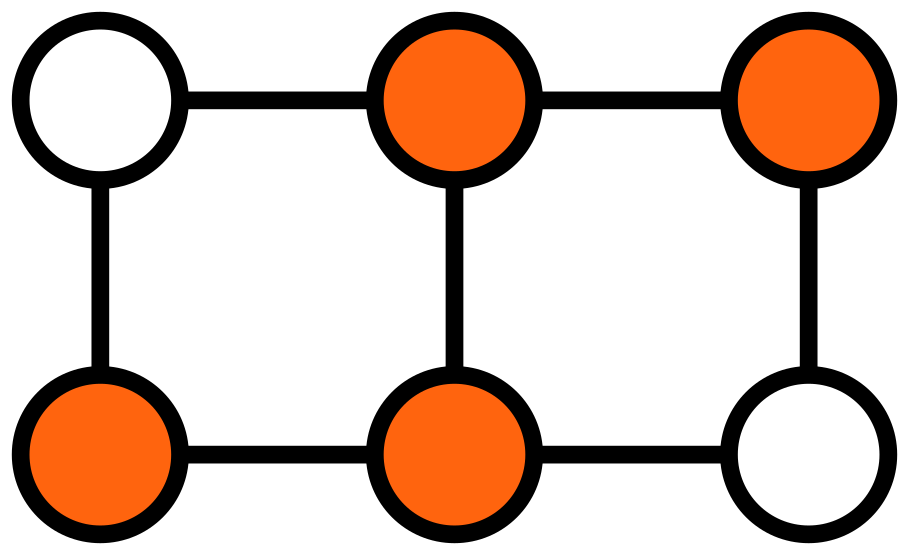} & \multicolumn{3}{l}{$ c_{4,4} = \frac{h_x( 2 c_{3,3} + 2 c_{2,1} c_{1,1} ) - J_x( 2 c_{2,1} + c_{1,1}^2) }{8 \abs{J_z}} $} \\
		\multirow{2}{*}{\Fig{0.08}{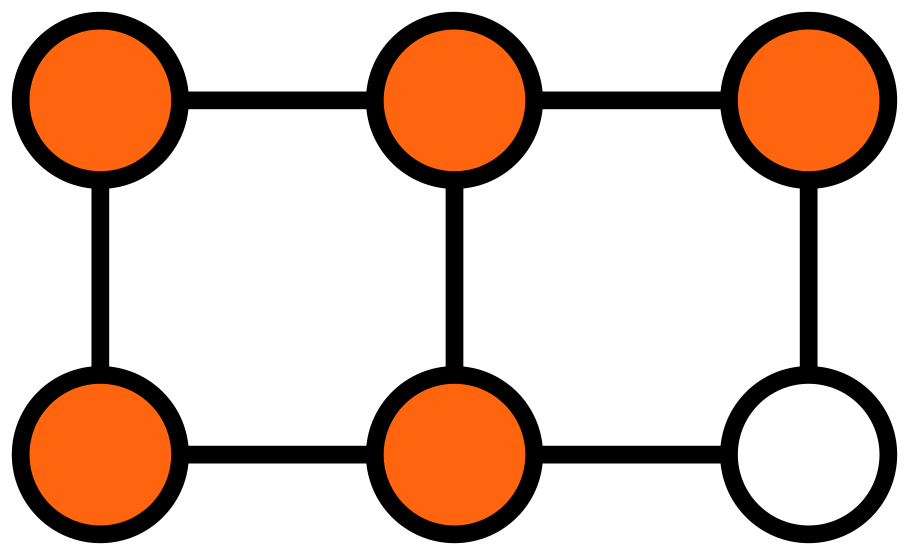}} & \multicolumn{3}{l}{$ c_{5,1} = \left[ h_x( c_{4,1} + c_{4,2} + c_{4,3} + c_{4,4} + c_{3,2} c_{1,1} ) \right. $} \\
													 & \multicolumn{3}{l}{$\left. -J_x( c_{3,1} + c_{3,2} + c_{3,3} + c_{2,2} c_{1,1} )\right] /(4 \abs{J_z})$} \\
		\Fig{0.08}{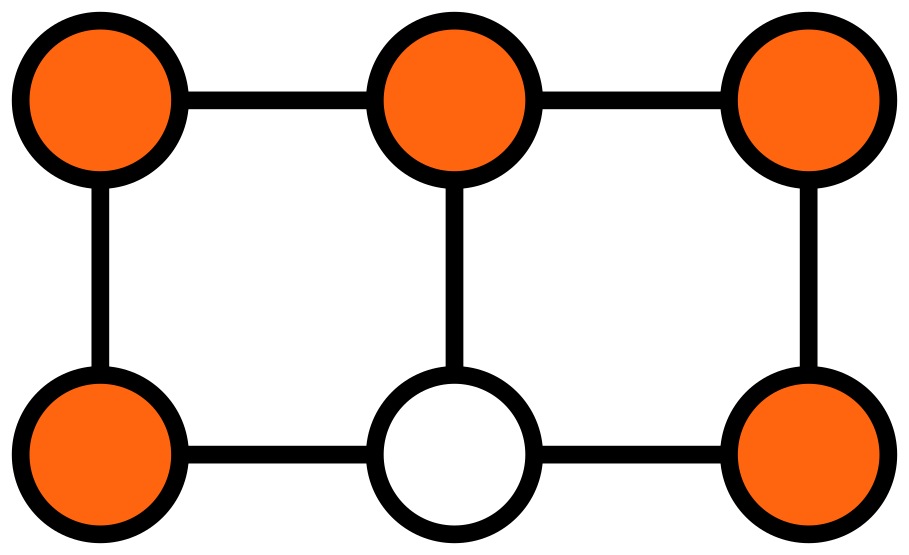} & \multicolumn{3}{l}{$ c_{5,2} = \frac{h_x( 2 c_{4,1} + c_{2,2}^2 + 2 c_{3,2} c_{1,1} ) -J_x( 2 c_{3,2} + 2 c_{2,2} c_{1,1} ) }{6 \abs{J_z}} $} \\
		\Fig{0.08}{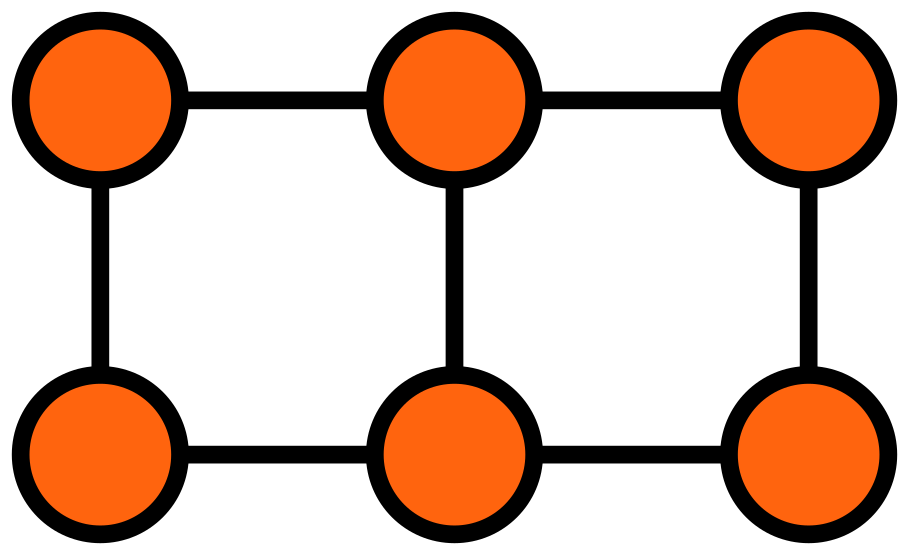} & \multicolumn{3}{l}{$ t_{3 \times 2} = - h_x( 4 c_{5,1} + 2 c_{5,2} ) + J_x( 2 c_{4,3} + c_{2,2}^2 + 4 c_{4,1} )$} \\
	\end{tabularx}
	\caption{Inequivalent connected clusters of a rectangular cluster of $6$ spins. Filled orange circles represent flipped spins. The intermediate tunneling coefficients satisfy the indicated recursion relations.}
	\label{tab:cluster_relations_rectangular}
\end{table}

\begin{table}
	\centering
	\begin{tabularx}{\columnwidth}{XlXl}
		\Fig{0.08}{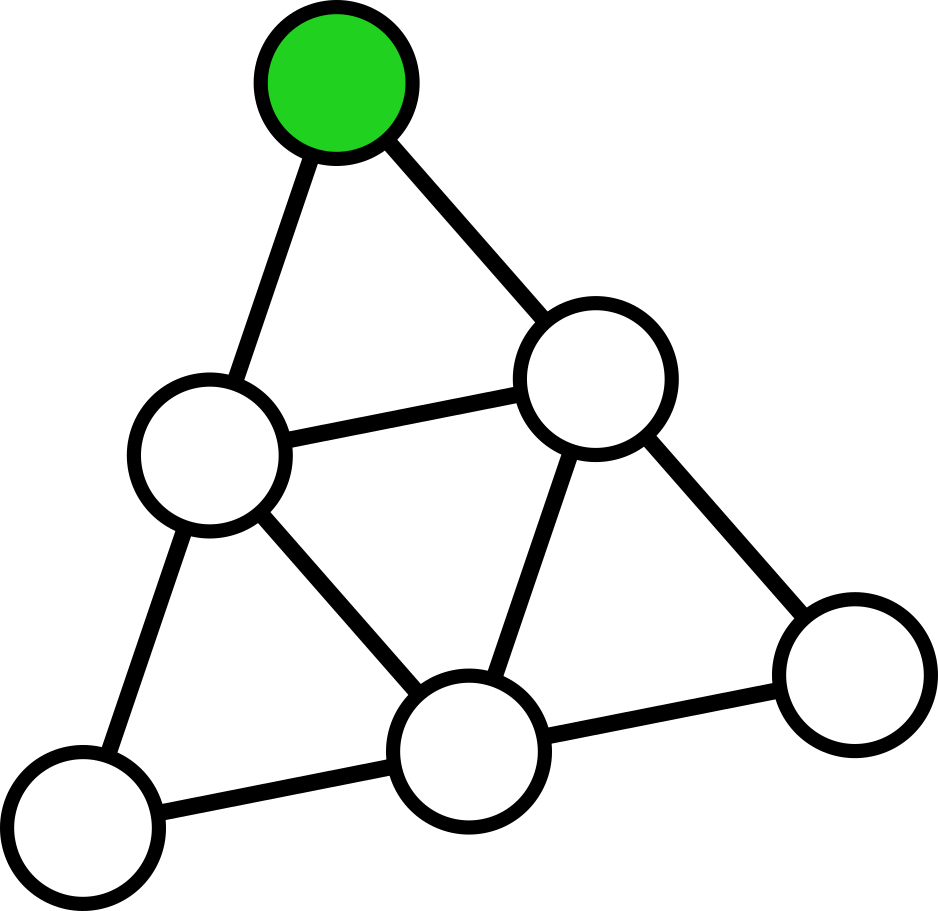} & $ c_{1,1} = \frac{ h_x }{4 \abs{J_z}} $ & \Fig{0.08}{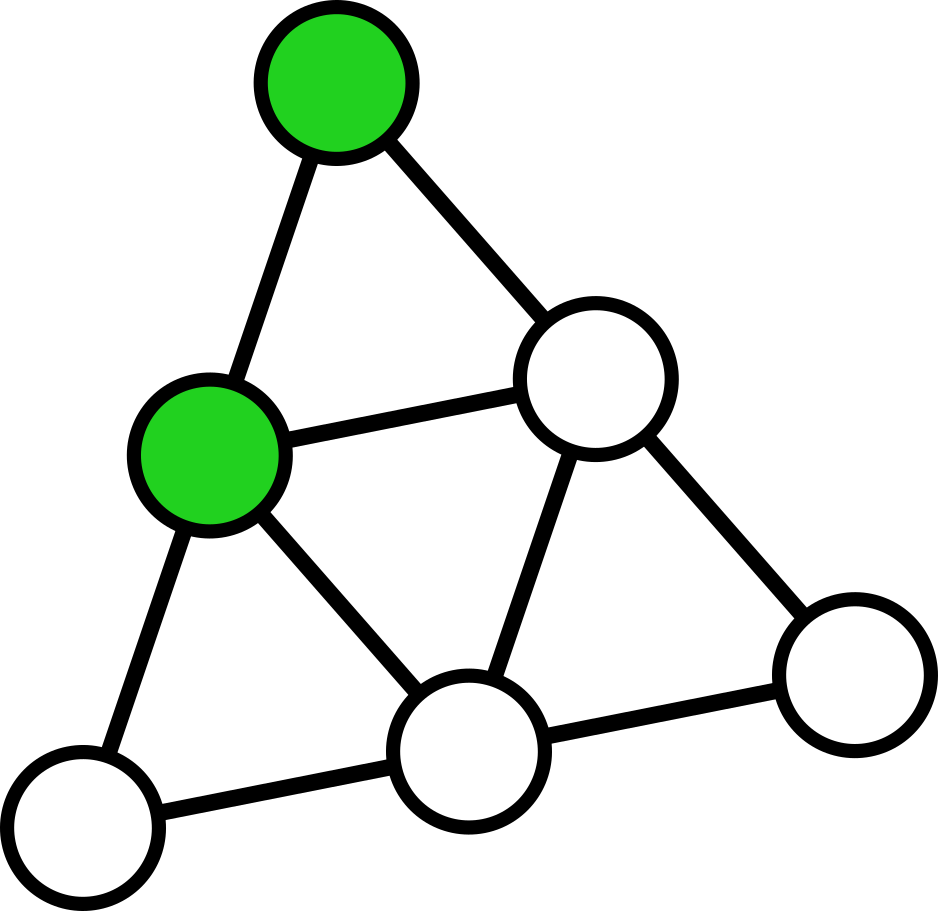} & $ c_{2,1} = \frac{ h_x( c_{1,1} + c_{1,2} ) -J_x }{8 \abs{J_z}} $ \\
		\Fig{0.08}{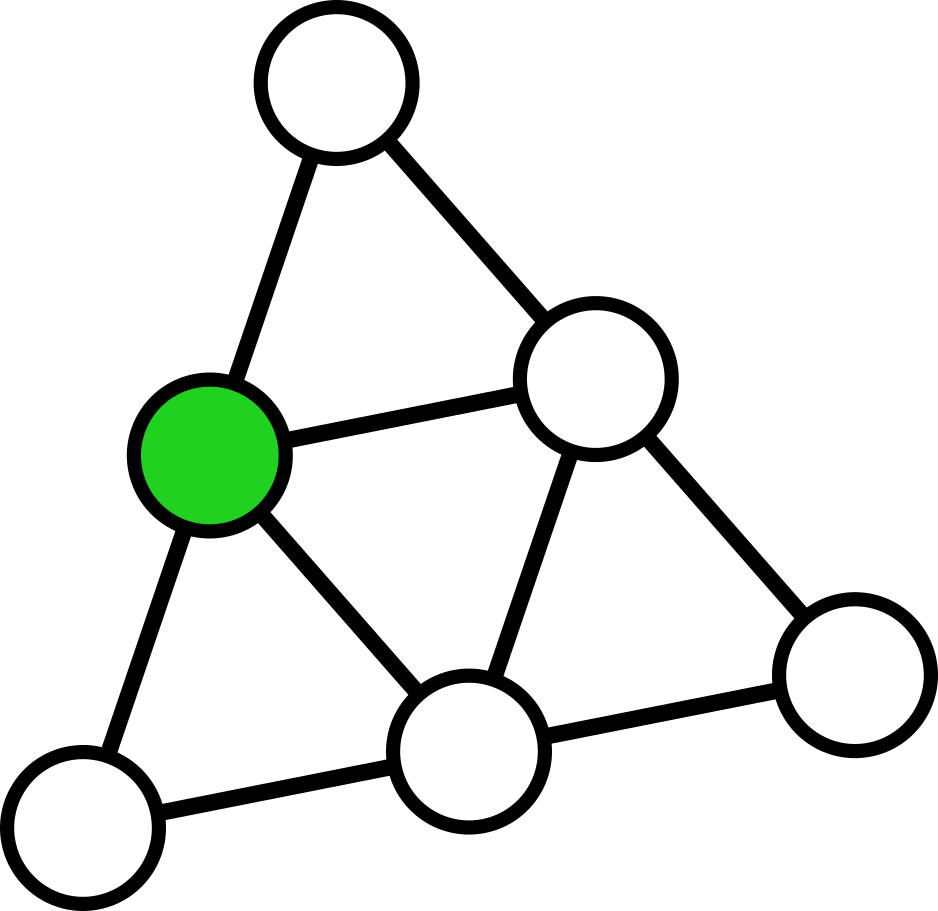} & $ c_{1,2} = \frac{ h_x }{8 \abs{J_z}} $ & \Fig{0.08}{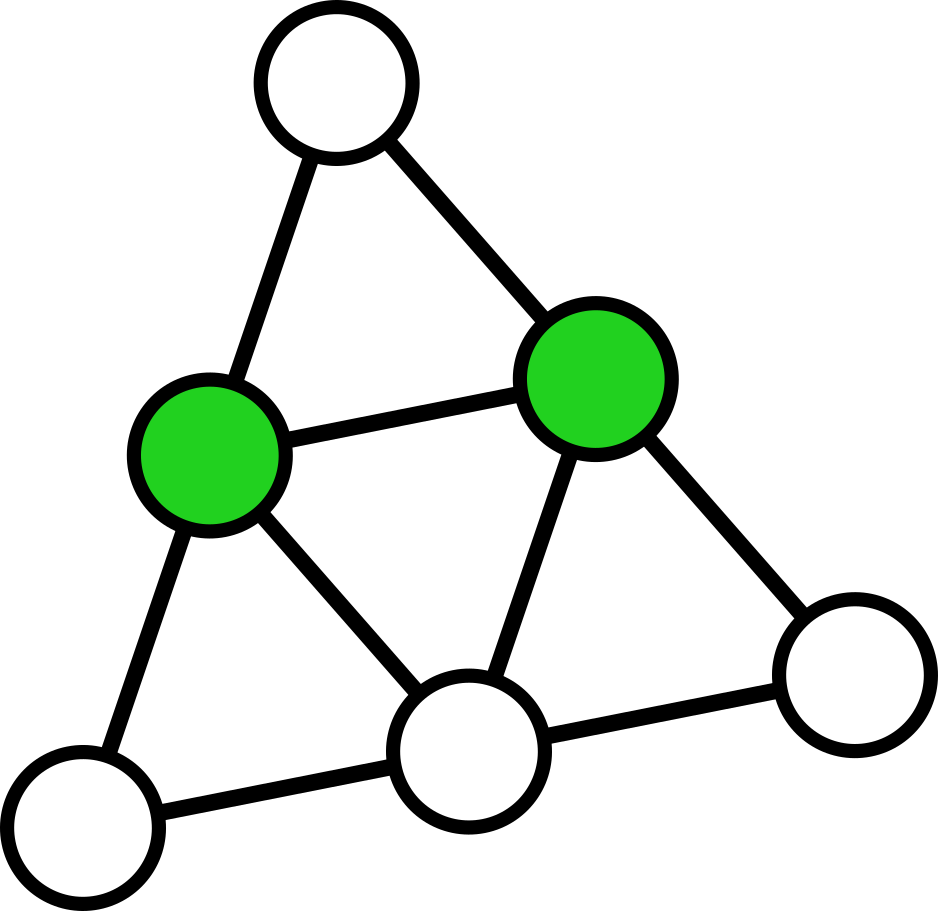} & $ c_{2,2} = \frac{ h_x( 2 c_{1,2} ) -J_x }{12 \abs{J_z}} $ \\
		\Fig{0.08}{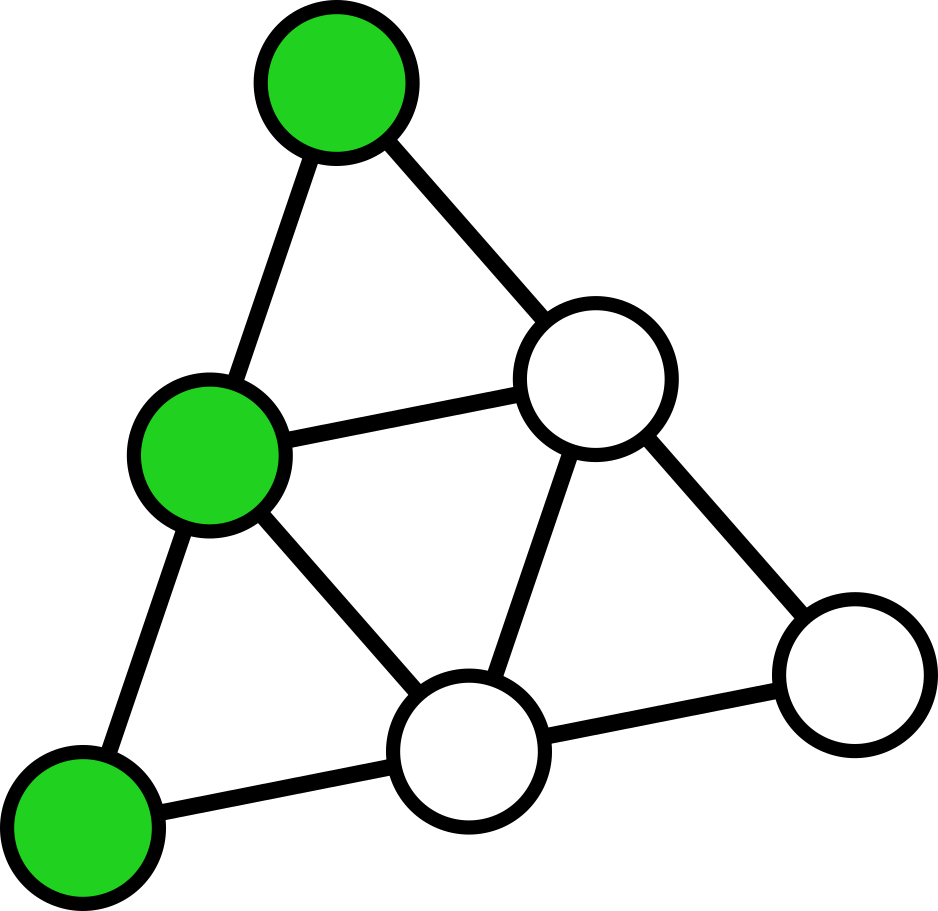} & \multicolumn{3}{l}{$ c_{3,1} = \frac{ h_x( 2 c_{2,1} + c_{1,1}^2 ) -J_x( 2 c_{1,1} ) }{8 \abs{J_z}} $} \\
		\Fig{0.08}{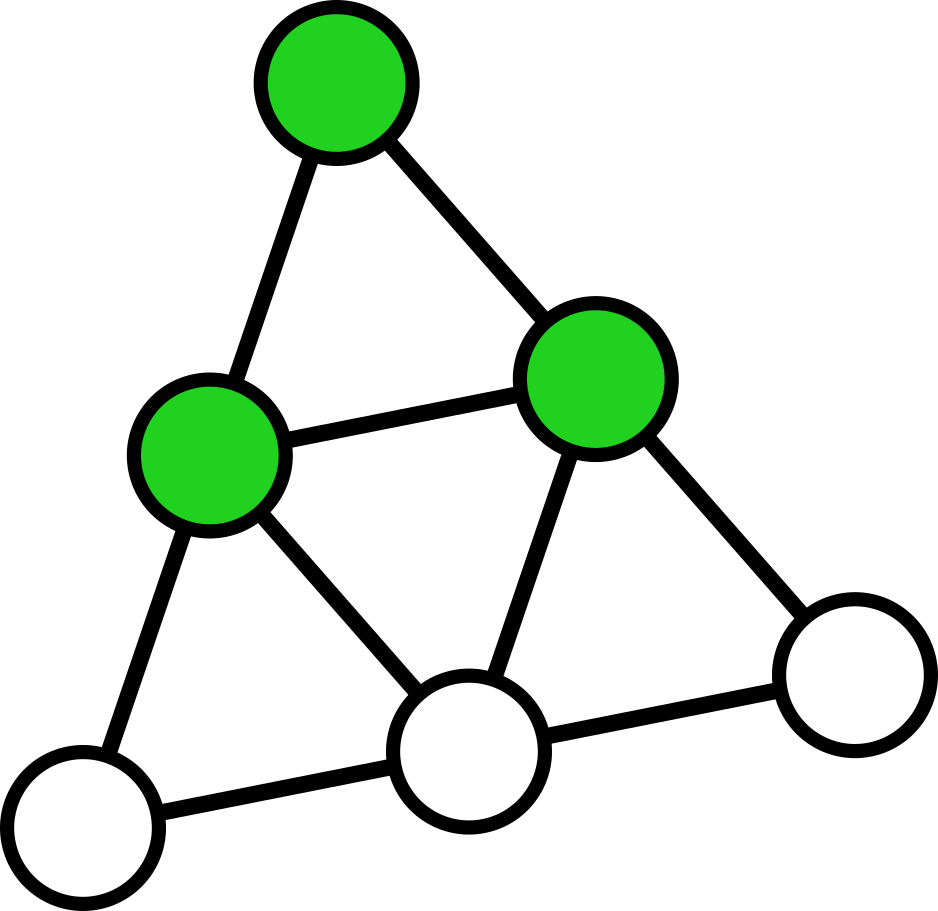} & \multicolumn{3}{l}{$ c_{3,2} = \frac{ h_x( 2 c_{2,1} + c_{2,2} ) -J_x( c_{1,1} + 2 c_{1,2} ) }{8 \abs{J_z}} $} \\
		\Fig{0.08}{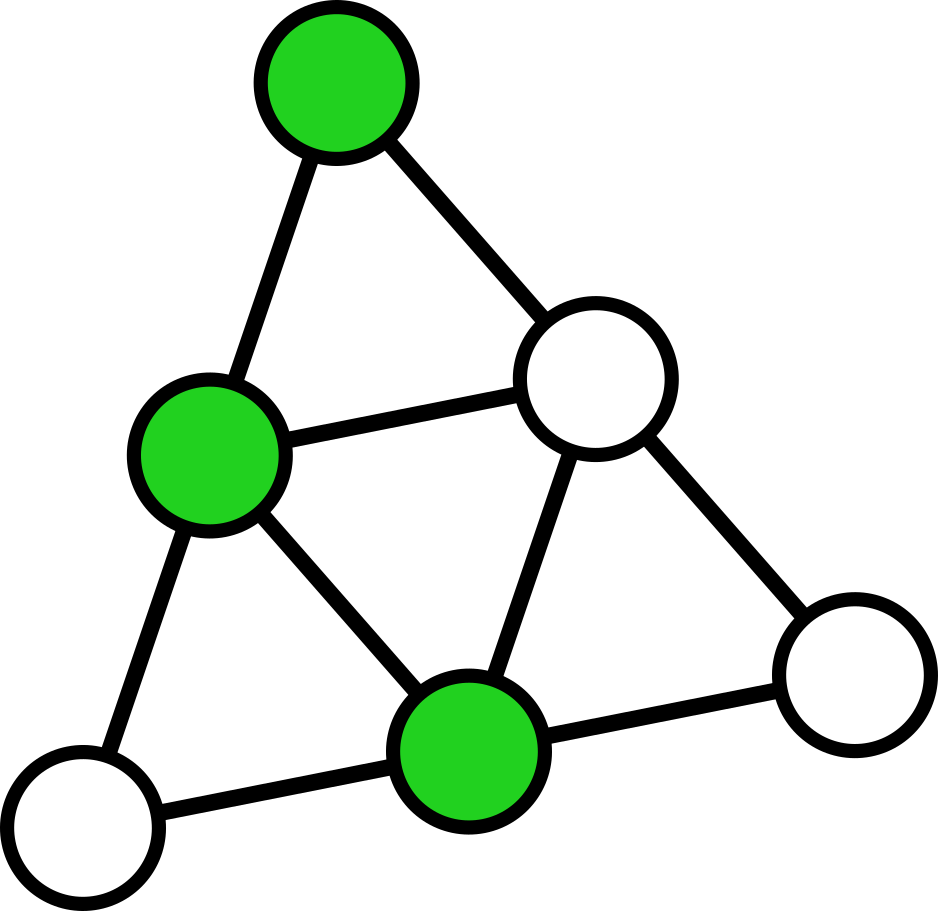} & \multicolumn{3}{l}{$ c_{3,3} = \frac{ h_x( c_{2,1} + c_{2,2} + c_{1,1} c_{1,2} ) -J_x( c_{1,1} + c_{1,2} ) }{12 \abs{J_z}} $} \\
		\Fig{0.08}{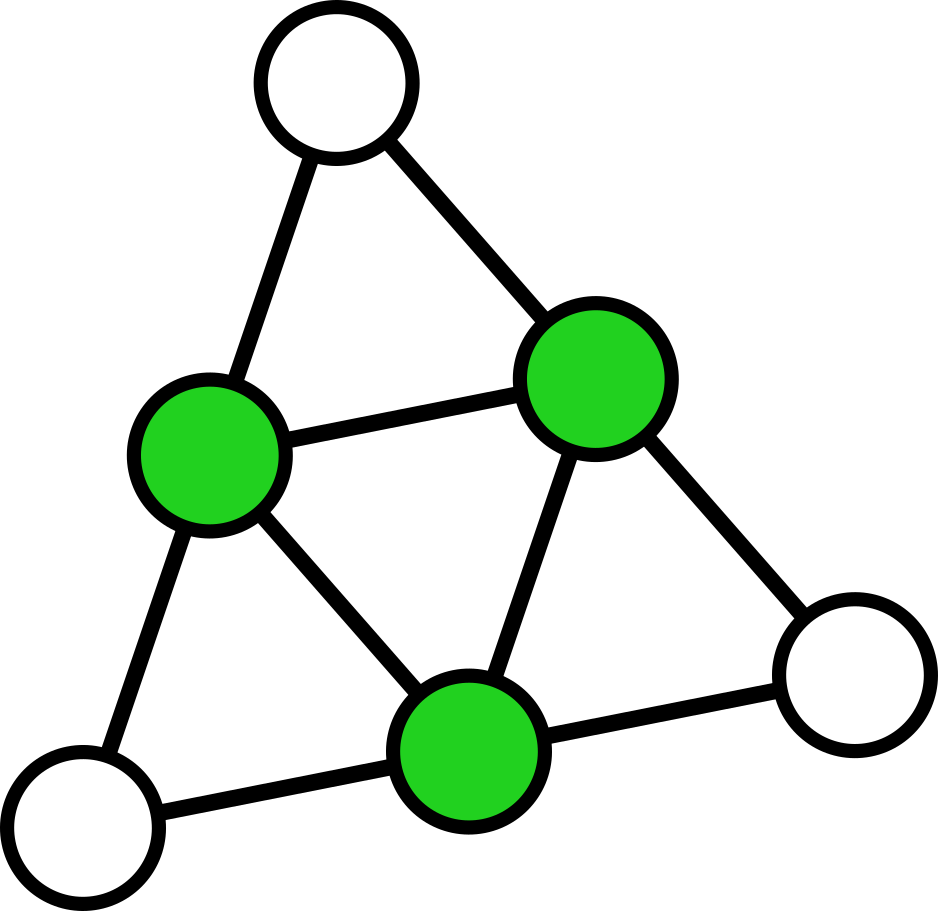} & \multicolumn{3}{l}{$ c_{3,4} = \frac{ h_x( 3 c_{2,2} ) -J_x( 3 c_{1,2} ) }{12 \abs{J_z}} $} \\
		\multirow{2}{*}{\Fig{0.08}{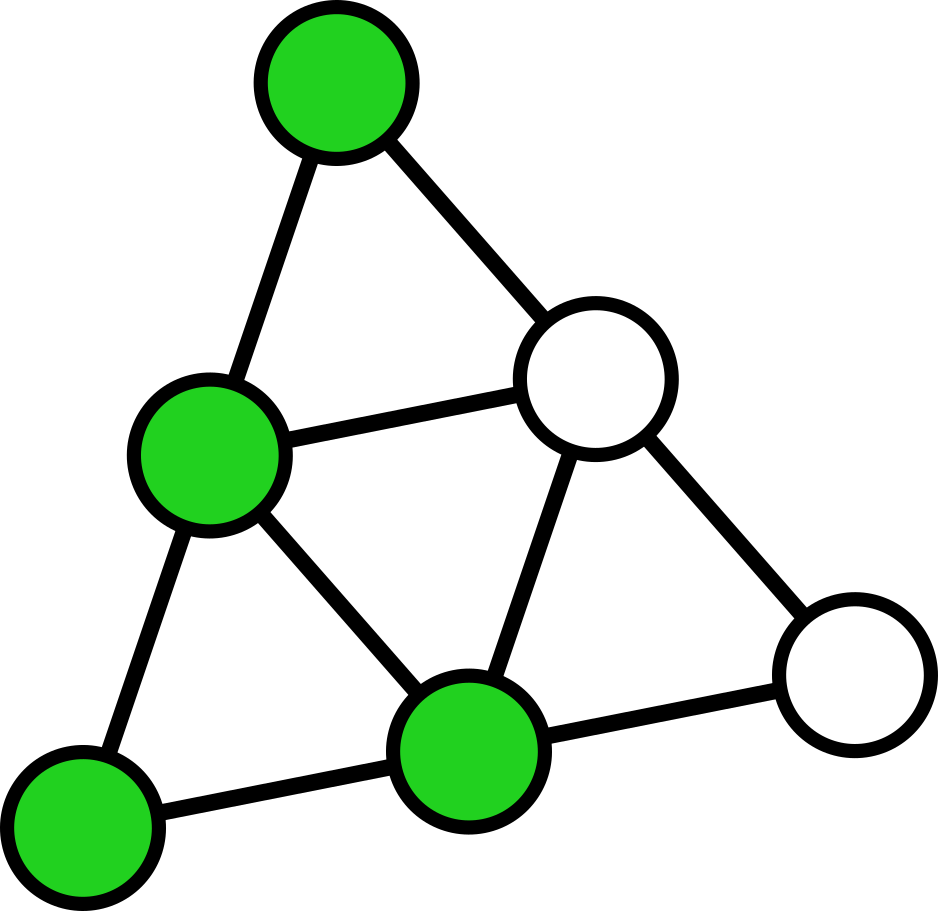}} & \multicolumn{3}{l}{$ c_{4,1} = \left[h_x( c_{3,1} + c_{3,2} + c_{3,3} + c_{2,1} c_{1,1} ) \right. $} \\
														 & \multicolumn{3}{l}{$ \left. -J_x( 2 c_{2,1} + c_{1,1} c_{1,2} + c_{1,1} c_{1,1} )\right]/(8 \abs{J_z}) $} \\
		\Fig{0.08}{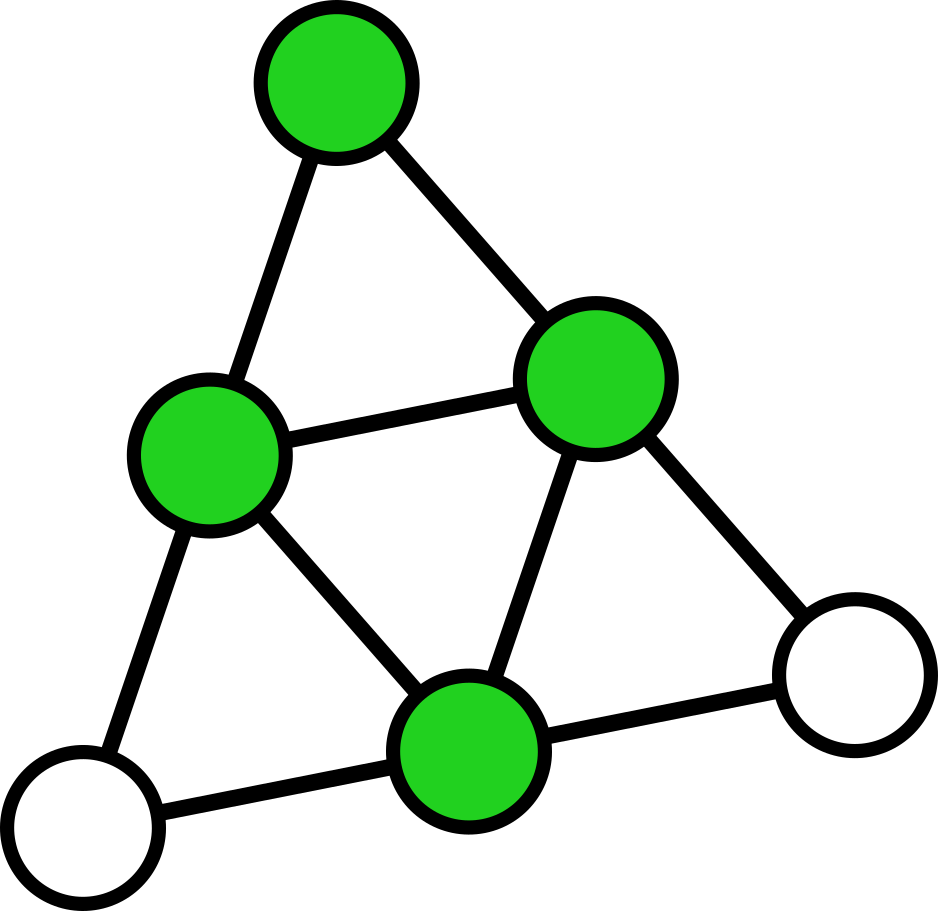} & \multicolumn{3}{l}{$ c_{4,2} = \frac{ h_x( c_{3,2} + 2 c_{3,3} + c_{3,4} ) -J_x( 2 c_{2,1} + 2 c_{2,2} + c_{1,1} c_{1,2} ) }{8 \abs{J_z}} $} \\
		\Fig{0.08}{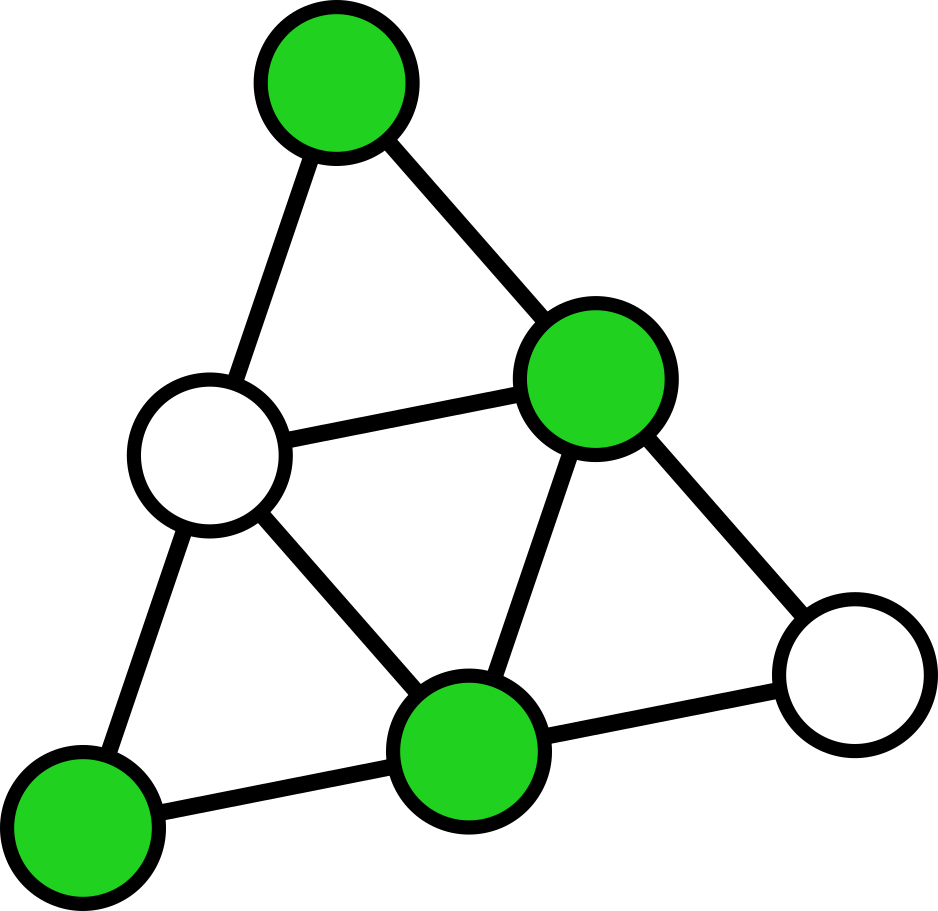} & \multicolumn{3}{l}{$ c_{4,3} = \frac{ h_x( 2 c_{3,3} + 2 c_{2,1} c_{1,1} ) -J_x( 2 c_{2,1} + c_{1,1}^2 ) }{12 \abs{J_z}} $} \\
		\Fig{0.08}{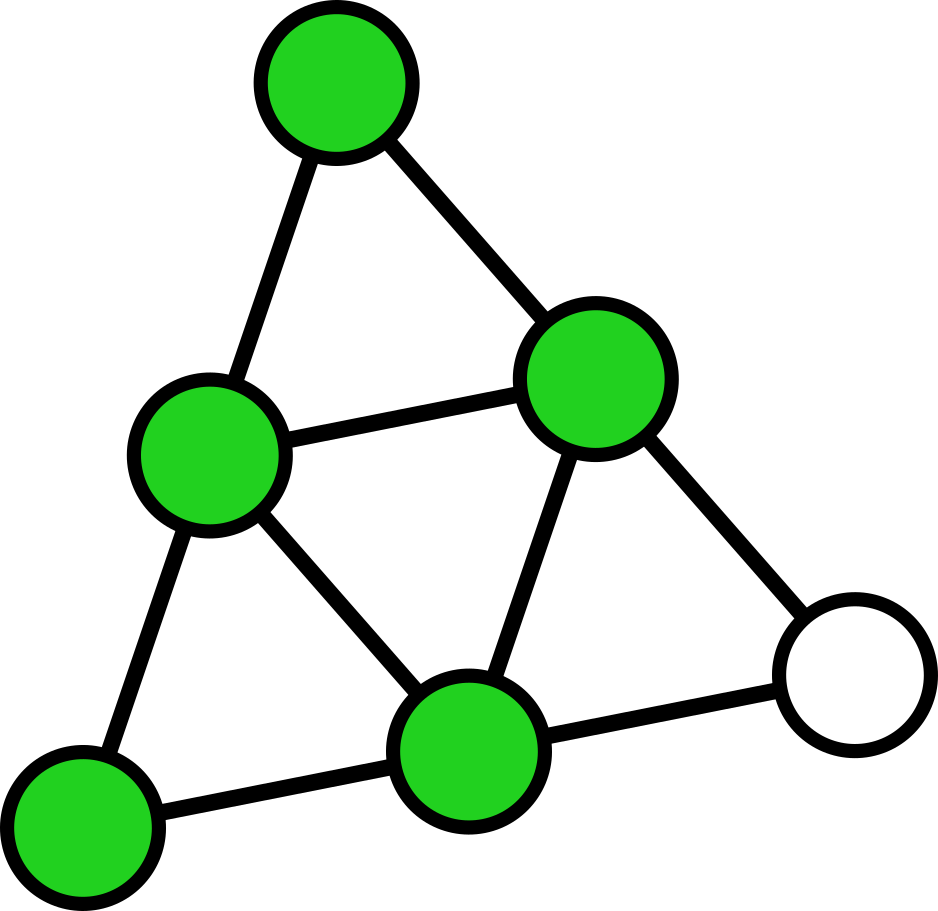} & \multicolumn{3}{l}{$ c_{5,1} = \frac{ h_x( 2 c_{4,1} + 2 c_{4,2} + c_{4,3} ) -J_x( 2 c_{3,2} + 2 c_{3,3} + c_{3,1} + 2 c_{2,1} c_{1,1} ) }{4 \abs{J_z}} $} \\
		\Fig{0.08}{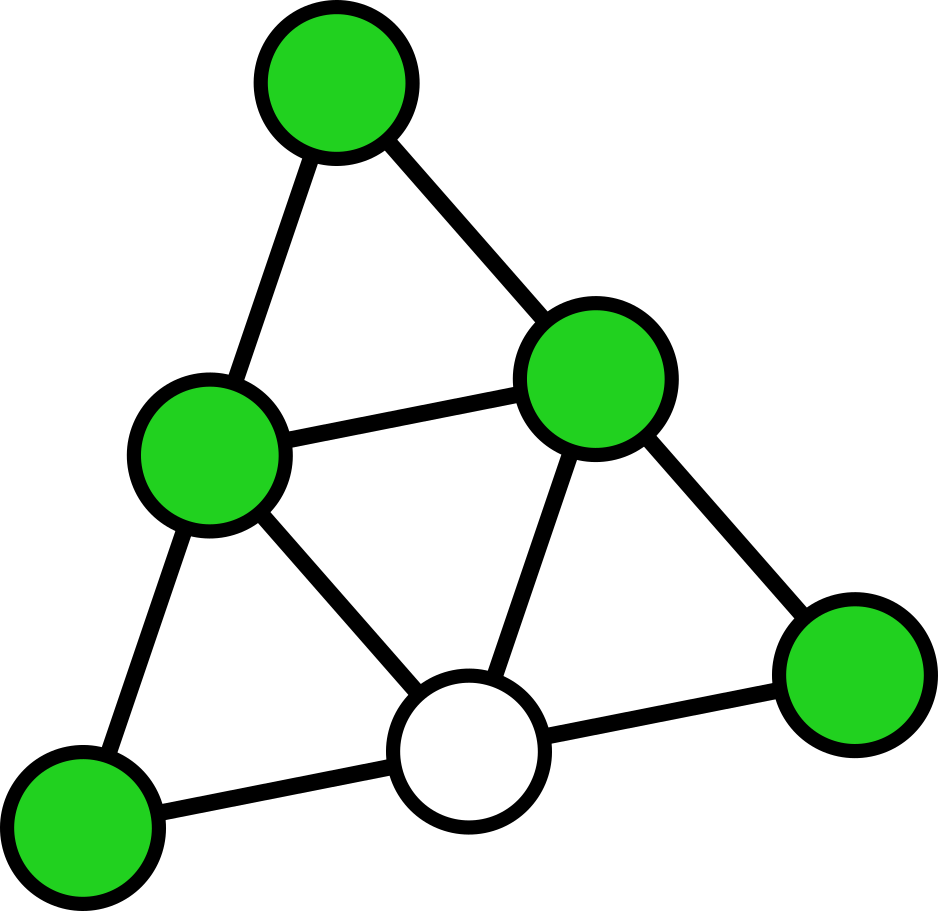} & \multicolumn{3}{l}{$ c_{5,2} = \frac{ h_x( 2 c_{4,1} + 2 c_{3,1} c_{1,1} + c_{2,2}^2 ) -J_x( 2 c_{3,1} + 2 c_{2,1} c_{1,1} ) }{8 \abs{J_z}} $} \\
		\Fig{0.08}{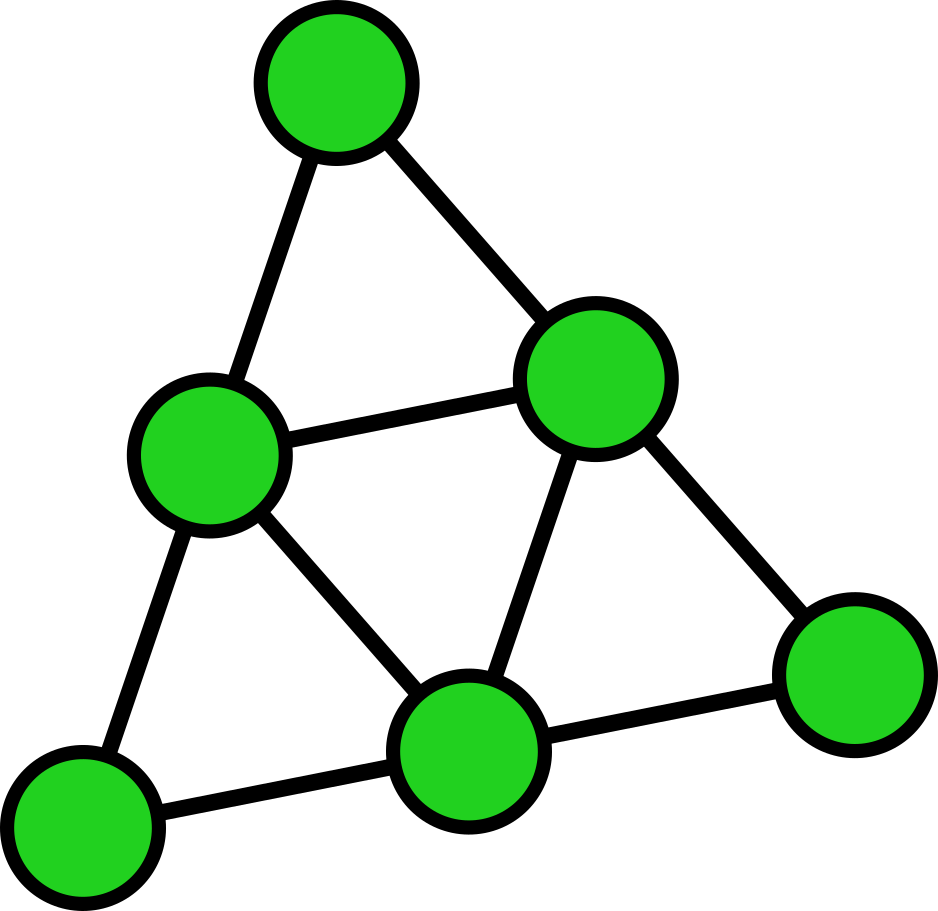} & \multicolumn{3}{l}{$ t_{\Delta} = J_x( 6 c_{4,1} + 3 c_{3,1} c_{1,1} ) - h_x( 3 c_{5,1} + 3 c_{5,2} ) $}
	\end{tabularx}
	\caption{Inequivalent connected clusters of a triangular cluster of $6$ spins. Filled green circles represent flipped spins. The intermediate tunneling coefficients satisfy the indicated recursion relations.}
	\label{tab:cluster_relations_triangular}
\end{table}

\begin{table}
	\centering
	\begin{tabularx}{\columnwidth}{XlXl}
		\Fig{0.08}{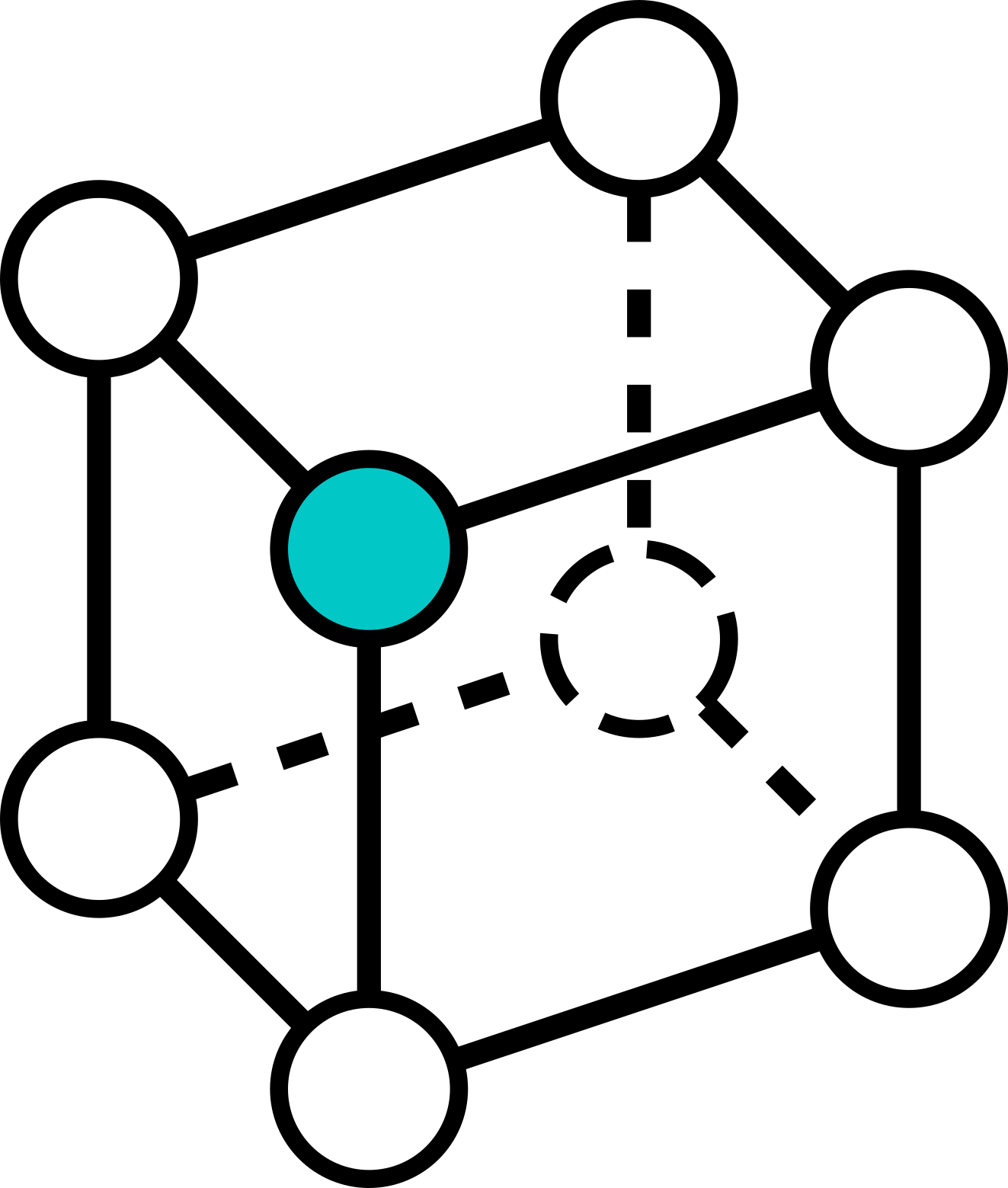} & $ c_{1,1} = \frac{ h_x }{6 \abs{J_z}} $ & \Fig{0.08}{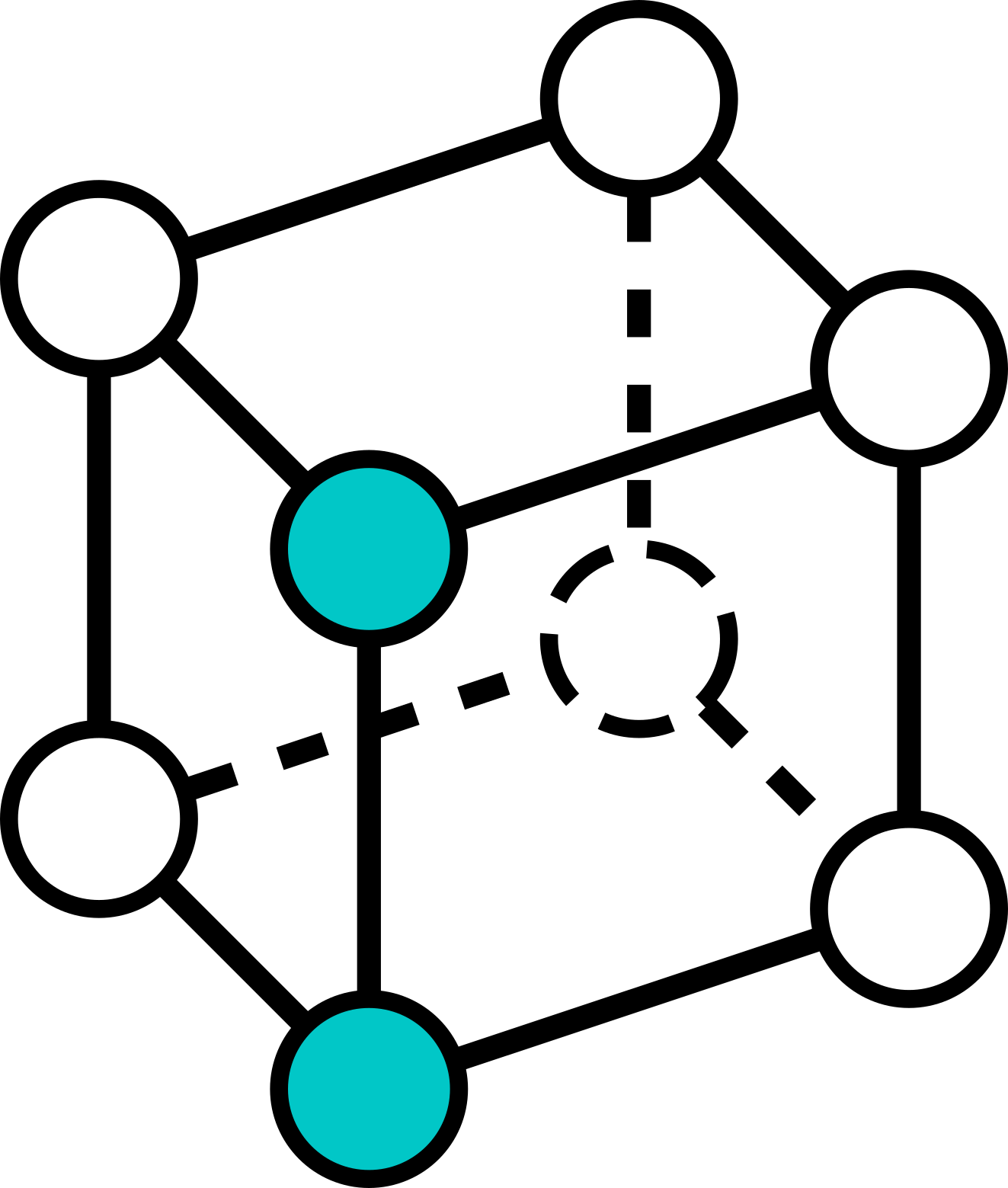} & $ c_{2,1} = \frac{ + h_x( 2 c_{1,1} ) -J_x( 1 ) }{8 \abs{J_z}} $ \\
		\Fig{0.08}{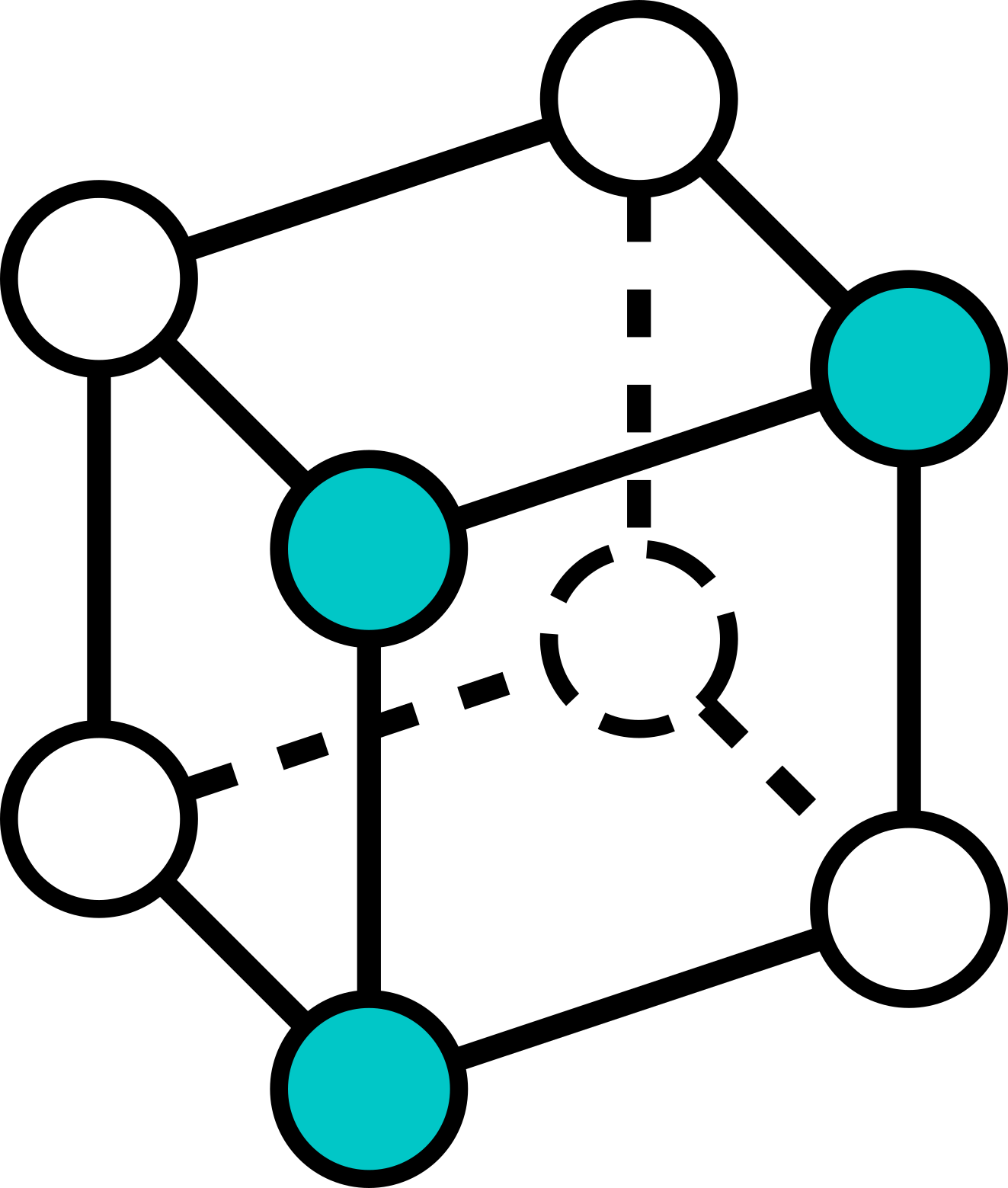} & \multicolumn{3}{l}{$ c_{3,1} = \frac{ h_x( 2 c_{2,1} + c_{1,1}^2 ) -J_x( 2 c_{1,1} ) }{10 \abs{J_z}} $} \\
		\Fig{0.08}{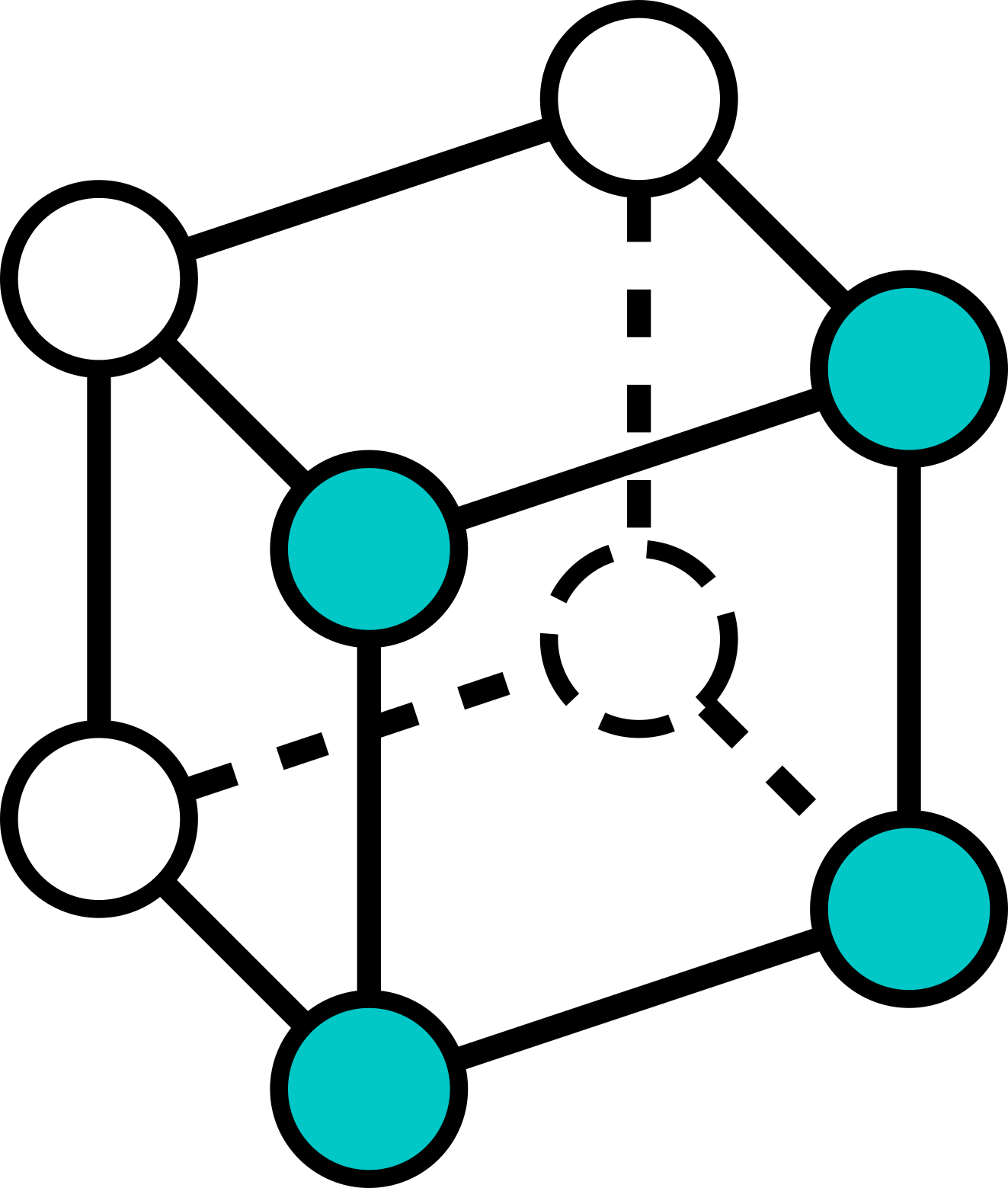} & \multicolumn{3}{l}{$ c_{4,1} = \frac{ h_x( 4 c_{3,1} ) -J_x( 4 c_{2,1} ) }{8 \abs{J_z}} $} \\
		\Fig{0.08}{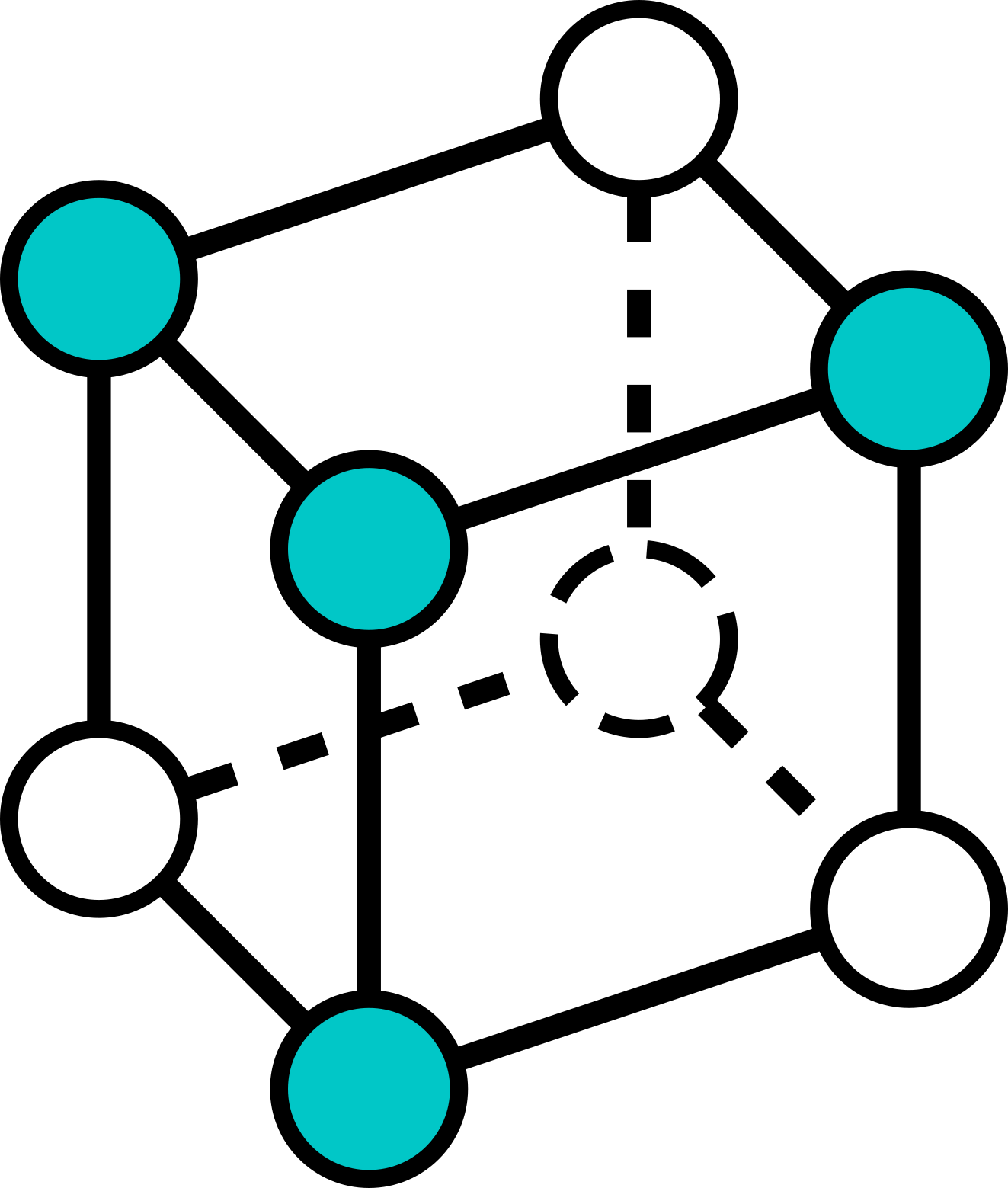} & \multicolumn{3}{l}{$ c_{4,2} = \frac{ h_x( 3 c_{3,1} + c_{1,1}^3 ) -J_x( 3 c_{1,1}^2 ) }{12 \abs{J_z}} $} \\
		\Fig{0.08}{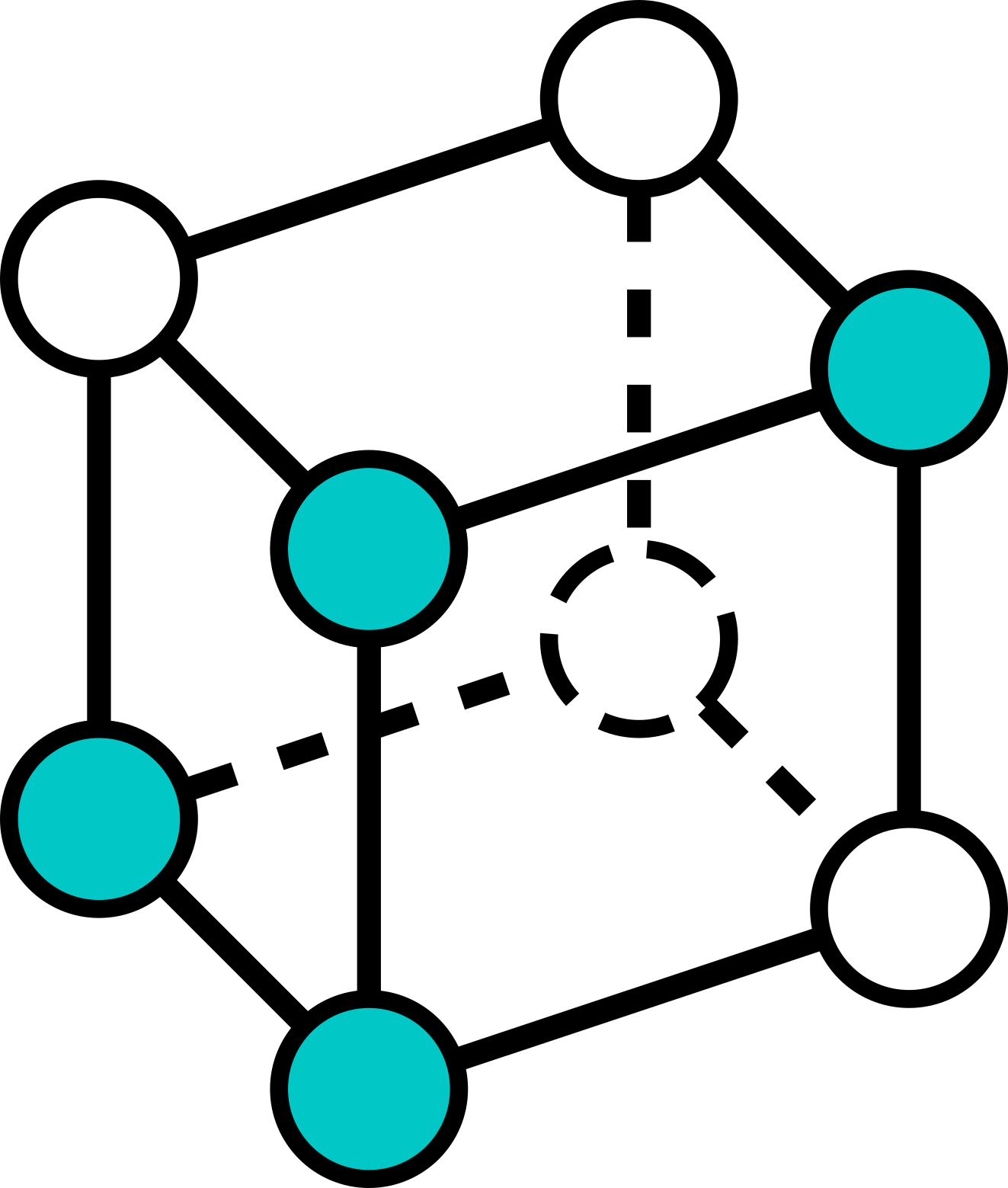} & \multicolumn{3}{l}{$ c_{4,3} = \frac{ h_x( 2 c_{3,1} + 2 c_{2,1} c_{1,1} ) -J_x( 2 c_{2,1} + c_{1,1}^2 ) }{12 \abs{J_z}} $} \\
		\Fig{0.08}{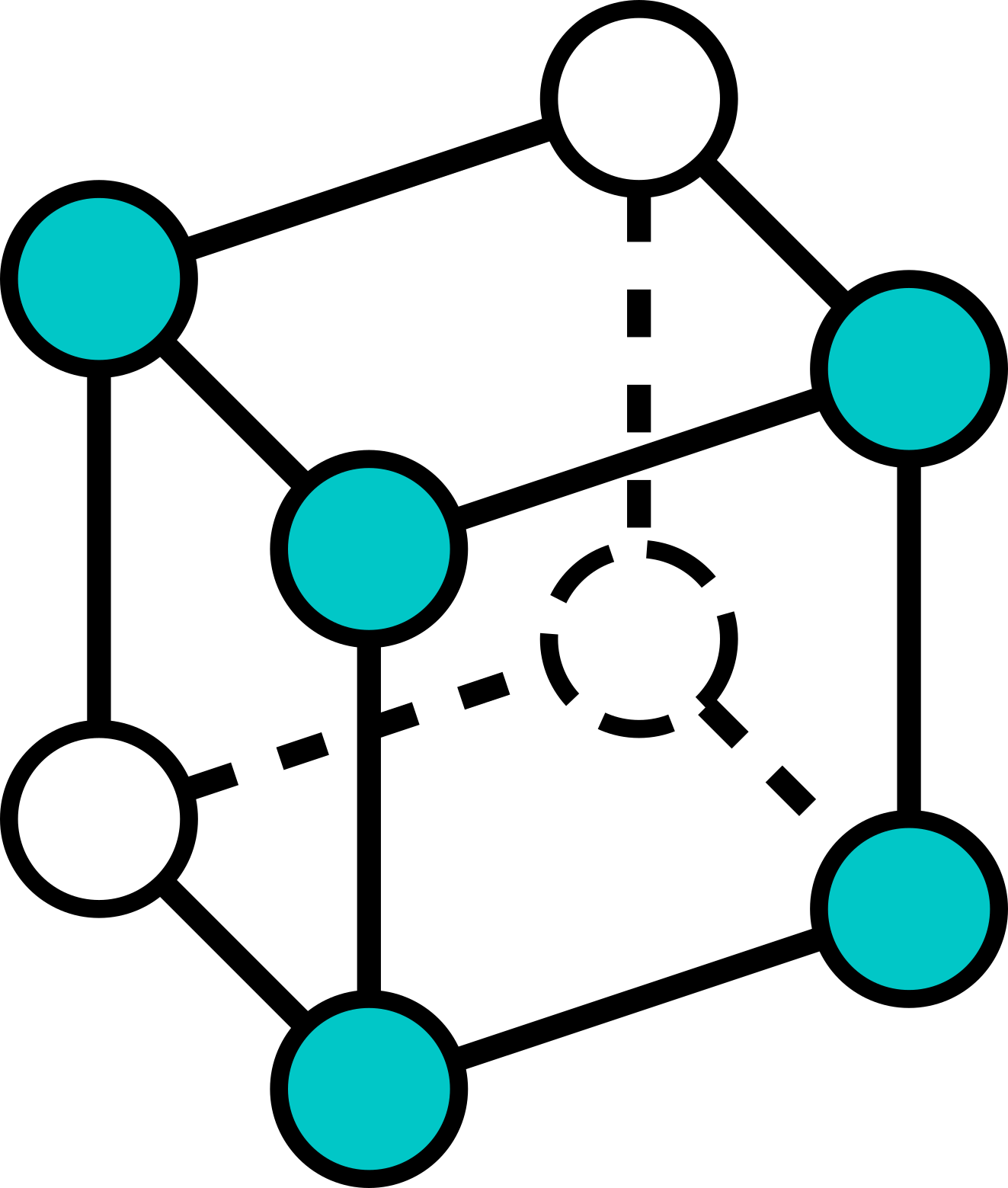} & \multicolumn{3}{l}{$ c_{5,1} = \frac{ h_x( c_{4,1} + c_{4,2} + 2 c_{4,3} + c_{3,1} c_{1,1} ) -J_x( 3 c_{3,1} + 2 c_{2,1} c_{1,1} ) }{10 \abs{J_z}} $} \\
		\Fig{0.08}{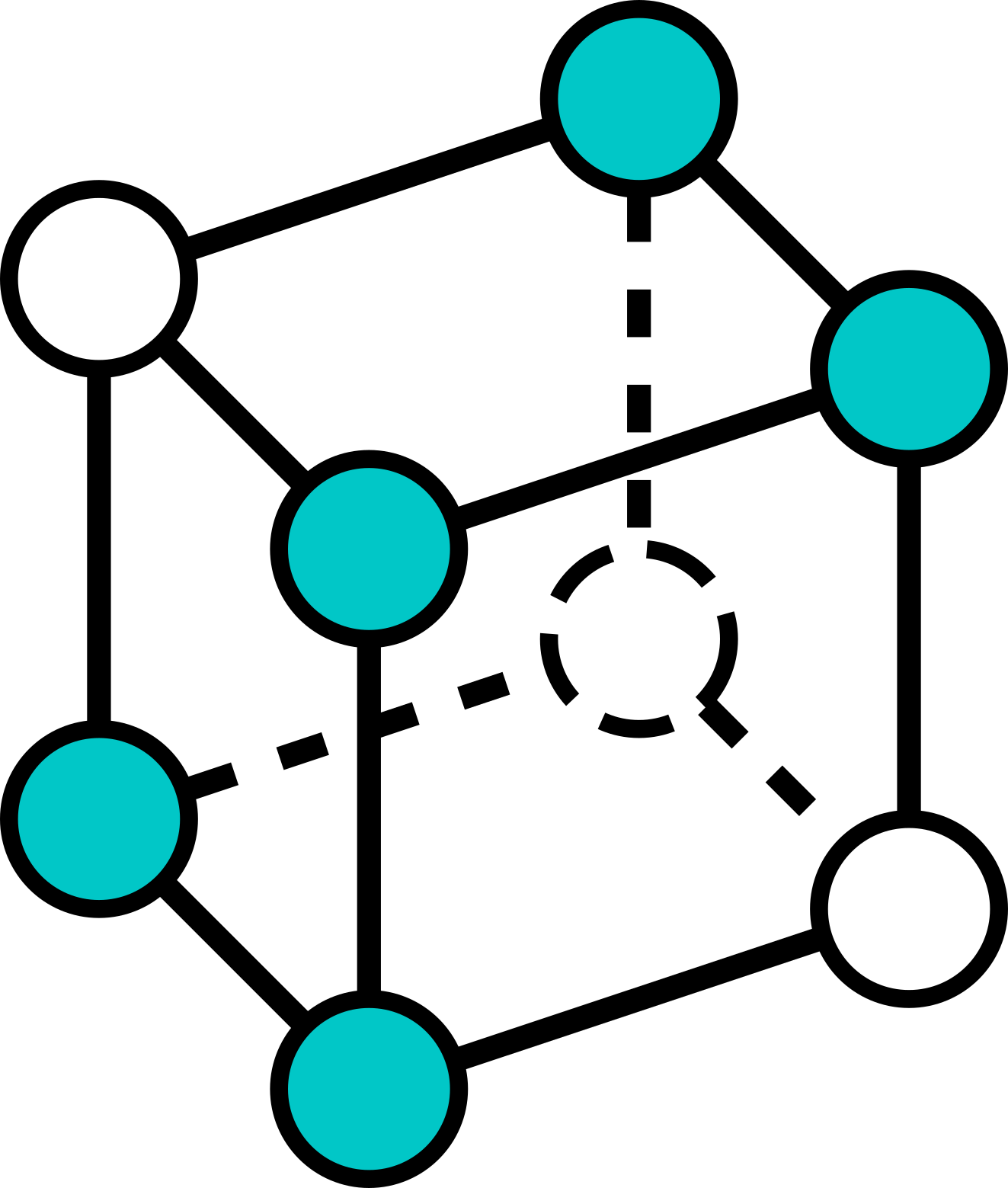} & \multicolumn{3}{l}{$ c_{5,2} = \frac{ h_x( 2 c_{4,3} + 2 c_{3,1} c_{1,1} + c_{2,1}^2 ) -J_x( 2 c_{3,1} + 2 c_{2,1} c_{1,1} ) }{14 \abs{J_z}} $} \\
		\Fig{0.08}{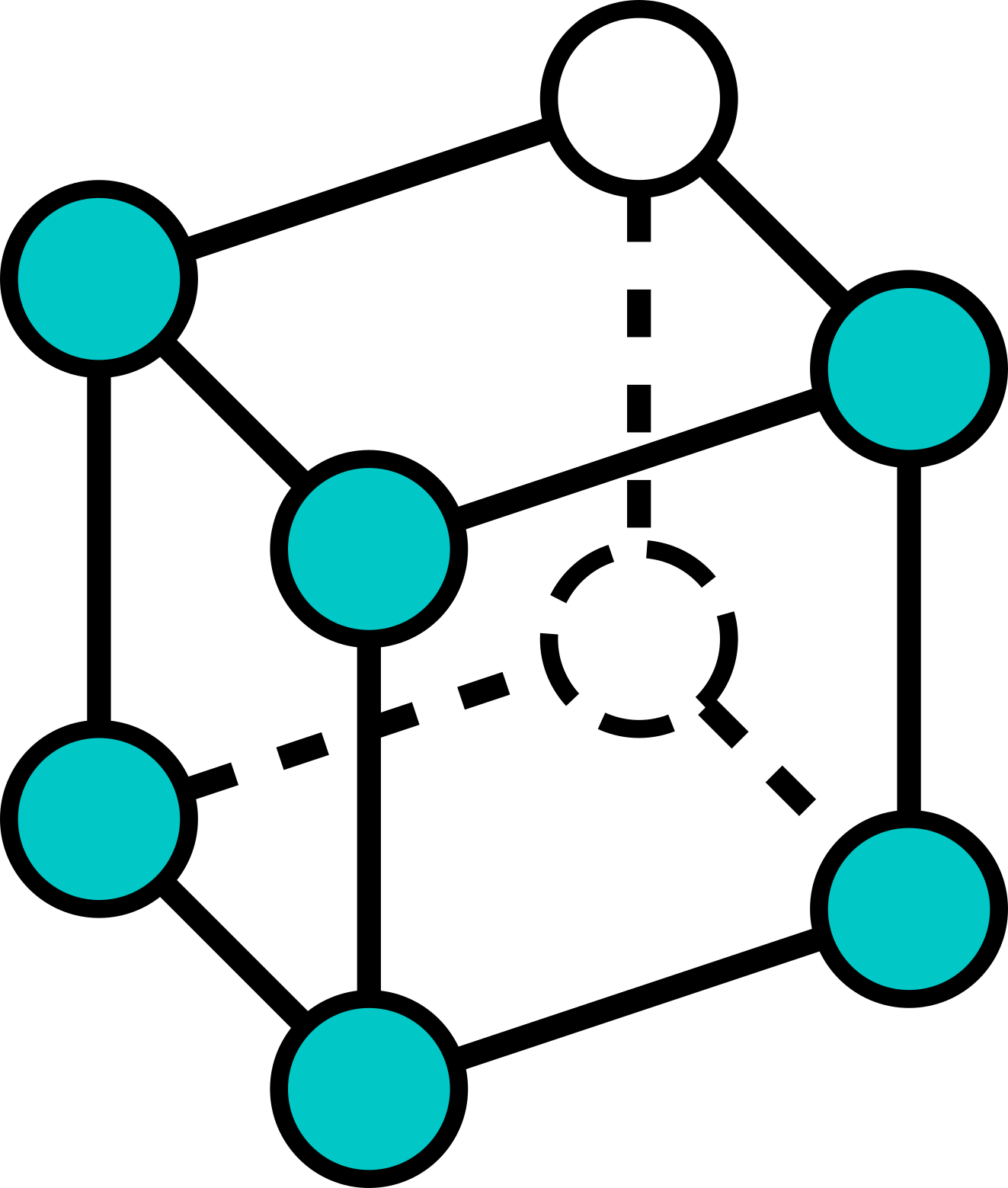} & \multicolumn{3}{l}{$ c_{6,1} = \frac{ h_x( 4 c_{5,1} + 2 c_{5,2} ) -J_x( 2 c_{4,1} + 4 c_{4,3} + c_{2,1}^2 ) }{8 \abs{J_z}} $} \\
		\Fig{0.08}{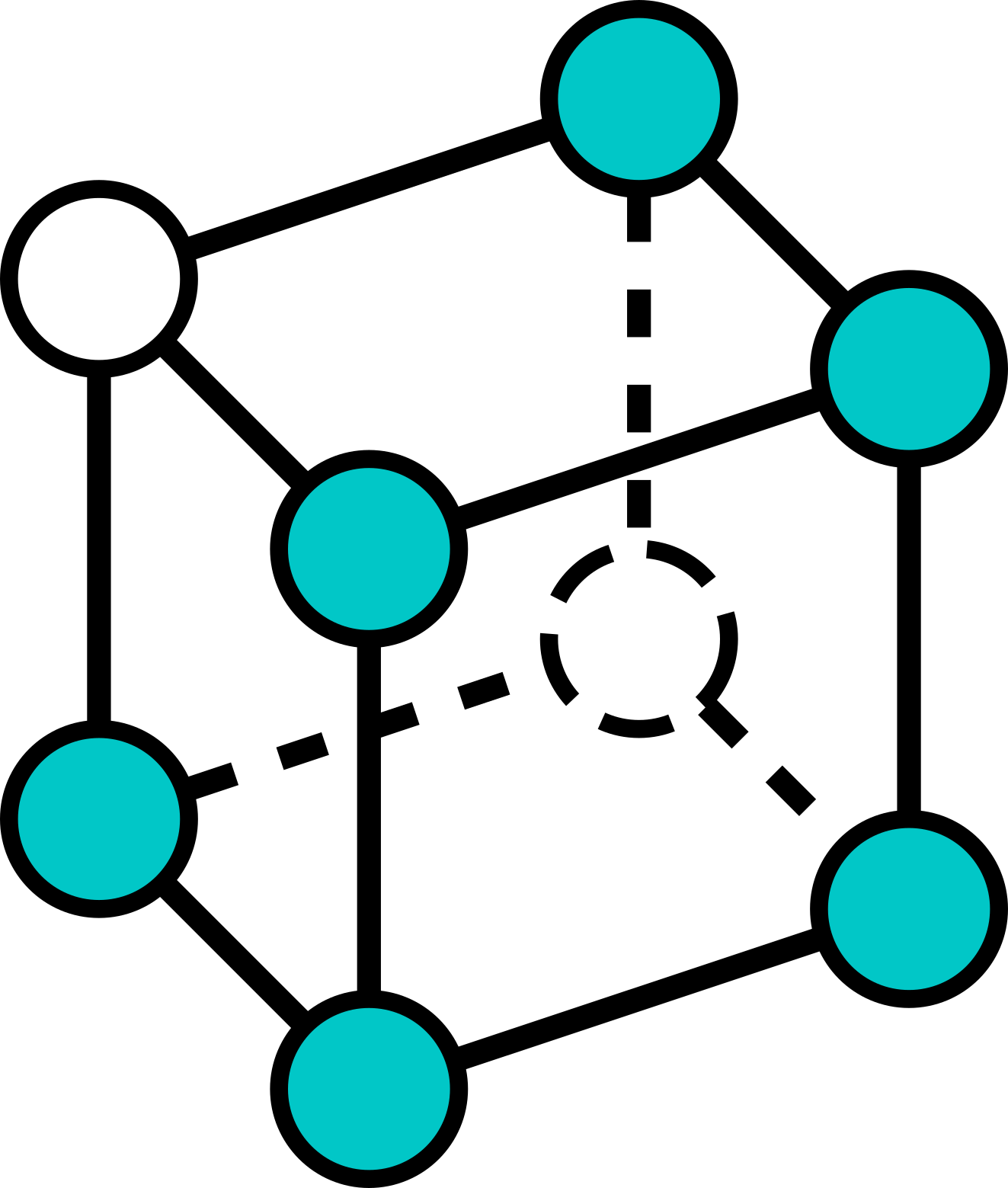} & \multicolumn{3}{l}{$ c_{6,2} = \frac{ h_x( 2 c_{5,1} + 2 c_{5,2} + 2 c_{4,2} c_{1,1} ) -J_x( 2 c_{4,2} + 4 c_{3,1} c_{1,1} ) }{12 \abs{J_z}} $} \\
		\Fig{0.08}{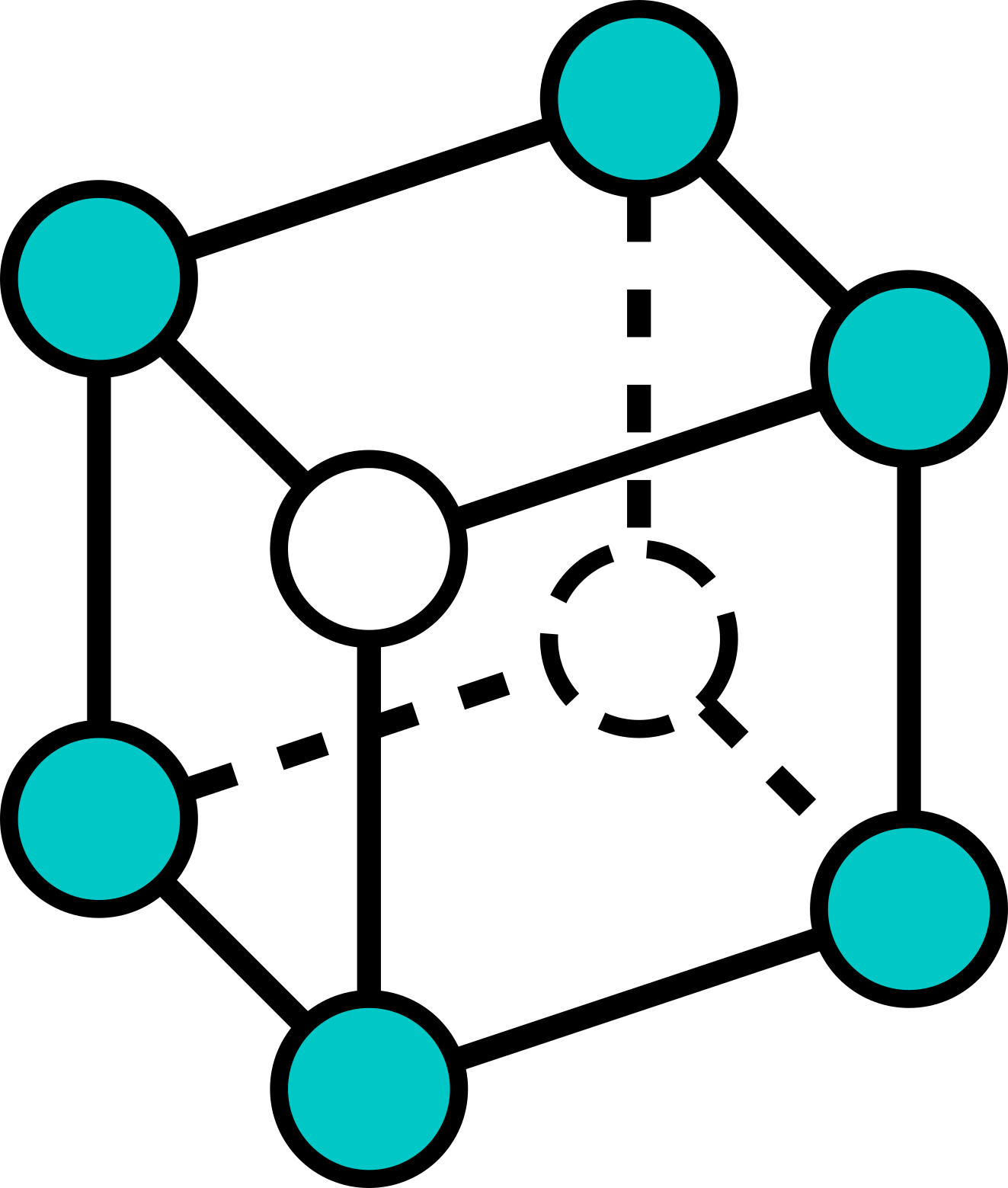} & \multicolumn{3}{l}{$ c_{6,3} = \frac{ h_x( 6 c_{5,2} ) -J_x( 2 c_{4,2} + 4 c_{3,1} c_{1,1} ) }{12 \abs{J_z}} $} \\
		\Fig{0.08}{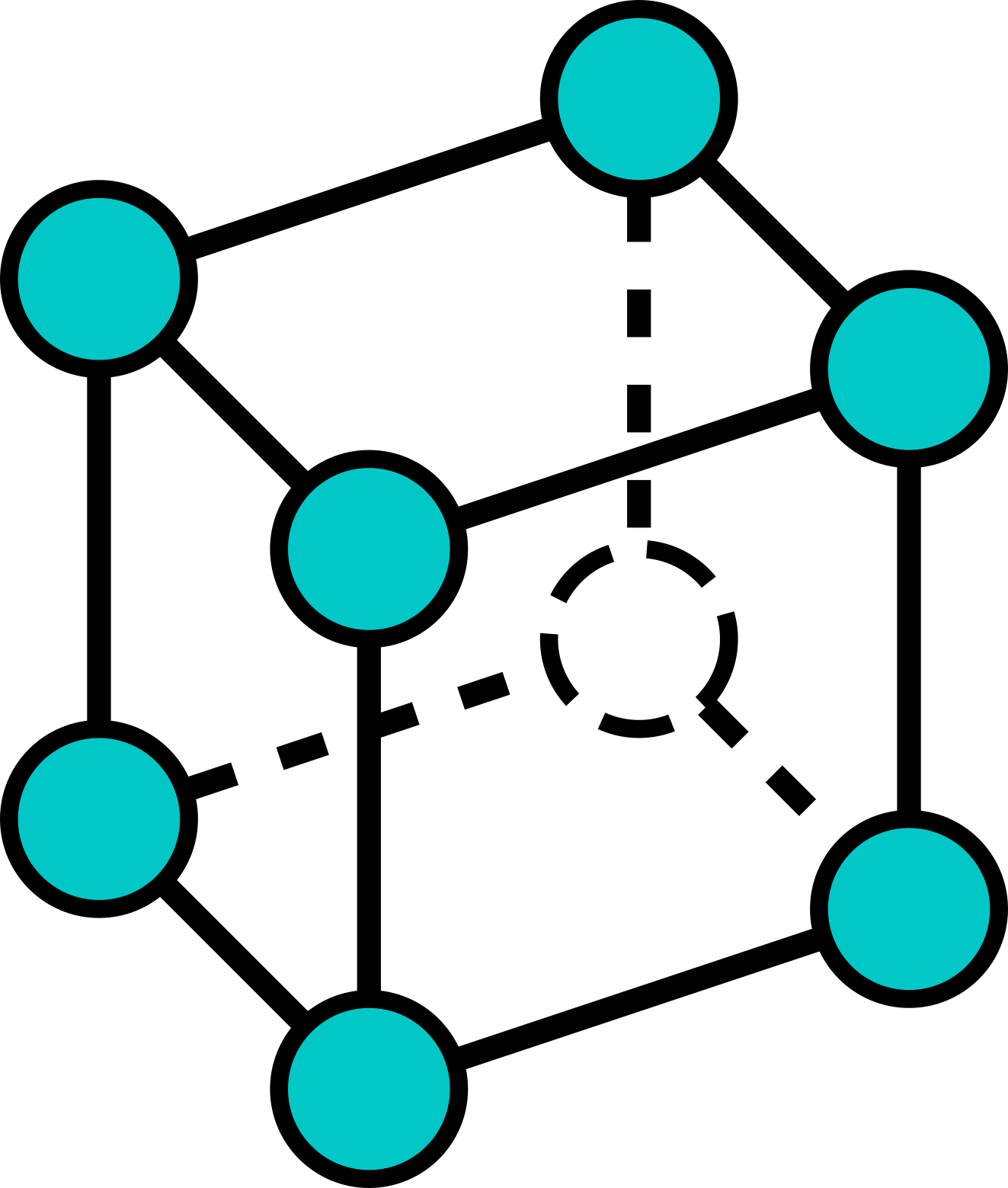} & \multicolumn{3}{l}{$ c_{7,1} = \frac{ h_x( 3 c_{6,1} + 3 c_{6,2} + c_{6,3} ) -J_x( 6 c_{5,1} + 3 c_{5,2} ) }{6 \abs{J_z}} $} \\
		\Fig{0.08}{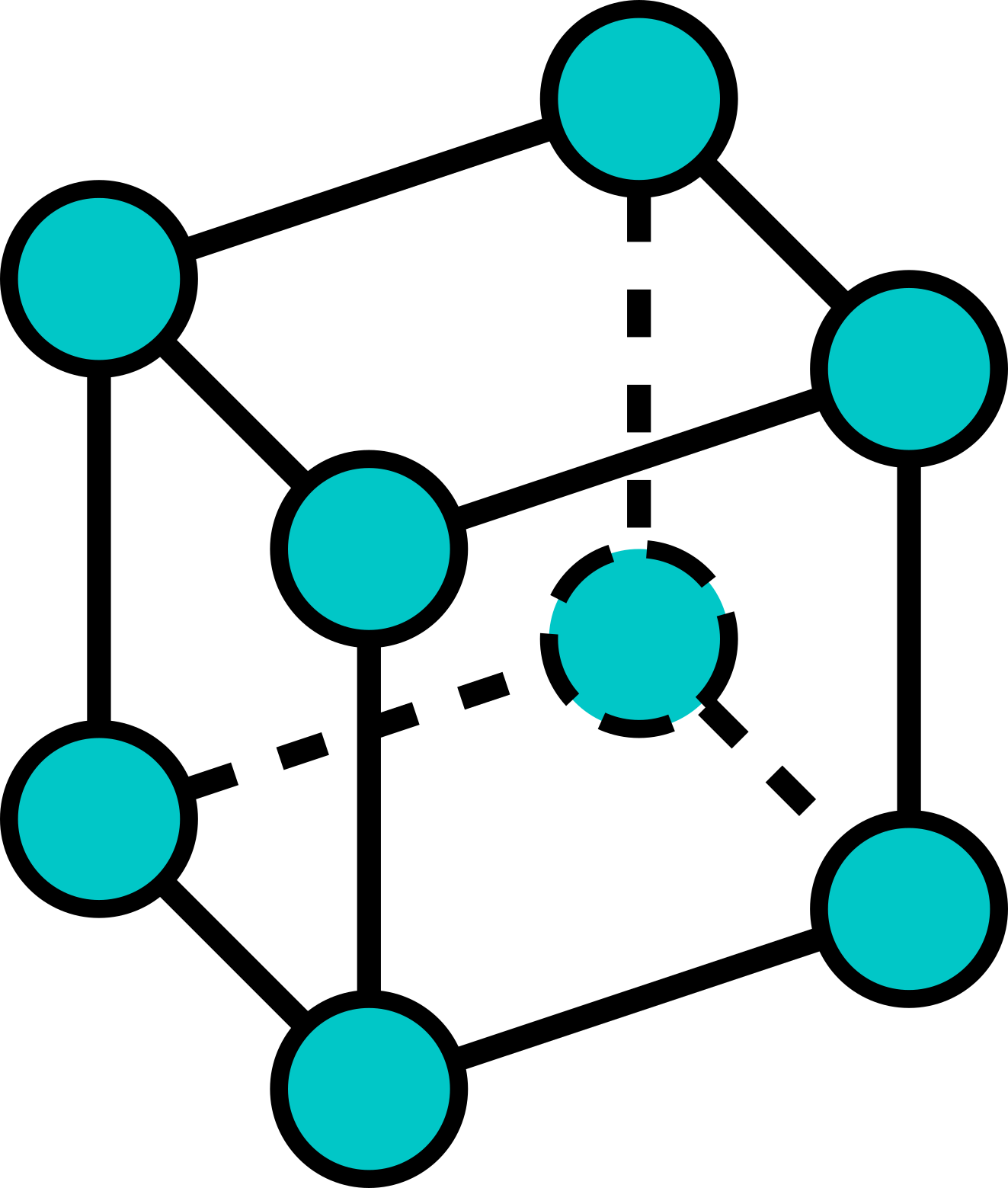} & \multicolumn{3}{l}{$ t_{2 \times 2 \times 2} = J_x( 12 c_{6,1} ) - h_x( 8 c_{7,1} ) $} \\
	\end{tabularx}
	\caption{Inequivalent connected clusters of a cubic cluster of $8$ spins. Filled blue circles represent flipped spins. The intermediate tunneling coefficients satisfy the indicated recursion relations.}
	\label{tab:cluster_relations_cubic}
\end{table}

While in 1D chains connected clusters only come in one shape (a connected stretch of spins), the length and position (edge or bulk) being their only  characteristics, in quasi 1D and in higher dimensions there are many more shapes of clusters we have to consider. Finding a general solution for the tunneling of any $N$-sized cluster therefore does not seem possible. However, we will calculate the polynomial $t_N(h_x,J_x)$ for small spin clusters to demonstrate the method and to show that the number of zeros still equals the number of spins $N$.

We first consider the model of Eq.~(\ref{eq:H_ising_model}) on a $3 \times 2$ cluster of spins with open boundary conditions, the smallest non-trivial 2D cluster. Besides the fully flipped cluster there are 14 inequivalent connected clusters, cf. Table~\ref{tab:cluster_relations_rectangular} for which we have to compute the intermediate tunneling coefficients. The $3 \times 2$ cluster differs from a ring of $6$ spins only by one additional bond. By showing that the tunneling polynomial still has $N$ zeros, we thus demonstrate the robustness of the number of zeros to certain perturbations. Upon gradually turning on the bond that transforms the ring into the $3 \times 2$ cluster, the zeros move towards higher fields, as one expects.
We further derive the tunneling polynomials $t_\Delta$ for an equilateral triangle made from $6$ spins, cf. Table~\ref{tab:cluster_relations_triangular}, and $t_{2 \times 2 \times 2}$ for a cube of 8 spins, cf. Table~\ref{tab:cluster_relations_cubic}.

\begin{figure}
	\centering
	\includegraphics[width=0.49\columnwidth]{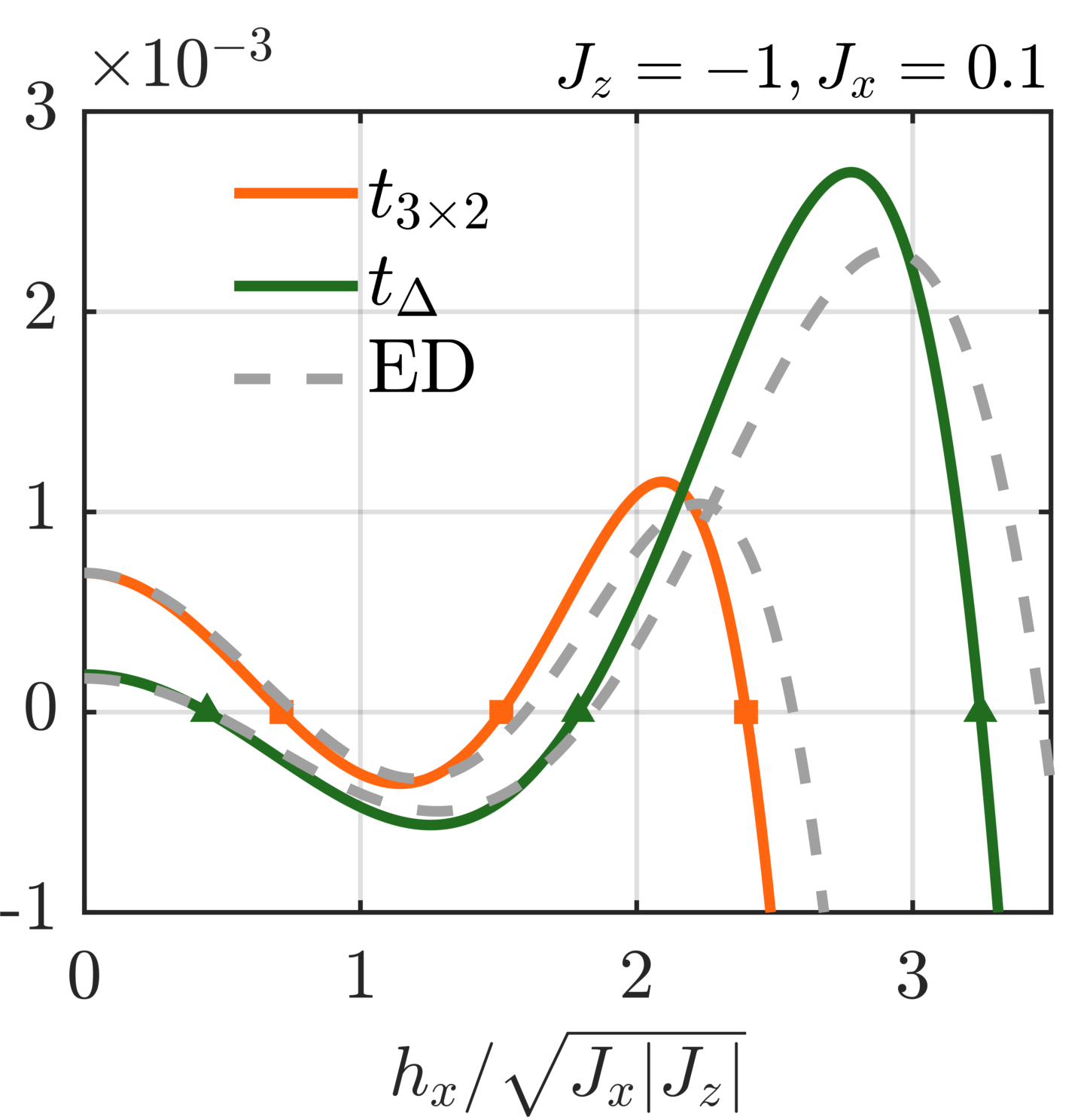}
	\includegraphics[width=0.49\columnwidth]{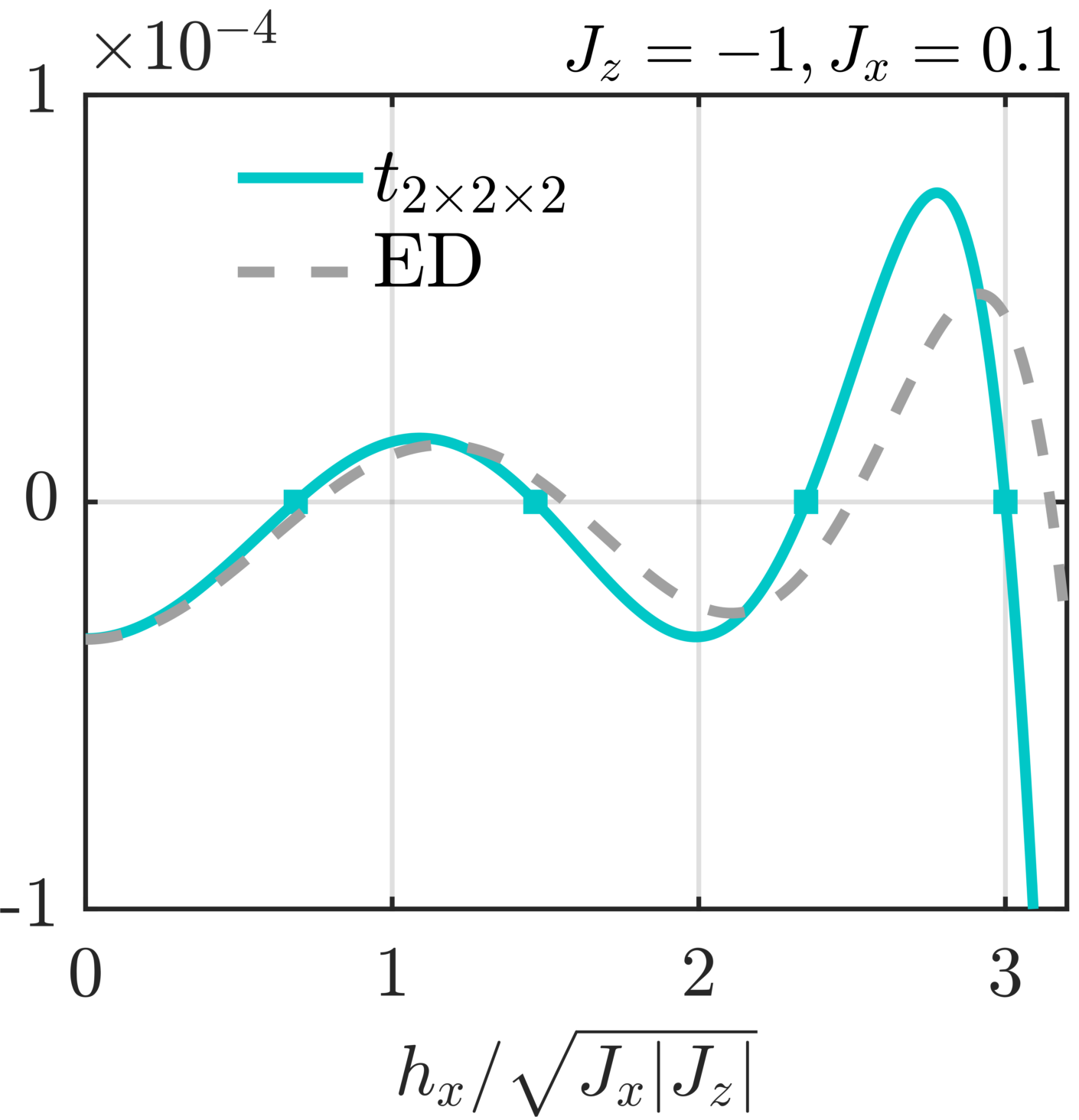}
	\caption{The tunneling amplitude $t_N$ for three small, higher-dimensional clusters, as a function of $h_x$, with the respective zeros marked. All these clusters have $N$ tunneling zeros, and the plot shows the $N/2$ positive ones.}
	\label{fig:tunneling_2D_3D_clusters}
\end{figure}

Solving the resulting recursion relations given in the Tables~\ref{tab:cluster_relations_rectangular},~\ref{tab:cluster_relations_triangular}~and~\ref{tab:cluster_relations_cubic}, we obtain the following polynomials, where we define the variable $x \equiv h_x^2/(J_x\abs{J_z})$:
\begin{align}
	t_{3 \times 2} &= \frac{J_x^3}{|J_z|^2} \left(-\frac{539 x^3 }{5184} + \frac{511 x^2 }{576} - \frac{193 x}{108} + \frac{25}{36}\right), \\
	t_{\Delta} &= \frac{J_x^3}{|J_z|^2} \left(-\frac{65 x^3}{2304} + \frac{905 x^2}{2304} - \frac{197 x}{192} + \frac{3}{16}\right), \\
			 &\begin{aligned}
				 \mathllap{t_{2 \times 2 \times 2}} &= \frac{J_x^4}{|J_z|^3}\left(-\frac{3119 x^4}{466560} + \frac{66839 x^3}{583200} - \frac{76921 x^2}{129600}\right. \\
													 &\quad + \left.\frac{6979 x}{7200} - \frac{43}{128}\right),
			 \end{aligned}
	\label{eq:2D_3D_polynomials}
\end{align}
which have the expected number of zeros, $N=6$ or $8$, as one can see in Fig.~\ref{fig:tunneling_2D_3D_clusters}.
%

\section{Weak Disorder}
\label{sec:low_disorder}

In this section we consider a 1D ring with weak disorder in the transverse fields and in the exchange in the form of
\begin{equation}
	h_{x,i} = h_x + \delta h_{x,i},\quad J_{x,i} = J_x +\delta J_{x,i},
\end{equation}
where $J_{x,i}$ connects spins $i$ and $i+1$ and where $J_x > 0$, such that there are zeros in the absence of disorder. We denote the disordered tunneling as $\td_N$, reserving $t_N$ for the tunneling in the disorder free limit. One can in principle determine the recursion relations of such a system and thus study disorder using explicit polynomials $\td_N$. However, we shall rather calculate the relevant average quantities to lowest order in an expansion in $\delta h_{x,i}$ and $\delta J_{x,i}$.

A quantity of particular interest is the typical finite tunneling induced by random fluctuations of the couplings when the average external field is held at one of the tunneling zeros $h_x^{(n)}$. Let us denote the disorder induced tunneling at a zero as
\begin{equation}
	T_n \equiv \left. \td_N \right\rvert_{h_x=h_x^{(n)}}.
\end{equation}
We are also interested in how much a ground state crossing shifts due to the presence of randomness:
\begin{equation}
	K_n \equiv \hd_x^{(n)} - h_x^{(n)},
	\label{eq:drift_zeros}
\end{equation}
where $\hd_x^{(n)}$ and $h_x^{(n)}$ are respectively the $n^{\rm th}$ zeros of the polynomials $\td_N$ and $t_N$. We can calculate the second moment of these random variables by considering a homogeneous ring where all spins and all bonds are equivalent and where in the absence of disorder all couplings are equal. However, explicit calculations of the second moment of $T_n$ and $K_n$ for open spin chains showed qualitatively very similar behavior as we find below for rings. For a ring, to first order in the perturbations, one has
\begin{equation}
	\td_N \approx t_N + \frac{1}{N}\frac{\partial t_N}{\partial J_x} \sum_{i=1}^{N} \delta J_{x,i} + \frac{1}{N}\frac{\partial t_N}{\partial h_x} \sum_{i=1}^{N} \delta h_{x,i}.
	\label{eq:t_perturbed_development}
\end{equation}
This follows since perturbations on different sites are equivalent, and thus all partial derivatives are equal:
\begin{equation}
	\left. \frac{\partial \td_N}{\partial (\delta h_{x,i})}\right\rvert_{\delta J_x = \delta h_x = 0} = \frac{1}{N}\frac{\partial t_N}{\partial h_x},
	\label{eq:t_perturbed_derivatives}
\end{equation}
and likewise for the exchange. For identically and independently distributed local disorder, the disorder-induced variance of the tunneling evaluated at a transverse field zero $h_x^{(n)}$ thus results as
\begin{equation}
	\begin{aligned}
		\braket{T_n^2} &= \frac{\braket{\delta J_i^2 }}{N}\left\lvert \frac{\partial t_N}{\partial J_x} \right\rvert^2_{h_x=h_x^{(n)}}  + \frac{\braket{\delta h_i^2 }}{N}\left\lvert \frac{\partial t_N}{\partial h_x} \right\rvert^2_{h_x=h_x^{(n)}} \\
					   &\quad +O(\braket{\delta J_i^4 },\braket{\delta h_i^4 }).
	\end{aligned}
	\label{eq:T_low_disorder}
\end{equation}
To leading order, the response of the tunneling amplitude $T_n$ to disorder is linear. The shift of the transverse field, $K_n$, necessary to compensate
for this disorder-induced tunneling is then given by the relation
\begin{equation}
	T_n + \left. \pdv[]{t_N}{h_x} \right\rvert_{h_x=h_x^{(n)}} K_n = 0.
	\label{eq:T_sigma_relation}
\end{equation}
From Eqs.~(\ref{eq:T_low_disorder}, \ref{eq:T_sigma_relation}), we see that in order to calculate $K_n$ we only need the ratio of the derivatives
$
\left.\left(\pdv[]{t_N}{h_x}/ \pdv[]{t_N}{J_x}\right)\right\rvert_{h_x=h_x^{(n)}}.
$
This ratio is easily obtained from the factorized polynomial form of $t_N$ (Eq.~(\ref{eq:t_factorized_form})) as
\begin{equation}
	\left.\left(\pdv[]{t_N}{h_x}/ \pdv[]{t_N}{J_x}\right)\right\rvert_{h_x=h_x^{(n)}}  = -\frac{h_x^{(n)}}{2J_x}.
\end{equation}
From this we deduce the average mean square of the drift in the zeros to leading order as:
\begin{align}
	\braket{K_n^2} &\approx \left(\frac{h_x^{(n)}}{2 J_x}\right)^2\frac{\braket{\delta J_{x,i}^2}}{N} + \frac{\braket{\delta h_{x,i}^2}}{N}\\
								&= \frac{1}{N} \left(\frac{h_x^{(n)}}{2}\right)^2 \left[\braket{\mu_i^2} + 4\braket{\eta_i^2}\right],
	\label{eq:sigma_solution}
\end{align}
where
\begin{equation}
	\mu_i \equiv \frac{\delta J_{x,i}}{J_x},
	\quad \eta_i \equiv \frac{\delta h_{x,i}}{h_x^{(n)}},
\end{equation}
quantify the relative fluctuations of the couplings. In many cases these are the most appropriate measures of the disorder strength. We see that the zeros corresponding to larger fields are more strongly affected by disorder. This certainly holds as long as the shifts are smaller than the typical spacing $O(1/N)$ between zeros.

One can obtain the standard deviation of the tunneling amplitude from Eqs.~(\ref{eq:T_sigma_relation}, \ref{eq:sigma_solution}), calculating $\pdv[]{t_N}{h_x}$ from the previously obtained expressions.  For a ring (Eq.~(\ref{eq:open_solution})) we get
\begin{equation}
	\braket{T_n^2} \approx \frac{1}{2 \pi}\left(\frac{J_x}{\abs{J_z}}\right)^{N-1} \frac{\left(h_x^{(n)}\right)^2 [\braket{\mu_i^2} + 4\braket{\eta_i^2}]}{\left[1 - \left(\frac{h_x^{(n)}}{2\sqrt{J_x \abs{J_z}}}\right)^2 \right]^{1/2}}.
\end{equation}
Again, we see that $T_n \sim h_x^{(n)}$ for small transverse field, while
it grows quickly as the largest zero $h_x^{(\floor{N/2})} = 2\sqrt{J_x \abs{J_z}}$ is approached.

\section{Tunneling in single spin models}%
\label{sec:tunneling_in_single_spin_models}

Here we consider the single spin models as presented in Sec.~\ref{sub:single_spin_model}. Similarly to an Ising ferromagnet, the unperturbed ($\lambda=0$) ground states correspond to the two states with $S_z=\pm S$. Applying our method to this Hamiltonian is rather straightforward: We simply calculate the tunneling matrix element from $-S$ to the $+S$ ground state to lowest order in $\lambda$, using intermediate tunneling coefficients $c_m$, where $m$ refers to the spin projection onto the $z$-axis, with eigenstates defined by
\begin{equation}
	S_z \ket{m} = m \ket{m}.
\end{equation}
If we use the Hamiltonian in the form of Eq.~(\ref{eq:H_single_spin}), the recursion for $c_m$ will involve both $c_{m-1}$ and $c_{m-2}$, but unlike in the problem treated in Sec.~\ref{sub:tunneling_in_the_ising_model_with_transverse_field_and_exchange}, the matrix elements and the denominators involved in the recursion depend themselves non-trivially on $m$. The resulting recursion is hard to solve analytically. We can, however, simplify the recursion greatly by first performing a rotation in the $x-z$ plane:
\begin{equation}
	\label{eq:rotation_single_spin}
	\begin{aligned}
		S_x &= \cos{\alpha} S_x' + \sin{\alpha} S_z',\\
		S_z &= \cos{\alpha} S_z' - \sin{\alpha} S_x',\\
		S_y &= S_y',
	\end{aligned}
\end{equation}
where we choose $\alpha$ to satisfy
$
\tan^2{\alpha} = \lambda^2 J_x/\abs{J_z},
$
such as to kill the matrix elements
$
\braket{k| H |k-2}
$
between $S_z'$-eigenstates,
$
S_z'\ket{k} = k \ket{k}.
$
This yields the Hamiltonian in the rotated basis
\begin{equation}
	\begin{aligned}
	&H' = (J_z + \lambda^2 J_x)S_z'^2 +\lambda\sqrt{J_x\abs{J_z}}(S_z'S_x'+S_x'S_z')\\
	&-\lambda h_x\left(\sqrt{\frac{\abs{J_z}}{\lambda^2 J_x + \abs{J_z}}} S_x' + \sqrt{\frac{\lambda^2 J_x}{\lambda^2 J_x + \abs{J_z}}} S_z'\right) - \lambda h_y S_y'.
	\end{aligned}
	\label{eq:rotated_H}
\end{equation}
where we consider dominant ferromagnetic Ising coupling $J_z<0$.
In the perturbative regime, the rotation angle $\alpha$ is small and the ground states of $H'$ still have a large overlap with the two $S_z'$ eigenstates $\ket{\pm S}$. We now deal with the problem of calculating the tunneling
\begin{equation}
	\braket{S|H'|-S} = t_{2S} \lambda^{2S} + O(\lambda^{2S+1})
\end{equation}
between these two states up to order $\lambda^{2S}$. Since the matrix form of $H$ in the $S_z'$ basis is tridiagonal, only the off-diagonal terms proportional to $\lambda$ contribute to $t_{2S}$. Thus we write the Hamiltonian in the form
\begin{equation}
	H' = H_0' + \lambda V_1' + O(\lambda^2),
	\label{eq:H_lambda_expansion}
\end{equation}
where
\begin{align}
	H_0' &= J_z S_z'^2,\\
	V_1' &= \sqrt{J_x\abs{J_z}}(S_z'S_x'+S_x'S_z') - h_x S_x' - h_y S_y'.
\end{align}
The matrix elements of these operators are
\begin{align}
	\braket{k|H_0'|k} =& J_z k^2,\\
	\frac{\braket{k \abs{V_1'} k-1}}{\braket{k \abs{S_x} k - 1}} =& \sqrt{J_x \abs{J_z}} (2k - 1) - (h_x -i h_y),
\end{align}
where we used
$
\braket{k \abs{S_x} k - 1} = -i \braket{k \abs{S_y} k - 1},
$
and
\begin{equation}
	\braket{k \abs{S_x} k - 1} = 1/2\sqrt{S(S+1) - k(k-1)}.
\end{equation}
Now that the Hamiltonian is tridiagonal, there is a single tunneling path between $\ket{\pm S}$ of order $\lambda^{2S}$. Its contribution to $t_{2S}$ is just the product of all off-diagonal matrix elements $\braket{k|V_1'|k-1}$ divided by (minus) the energies of all intermediate excited states. The total tunneling amplitude is
\begin{align}
&t_{2S} = \frac{\prod\limits_{k=-S+1}^{S}\braket{k|V_1'|k-1}}{\prod\limits_{k=-S+1}^{S-1}\left[ \braket{S|H_0'|S}-\braket{k|H_0'|k} \right]}\\
	\nonumber &= \frac{ \prod\limits_{k=-S+1}^{S} \left[ \sqrt{J_x \abs{J_z}} (2k - 1) - (h_x - i h_y) \right] \braket{k \abs{S_x} k - 1} }{\prod\limits_{k=-S+1}^{S-1} J_z (S^2 - k^2)}.
\end{align}
Note that this already takes the form of a factorized polynomial, with zeros given by the equation
\begin{equation}
	h_x -i h_y = \sqrt{J_x \abs{J_z}} (2n - 1),
	\label{eq:single_spin_zeros_condition}
\end{equation}
where
$
n = -S+1,-S+2,\dots,S.
$
Recall that  we took the $x-$axis to be the hard axis $(J_x>0)$. We find $2S$ equally spaced ground state crossings for $h_y=0$ and transverse fields
\begin{equation}
	h_x^{(n)} = \sqrt{J_x \abs{J_z}} (2n - 1).
	\label{eq:single_spin_zeros_perturbation}
\end{equation}
This is the same qualitative behavior as in a ferromagnetic Ising cluster as seen in Sec.~\ref{sub:generic_transverse_couplings}, cf. Eqs.~(\ref{eq:zeros_ising_ferro_x}, \ref{eq:zeros_ising_ferro_y}). This result coincides in lowest order with the zero positions as calculated by A. Garg~\cite{Garg1993} and as demonstrated experimentally by Wernsdorfer~\cite{Wernsdorfer1999}. However, in the rotated frame it becomes quite simple to obtain the exact degeneracies of this model, as we show in the next subsection.

\subsubsection{Exact degeneracies of single spin model}%
\label{sub:exact_degeneracies_of_rotated_hamiltonian}

If we are only interested in the location of the zeros, the rotated Hamiltonian can be used beyond perturbation theory to determine the position of the zeros exactly. Indeed, a zero of the tunneling matrix element $\braket{S|H'|-S}$ occurs whenever one of the off-diagonal entries of $H'$ becomes zero. At that point, the Hamiltonian splits into two uncoupled blocks for $S_z\geq k$ and $S_z\leq k-1$, implying that the up-state is strictly decoupled from the down-state to all orders. This entails an exact double degeneracy of the ground state (which map onto each other upon rotation by $\pi$ around the $x$-axis). Now, the off-diagonal matrix element is given by
\begin{equation}
	\frac{\braket{k \abs{H'} k-1}}{\braket{k \abs{S_x} k - 1}} = \sqrt{J_x \abs{J_z}} (2k - 1) - \left( h_x \sqrt{\frac{\abs{J_z}}{J_x+\abs{J_z}}} -i h_y \right).
	\label{eq:aa}
\end{equation}
Since $J_x>0$, we can again have a ground state degeneracy only for a transverse field along $x$. The critical fields are determined by the exact condition
\begin{equation}
	h_x^{(k)} = \sqrt{J_x(J_x + \abs{J_z})} (2k - 1),
	\label{eq:single_spin_zeros_exact}
\end{equation}
which agrees with the perturbative result of Eq.~(\ref{eq:single_spin_zeros_perturbation}) to lowest order in $J_x$, and reproduces the non-perturbative path integral results by Garg~\cite{Garg1993}. Here we have shown that this yields the location of the zeros exactly, independently of the size $S$ of the spin.

The result that the transverse field has to be applied along the hard axis is fully consistent with what we found for FM spin clusters in Sec.~\ref{sub:generic_transverse_couplings}. The main difference between the exact cluster calculation and the single spin model consists, however, in the precise location of the zeros. For the single spin model, the zeros are equally spaced, while for clusters they are spaced more and more densely the larger the transverse field, as one can see, e.g. in Fig.~\ref{fig:DPT_int_periodic}, or read off from the analytical result in Eq.~(\ref{eq:periodic_zeros}).

\section[Other systems with competing tunneling channels]{Other systems with competing tunneling channels}%
\label{sec:other_systems_with_competing_tunneling_channels}

The mechanism we have studied here, namely the interference of parallel multi-step tunneling channels between an initial and a final state is very general in nature and appears in various physical contexts.

A famous example is the case of resonant single-particle tunneling via several intermediate sites, a problem introduced by Nguyen, Spivak and Shklovskii~\cite{Shklovskii1985_1,Shklovskii1985_2}, with comprehensive reviews given in Refs.~\cite{Shklovskii1991,Kardar2007}.

For free particles (non-interacting fermions), different paths from an initial to a final site contribute with an amplitude whose sign alternates with the number of intermediate sites whose energy is above the chemical potential. This leads to negative interference between alternative paths. A magnetic field introduces additional Aharonov-Bohm phases and decreases the likelihood of full negative interference, resulting in increased transmission, that is, negative magnetoresistance~\cite{Sivan1988,Zhao1991}. Recently, it was found that the equivalent question for hard core interacting bosons leads to a similar interference problem, where, however, at energies close to the chemical potential all path amplitudes contribute with the same sign, leading to maximally constructive interference.~\cite{Syzranov2012,Muller2013} This situation resembles that of a ferromagnetic cluster with transverse field applied in the direction in which the transverse exchange is more ferromagnetic (i.e., the softer axis). In contrast to the magnetic clusters, however, in these hopping problems it is very hard or even impossible to tune a parameter (e.g., the magnetic field or the chemical potential) to suppress the tunneling completely.

Competing tunneling terms also arise in more general magnetic clusters composed of electronic and nuclear spins, a situation that frequently occurs in rare earth compounds. The magnetic ions are coupled to their nuclear spins, while the electronic spins couple to each other via dipolar and/or exchange couplings. Clusters of such ions often have doubly degenerate ground or excited states, which are only split by higher order tunneling processes that involve the interference of transverse fields, exchange/dipolar interactions and hyperfine couplings, that generically contribute with competing signs. Tuning the transverse field often allows to induce zeros in the corresponding collective tunneling. Similarly, the tunneling of the spin associated with a crystal field doublet of a magnetic ion can under certain circumstances be suppressed by a transverse field applied at specific angles, if different channels involving the magnetic field and transverse crystal field terms compete.

Ground state crossings have also been reported in SU(2) invariant, gapped frustrated spin chains~\cite{Natalia2017,Natalia2018}. In that case, the crossings are related to the interaction between the edge states of the chain. This is reminiscent of the explanation of the level crossings in the model of Eq.~(\ref{eq:H_ising_model}) in terms of Majorana edge states~\cite{Gregoire2017}, and it is natural to ask whether these crossings can also be seen as a consequence of destructive interferences between different channels. For that purpose, let us consider the level crossings in the bilinear-biquadratic spin-1 chain~\cite{Natalia2018},
$
H=\sum_i J_1 (S_i \cdot S_{i+1}) + J_b (S_i \cdot S_{i+2})^2.
$
If one adds a strong uniaxial anisotropy along $z$, one may work with respect to an AF ground state, and the transverse terms with $\Delta S_z = 2$ and $4$ have competing signs if $J_b>0$, presumably leading to level crossings similar to those of the isotropic case. It would be interesting to see if a more direct connection can be established by studying the effective coupling between the edge states starting from the AKLT model $J_b=J_1/3$ for which the edge states are fully decoupled in the ground state~\cite{AKLT}. This goes beyond the scope of the present paper, however.

\section[Final Remarks]{Summary and outlook}%
\label{sec:final_remarks}

High order degenerate perturbation theory allows us to understand transverse field zeros in terms of negatively interfering tunneling paths, which in turn is tied to the presence of competing quantum fluctuations in the Hamiltonian. Our method nicely applies to 1D systems, where the tunneling can be obtained for any system size exactly, in contrast with 2D and 3D clusters where the number of different connected clusters grows exponentially with system size. Overall, the results support the existence of $N$ zeros in some region of the parameter space independently of the geometry.

The original model (Eq.~(\ref{eq:H_ising_model})) can be further extended by staggering the field or by adding exchange couplings along the $y$-axis while keeping the crossings. As we saw in Sec.~\ref{sub:generic_transverse_couplings}, systems with FM ground states exhibit zeros only when the field is applied along the "hard axis" (the one with the strongest antiferromagnetic or the weakest ferromagnetic coupling). In contrast, AFM clusters on a bipartite lattice exhibit suppressed tunneling on approximate circles in the transverse field plane. This may make AFM cluster ground states more attractive since the tunneling suppression is more resistant to fluctuations in the orientation of the applied field. This ability to control and suppress the quantum fluctuations in small magnetic clusters or single molecule magnets is indeed considered an important goal~\cite{Sorensen2018}.

Introducing disorder in the exchange couplings and in the transverse fields, the crossings change position but do not disappear. The latter only happens when in ferromagnets $J_x-J_y$, or in antiferromagnets $J_x+J_y$, starts to change sign and turn negative. The relation between a set of $J_x$, either randomly generated or carefully chosen, and the resulting number of crossings remains to be studied more deeply.

A certain amount of disorder in the exchange is always to be expected from static sources such as lattice imperfections, strain, or dynamically due to slow phonons. Spatial inhomogeneities can also induce $g$-factor variations that lead to an effective disorder in the transverse field. In an ensemble of weakly disordered clusters it is thus impossible to suppress the tunneling simultaneously in all clusters, and even in a single cluster temporal fluctuations of the parameters will destroy the perfect negative interference of competing tunneling channels. The best strategy to suppress the tunneling as much as possible consists then in tuning the transverse field to the first (smallest) zero, $h_x^{(n=1)}$, corresponding to the average exchange coupling in the system. The disorder-induced fluctuations away from vanishing tunneling turn out to be smallest under those conditions. This is closely related with the fact that the location of this smallest transverse field zero moves the least as the parameters of the Hamiltonian are slightly perturbed. Hence this zero seems to be the most interesting one for most applications.

Our recursive calculation of collective tunneling amplitudes generalizes nicely to simpler single spin models, and the ground state crossings in this model can be interpreted with the same tunneling interference argument. Given the $2S$ crossings of a single spin, one may expect that an appropriately chosen spin-S model on a lattice of $N$ spins will exhibit $2SN$ crossings.

It is an interesting question to ask what happens to the zeros as one leaves the perturbative regime. In the ferromagnetic single spin model we can trace them easily, since we can obtain them exactly. If the hard axis is along the $x$-axis and one tunes $J_y$ up to and beyond $J_z$ for example, the number of $2S$ zeros remains intact, even though the easy axis has undergone a flop from the $z$- to the $y$-axis.
In lattice models, the zero lines in the $h_x$-$J_x$ plane do not seem to disappear either. They even may cross quantum phase transition lines, as long as they enter a new phase with a degenerate ground state or a gapless phase. The study of the related phenomena and implications is left for future work.

\appendix

\section[Deduction of method]{Deduction of method}%
\label{app:deduction_of_method}

Following Bloch's recipe~\cite{Bloch1958} we consider the effective Hamiltonian $\Heff$ projected onto the unperturbed ground state subspace $g = \{\emptyset,\Sigma\}$ by the projection operator $P$. It takes the form
\begin{equation}
	\Heff = PHP + P \left( \sum_{n=2}^{\infty} \sum_{\{k_i\}} V S_{k_1} \dots V
	S_{k_{n-1}} \right) V P,
	\label{eq:Heff_definition}
\end{equation}
where $n$ specifies the order in $V$ of the term. For a given $n$, we sum over all $(n-1)-$tuples of $k_i = 0,1,\dots$ that obey
\begin{equation}
	k_1 + k_2 + \dots + k_{s} \ge s,
	\label{eq:Heff_k_condition_1}
\end{equation}
for all $s = 1,\dots, n-2$, and
\begin{equation}
	k_1 + k_2 + \dots + k_{n-1} = n-1.
	\label{eq:Heff_k_condition_2}
\end{equation}
The operator $S_k$ is defined as
\begin{equation}
	S_k = \left\{
		\begin{array}{lr}
			-P = -\sum_{m \in g}\ket{m}\bra{m}, & k = 0,\\
			\frac{1-P}{{(\en_{\emptyset} - H_0)}^k} =
			\sum_{m \not\in g} \frac{\ket{m}\bra{m}}{{(-\Den_m)}^k}, & k \geq 1, \\
	\end{array} \right.
\end{equation}
where
$
\Den_m = \en_m - \en_{\emptyset}.
$
The eigenvalue equations read
\begin{equation}
	\Heff P \ket{\Psi_{\pm}} = E_{\pm} P\ket{\Psi_{\pm}},
\end{equation}
where $\ket{\Psi_{\pm}}$ are the lowest energy eigenstates of $H$. Due to the symmetry $R$ (Eq.~(\ref{eq:R_symmetry_definition})) we may write the eigenstate projections to leading order as
\begin{equation}
	P\ket{\Psi_{\pm}} = \ket{\emptyset} \pm \ket{\Sigma} +O(\lambda).
	\label{eq:Psi_i_projections}
\end{equation}
The relevant matrix element to calculate is
\begin{equation}
	t\equiv \bra{\emptyset} \Heff \ket{\Sigma}.
\end{equation}
Upon expanding
$
V = \lambda V_1 + \lambda^2 V_2
$
in Eq.~(\ref{eq:Heff_definition}) and substituting in $t$, we group terms according to their powers of $\lambda$. Taking $n_1$ ($n_2$) to be the number of $V_1$ ($V_2$) operators present in a term of orders $n$ and $u$, we have
\begin{equation}
	n_1 + n_2 = n,\quad n_1 + 2n_2 = u,
\end{equation}
from which we find what orders of $n$ contribute to $u$ by taking the limiting cases of $n_1 = {\rm mod}(u,2)$ and $n_2 = 0$. Finally, we only need to sum over the permutations of $V_1$ and $V_2$ that respect the order $u$. Applying this to $t$, we have
\begin{equation}
	t = \sum_{u=1}^{\infty} \lambda^u \sum_{n=\ceil{\frac{u}{2}}}^{u}
	\sum_{\{k_i\},\{l_i\}} \bra{\emptyset}V_{l_1} S_{k_1} \dots V_{l_{n-1}} S_{k_{n-1}}
	V_{l_{n}} \ket{\Sigma},
	\label{eq:t_full_development}
\end{equation}
where the $l_i = 1,2$ obey
$
l_1 + l_2 + \dots l_n = N.
$
It helps to look at the calculation of the matrix element in Eq.~(\ref{eq:t_full_development}) sequentially; that is, starting with the extreme left operator, we apply each operator to the states on its left. While $V_1$ and $V_2$ always transform the states $\bra{m}$ they act on, $S_k$ mainly acts as a projector onto a subspace of states, either the ground state $g$ (if $k=0$), or the excited states (if $k>0$). Now, the treatment of $V_2$ as a second order perturbation is crucial for our method and physically justified by the fact that the basic action of $V_2$ on $\ket{m}$ is to flip pairs of neighboring spins while $V_1$ flips single spins. In the term of order $\lambda^u$, $\bra{\emptyset}$ is acted on with enough $V_1$'s and $V_2$'s to at most flip $u$ spins. Since we need a minimum of $N$ spin flips to transform $\emptyset$ into $\Sigma$, it follows that the lowest order is $\lambda^N$. After applying a $V_l$ to the states on its left, the resulting states must have $l$ more spin(s) flipped than before for such terms to yield a non-zero contribution to order $\lambda^N$. In particular, this means a projection onto $g$ by $S_0 = -P$ would only give terms that will eventually have zero contribution. This imposes $k_i>0$ in the leading term $O(\lambda^N)$. However, the constraint~(\ref{eq:Heff_k_condition_2}) only allows for one single choice of the $k_i$, namely, $k_i = 1$ for all $i$. Using this information in Eq.~(\ref{eq:t_full_development}), and writing $t = t_N \lambda^N + O(\lambda^{N+1})$, we now have
\begin{equation}
	t_N = \sum_{n=\ceil{\frac{N}{2}}}^{N} \sum_{\{l_i\}} \bra{\emptyset}V_{l_1}
	S \dots V_{l_{n-1}} S V_{l_{n}} \ket{\Sigma}
	\label{eq:t_development_appendix}
\end{equation}
where $S \equiv S_1$. This proves Eq.~(\ref{eq:t_development}) in the main text.

Now, we shall prove the recursion relations in Eq.~(\ref{eq:c_m_recursion}), starting from the definition of the intermediate tunneling coefficients
\begin{equation}
	c_{m} \equiv \sum_{n=\ceil{\frac{\abs{m}}{2}}}^{\abs{m}} \sum_{\{l_i\}} \bra{\emptyset}V_{l_1} S \dots V_{l_{n-1}} S V_{l_{n}} S \ket{m},
	\label{eq:c_m_definition_appendix}
\end{equation}
where
\begin{equation}
	l_1 + l_2 + \dots l_n = \abs{m},
\end{equation}
and we remind the reader that $\abs{m}$ is the number of spins of $\ket{m}$ that are flipped relative to $\ket{\emptyset}$. Summing over $l_n$ and redefining
$
n \rightarrow n-1,
$
we have
\begin{equation}
	\begin{aligned}
		c_{m} = &\sum_{n=\ceil{\frac{\abs{m}}{2}} -1}^{\abs{m}-1} \sum_{\{l_i\}} \bra{\emptyset}V_{l_1} S \dots V_{l_{n}} S V_1 S\ket{m}\\
				+ & \sum_{n=\ceil{\frac{\abs{m}}{2}} -1}^{\abs{m}-1} \sum_{\{l_i'\}} \bra{\emptyset}V_{l'_1} S \dots V_{l'_{n}} S V_2 S\ket{m},
	\end{aligned}
	\label{eq:c_m_expansion_1}
\end{equation}
where
\begin{equation}
	l_1 + \dots l_n = \abs{m}-1,\quad l'_1 + \dots l'_n = \abs{m}-2.
\end{equation}
By expanding $V_{1,2} S\ket{m}$ in Eq.~(\ref{eq:c_m_expansion_1}), we shall see that we recover the cluster coefficients of smaller clusters. Consider first $V_1 S \ket{m}$. We have that
\begin{equation}
	S\ket{m} = \frac{1}{-\Den_m}\ket{m},
\end{equation}
which follows from the definition of $S$. Then, applying $V_1$ to the state $\ket{m}$, we get a sum over states $m'$ which differ by one spin flip from $m$. However, only $m'$ clusters with $|m'|=|m|-1$  yield a non-zero contribution to $c_m$. Thus,
\begin{align}
&\sum_{n=\ceil{\frac{\abs{m}}{2}} -1}^{\abs{m}-1} \sum_{\{l_i\}} \bra{\emptyset}V_{l_1} S \dots V_{l_{n}} S V_1 S\ket{m} \nonumber\\
	=& \sum_{n=\ceil{\frac{\abs{m}}{2}} -1}^{\abs{m}-1} \sum_{\{l_i\}} \bra{\emptyset} V_{l_1} S \dots V_{l_{n}} S \sum_{\mathclap{\underset{|m'|=|m|-1}{m'}}} \ket{m'}\bra{m'} V_1 S\ket{m} \nonumber\\
	=& \sum_{\underset{|m'|=|m|-1}{m'}} c_{m'}\bra{m'} V_1 S\ket{m},
\end{align}
With an analogous argument applied to $V_2 S\ket{m}$ we obtain the recursion relations~(\ref{eq:t_N_general_recursion}, \ref{eq:c_m_general_recursion}) in the main text. The expression for $t_N$ follows from an analogous derivation, the only difference being that there is no insertion of the operator $S$ at the last step, as one can note from comparing Eq.~(\ref{eq:t_development_appendix}) and Eq.~(\ref{eq:c_m_definition_appendix}). This eliminates the corresponding energy denominator.

\section[Cluster independence]{Cluster independence}%
\label{app:cluster_independence}

Consider a cluster $C$ which is composed of two (dis)connected clusters of flipped spins $A$ and $B$, by which we mean that the excitation energy of cluster $C$ is the sum of independent excitation energies,
\begin{equation}
	\label{eq:Esum}
	\Den_C = \Den_A + \Den_B,
\end{equation}
Here we want to prove the relation among intermediate tunneling coefficients:
\begin{equation}
	c_{C} = c_{A}c_{B},
	\label{eq:2_cluster_independence}
\end{equation}
which one should expect to hold because to leading order we can simply reduce the Hamiltonian to the parts acting on either $A$ or $B$ and drop all other terms, so that the flipping of $A$ and $B$ are independent processes.

This assertion is conveniently proved by induction on the size of the cluster $C$. We suppose that we have proved it for a small cluster of size up to $|C|-1$. For size $0$ and $1$ the assertion is trivial. The general recursion formula (\ref{eq:c_m_general_recursion}) shows that
\begin{equation}
\label{genrec}
	c_C = -\frac{1}{\Delta\en_C} \sum_{m'} c_{m'} \bra{m'} V \ket{C},
\end{equation}
where we sum over the $m'$ obeying $\abs{m'} = \abs{C}-1$ or $\abs{m'} = \abs{C}-2$. Let us now write $C= A\cup B$, and $m' = A'\cup B$ or $m' = A \cup B'$, depending on where $V$ acts. Thus:
\begin{eqnarray}
	c_C &= -\frac{1}{\Delta\en_C} &\left(\sum_{A'\subset A} c_{A'\cup B} \bra{{A'}\cup B} V \ket{A \cup B} \right. \\
		&& \left. +\sum_{B' \subset B} c_{A\cup B'} \bra{{A}\cup B'} V \ket{A \cup B} \right)\nonumber\\
		&=- \frac{1}{\Delta\en_C} &\left(\sum_{A' \subset A} c_{A'} c_B \bra{{A'}} V \ket{A} \right. \\
		&& \left. + \sum_{B' \subset B} c_{A} c_{B'} \bra{B'} V \ket{ B} \right),\nonumber
\end{eqnarray}
where in the second line we used the induction hypothesis for smaller clusters, which implies that $c_{A'\cup B} = c_{A'} c_B$. Now we use the relation (\ref{genrec}) in the form
\begin{equation}
	\sum_{A'} c_{A'} \bra{{A'}} V \ket{A} = -\Delta\en_A c_A,
\end{equation}
and an analogous expression for $c_B$. Together with (\ref{eq:Esum}), this proves the relation (\ref{eq:2_cluster_independence}).


\acknowledgements{} This work has been supported by the Swiss National Science Foundation and the Portuguese Science and Technology Foundation through the grant SFRH/BD/117343/2016.

\bibliographystyle{apsrev4-1}
\bibliography{bib}

\end{document}